\documentclass[iop]{emulateapj}

\usepackage{subfigure,footmisc,array}

\newcommand{\mum}{$\mu$m}

\shorttitle{The RCS231953 Supercluster}
\shortauthors{Faloon et al.}

\begin{document}

\title{The structure of the merging RCS 231953+00 Supercluster at $z \sim 0.9$}

\author{A.J. Faloon\altaffilmark{1}, T.M.A. Webb\altaffilmark{1},  E. Ellingson\altaffilmark{2}, R. Yan\altaffilmark{3}, David G. Gilbank\altaffilmark{4}, J.E. Geach\altaffilmark{1}, A.G. Noble\altaffilmark{1}, L.F. Barrientos\altaffilmark{5}, H.K.C. Yee\altaffilmark{6}, M. Gladders\altaffilmark{7} and J. Richard\altaffilmark{8}}

\altaffiltext{1}{Department of Physics, McGill University, 3600 Rue University, Montr\'eal, QC, H3A 2T8, Canada}
\altaffiltext{2}{Center for Astrophysics and Space Astronomy, Department of Astrophysical and Planetary Science, UCB-389, University of Colorado, Boulder, CO, 80309, USA}
\altaffiltext{3}{Center for Cosmology and Particle Physics, Department of Physics, New York University, 4 Washington Place, New York, NY, 10003, USA}
\altaffiltext{4}{South African Astronomical Observatory, P.O.\,Box 9, Observatory, 7935, South Africa}
\altaffiltext{5}{Departamento de Astronom\'ia y Astrof\'isica Pontifica Universidad Cath\'olica de Chile, Vicu\~na MacKenna 4860, 7820436 Macul, Santiago, Chile}
\altaffiltext{6}{Department of Astronomy \& Astrophysics, University of Toronto, 50 St.\,George St., Toronto, ON, M5S 3H4, Canada}
\altaffiltext{7}{Department of Astronomy \& Astrophysics, University of Chicago, 5640 S.\,Ellis Ave, Chicago, IL, 60637, USA}
\altaffiltext{8}{Observatoire de Lyon, Centre de Recherche Astrophysique de Lyon, Universit\'e Lyon, 19 Avenue Charles Andr\'e, 69561 Saint-Genis-Laval, France}

\begin{abstract}
The RCS\,2319+00 supercluster is a massive supercluster at $z=0.9$ comprising three optically selected, spectroscopically confirmed clusters separated by $<3$ Mpc on the plane of the sky.  This supercluster is one of a few known examples of the progenitors of present-day massive clusters ($10^{15}\,M_\odot$ by $z\sim0.5$).  We present an extensive spectroscopic campaign carried out on the supercluster field resulting, in conjunction with previously published data, in 1961 high confidence galaxy redshifts.  We find 302 structure members spanning three distinct redshift walls separated from one another by $\sim65$ Mpc ($\Delta\,z=0.03$).  The component clusters have spectroscopic redshifts of 0.901, 0.905 and 0.905.  The velocity dispersions are consistent with those predicted from X-ray data, giving estimated cluster masses of $\sim10^{14.5}\,-\,10^{14.9}\,M_\odot$.  The Dressler-Shectman test finds evidence of substructure in the supercluster field and a friends-of-friends analysis identified 5 groups in the supercluster, including a filamentary structure stretching between two cluster cores previously identified in the infrared by \citet{Coppin2012a}.  The galaxy colors further show this filamentary structure to be a unique region of activity within the supercluster, comprised mainly of blue galaxies compared to the $\sim$43-77\% red-sequence galaxies present in the other groups and cluster cores.  Richness estimates from stacked luminosity function fits results in average group mass estimates consistent with $\sim10^{13}\,M_\odot$ halos.  Currently, 22\% of our confirmed members reside in $\gtrsim10^{13}\,M_\odot$ groups/clusters destined to merge onto the most massive cluster, in agreement with the massive halo galaxy fractions important in cluster galaxy pre-processing in N-body simulation merger tree studies.

\end{abstract}


\section{Introduction}

In a universe dominated by hierarchical structure formation, galaxies accrete along high density filaments and sheets to form groups and clusters.  Superclusters represent the largest density enhancements in this cosmic web and can reach scales of $\sim 100-200~h^{-1}$ Mpc, with close, aligned clusters having the densest filamentary cluster-cluster bridges \cite[e.g.][]{Bond1996, Bahcall1983, Wray2006}.

As superclusters represent the largest component of the large-scale structure in the universe they can be used to test cosmological models \citep[e.g.][]{Bahcall1988, Kolokotronis2002, Einasto2011}.  They also offer a range of spatial and dynamical sub-environments, from infalling galaxies, groups and filaments to the high-density cluster cores, and are therefore ideal laboratories in which to study the effects of the environment on galaxy evolution.  In the nearby universe superclusters have long been accessible for study \citep[e.g.][]{Gregory1978, Gregory1981, Oort1983} and catalogs of superclusters in the low-redshift universe have been compiled through the large spectroscopic campaigns done by the Sloan Digital Sky Survey (SDSS) \citep[e.g.][]{Einasto2002, Liivamagi2012} and the 2 degree Field Galaxy Redshift Survey (2dFGRS) \citep[e.g.][]{Einasto2008}.  

The study of superclusters in the high redshift universe, however, remains elusive.  In N-body simulations of hierarchical structure formation, the density of superclusters in the universe decreases with redshift, with few rich, compact superclusters found at high redshift \citep[e.g.][]{Wray2006}.  Though the number of superclusters found by \citet{Wray2006} decreased from $z \sim 0$ to $z \sim 1$ by only a factor of a few, their mean cluster separation increases, making them difficult to find observationally since typical high redshift survey sizes are not large enough to recognize these immense structures and superclusters on smaller scales become more rare. 

To date, only a handful of spectroscopically confirmed superclusters have been found at high redshift.  The CL0016 supercluster at $z = 0.55$ extends over 20 Mpc and includes three clusters along with companion groups and filamentary structure \citep{Tanaka2009}.  At $z \sim 1$ three superclusters of varying scales have been reported.  \citet{Lubin2000} identified a supercluster containing ten spectroscopically confirmed cluster and group candidates within a radial length of $\sim100~h^{-1}_{70}$ Mpc and transverse length of $\sim 13~h^{-1}_{70}$ Mpc at $z = 0.9$ \citep{Gal2004, Gal2008}.  The \citet{Swinbank2007} supercluster at $z = 0.89$, found in the UK Deep Infrared Sky Survey (UKIDSS) Deep eXtragalactic Survey (DXS), contains five clusters ($\sim 10^{13.5-14.0}\,M_{\odot}$) spread over 30 Mpc on the sky.  The RCS\,2319+00 supercluster was first spectroscopically confirmed by \citet{Gilbank2008} and contains three massive clusters at $z = 0.9$ with X-ray masses of $M_{X,tot} \sim 4.7-6.4\times10^{14}~\textup{M}_\odot$ \citep{Hicks2008}.  The highest redshift supercluster confirmed to date is the Lynx Supercluster of two X-ray emitting clusters at $z = 1.26$ and $z = 1.27$, first discovered by \citet{Rosati1999}, with seven additional candidate groups and clusters identified through photometric redshifts \citep{Nakata2005}.

In this paper we present the results of a large spectroscopic campaign on the RCS\,2319+00 supercluster field at $z \sim 0.9$.  These data will form the basis of a series of upcoming papers studying the effects of the high redshift supercluster environment on its member galaxies.  We aim to map out the substructure of the supercluster to provide a spectroscopically confirmed catalog of members in the regions of various densities in the cluster field for further studies on the galaxy populations. The paper is organized as follows.  In \S\ref{sec:2319} we outline the currently published information available on the RCS\,2319+00 supercluster.  Section \ref{sec:spec} details the extensive spectroscopic campaigns that have been completed on the supercluster while \S\ref{sec:Cat} summarizes the output of the redshift catalog.  In \S\ref{sec:clusterProps} we attempt to quantify the cluster properties including the redshifts, velocity dispersions, virial radii and masses of the three main component clusters.  We further trace the substructure of the supercluster in \S\ref{sec:structure}.  The recovered structure is used to investigate the color-magnitude relation and galaxy color distribution in different density environments in \S\ref{sec:CMR}.  In \S\ref{sec:LF} we attempt to place constraints on the masses of the potential infalling groups identified in \S\ref{sec:structure} and put them in context with current theories of group infall in clusters.  Finally, a summary is offered in \S\ref{sec:summary}.
Throughout this paper we assume cosmological parameters of $\Omega_{\Lambda}$ = 0.73, $\Omega_m$ = 0.27, and $H_o$ = 71~km~s$^{-1}$ \citep{Larson2011}.

\section{The RCS\,2319+00 Supercluster}\label{sec:2319}

RCS\,231953+0038.1 (hereafter referred to as cluster A) at $z \sim 0.9$ was identified in the first Red-Sequence Cluster Survey \cite[RCS-1;][]{Gladders2005} as a remarkable strong-lensing cluster \citep{Gladders2003}.  Subsequent investigation of the RCS-1 catalog identified two potential companion clusters (RCS\,232003+0033.5 -- cluster B and RCS\,231946+0030.6 -- cluster C) whose red-sequence peaks fall at a similar photometric redshift to cluster A.  The density profiles of the red-sequence galaxies indicate that the three clusters are separated by $<$\,3\,Mpc in the plane of the sky with each cluster core seemingly aligned towards its closest neighbor, further substantiating the claim that these clusters are bound to each other in a supercluster (as found by \citet{Binggeli1982} in a study of Abell clusters).

\textit{Chandra} X-ray imaging confirmed the presence of three X-ray luminous cluster cores with $L_X \sim 3.6-7.6\times10^{44}~\textup{erg s}^{-1}$ and $M_{X,tot} \sim 4.7-6.4\times10^{14}~\textup{M}_\odot$ \citep{Hicks2008}.  Spectroscopic confirmation that the three clusters all lie at $z \sim 0.9$ was first shown in \citet{Gilbank2008}.  Using the methods of \citet{Sarazin2002} to estimate the merger rate, \citet{Gilbank2008} postulate that the three clusters will merge to form a Coma-like cluster ($\sim 10^{15}\,M_{\odot}$) by $z \sim 0.5$.

As the two other confirmed $z \sim 0.9$ superclusters of \citet{Swinbank2007} and \citet{Lubin2000} have lower mass cluster components that are spread over larger scales, they are therefore less likely to merge into as massive a cluster as RCS\,2319+00 \citep{Gilbank2008}. Thus RCS\,2319+00 is a rare example of a massive cluster caught in the process of forming through the merging of multiple cluster and group components and is an ideal laboratory in which to study the role of the environment and hierarchical processes in galaxy evolution.

Extensive follow-up is underway on the RCS\,2319+00 supercluster to study the effect of this unique environment on its member galaxies.  Cluster A has already been studied in surveys of weak-lensing in massive galaxy clusters \citep{Jee2011}, the Sunyaev-Zel'dovich (SZ) effect of strong lensing clusters \citep{Gralla2011} and the submillimetre source counts of high-redshift clusters \citep{Noble2012}.  The X-ray properties of the three component clusters of the RCS\,2319+00 supercluster have been investigated by \citet{Hicks2008}.  \citet{Coppin2012a} discovered an infrared-bright filament of galaxies connecting component clusters A and B with a total SFR\,$\simeq$\,900\,$M_\odot$\,yr$^{-1}$.  This data paper is the first in a series to further investigate the hierarchical build-up of large scale structure at $z \sim 1$ through the RCS\,2319+00 supercluster environment.  

\section{The Spectroscopic Data}\label{sec:spec}

\begin{figure}[htb]
\centering
{\includegraphics[width=9 cm, angle=0, trim=0cm 1cm 1cm 0.5cm, clip=true]{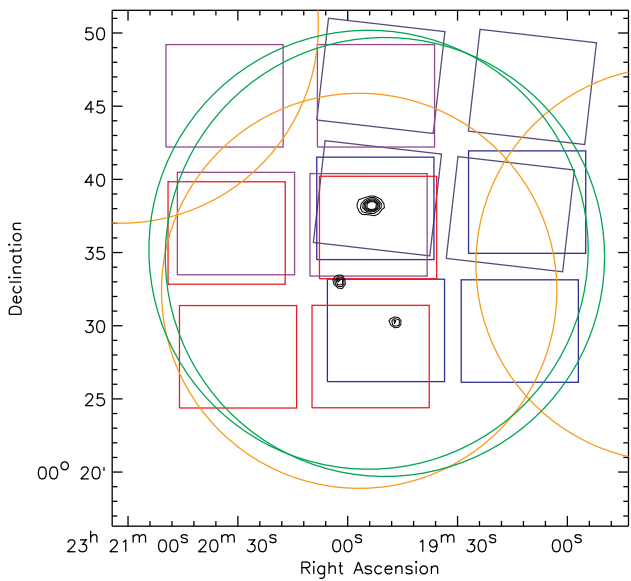}}
\vspace{-0.5cm}
\caption{\small Spectroscopic mask coverage centered on $23^\textup{h}19^\textup{m}52^\textup{s}.8 +00^{\circ}34'12''$ over a 35 $\times$ 35 arcminute region with the three component cluster positions identified by the X-ray contours from \citet{Hicks2008}.  Rectangles denote the four VIMOS pointings with each color representing the four quadrants of a single VIMOS observation.  The orange circles show the three IMACS positions and green circles represent the FMOS fov.  The coverage with GMOS and FORS2 is not shown.  Both instruments were centered on cluster A, with GMOS targets covering $\sim$1 square arcminute and FORS2 targets over a $\sim$6 square arcminute region.\label{fig:specLayout}}
\vspace{0.1cm}
\end{figure}

\begin{table*}
\begin{center}
\caption{\small Spectroscopic mask observations on the RCS\,2319+00 supercluster field. \label{tab:spec}}
\begin{tabular}{ccccccc} 
\tableline
Instrument	&	Mask	& RA		&	DEC	&	Date/ID 	&	Total Int time (s)	&	\# slits        			\\
\tableline					
IMACS	&	231953+A	&	23 19 53.00	&	+00 38 00.00	&	October 29, 2006	&		7200		&	722	\\
IMACS	&	231953+B	&	23 19 53.00	&	+00 38 00.00	&	October 9, 2007	&		7200		&	708	\\
IMACS	&	231953+C	&	23 19 53.00	&	+00 38 00.00	&	October 10, 2007			&		7200		&	644	\\
IMACS	&	2319NE		&	23 18 31.00	&	+00 34 12.00	&	October 9, 2007			&		7200		&	751	\\
IMACS	&	2319W		&	23 21 02.00	&	+00 50 30.00	&	October 10, 2007			&		7200		&	740	\\
VIMOS	&	RCS2319P1			&	23 19 32.93	&	+00 42 26.10	&	July 9, 2005		&		2500		&	341	\\
VIMOS	&	RCS2319P2			&	23 19 32.43	&	+00 33 55.44	&	August 13, 2005	&		2500		&	337	\\
VIMOS	&	RCS2319P3			&	23 20 13.13	&	+00 32 11.90	&	August 22, 2006	&		1920		&	355	\\
VIMOS	&	RCS2319P4			&	23 20 13.70	&	+00 41 09.60	&	October 16, 2006	&		2880		&	339	\\
FMOS	&	RCS2319P1			&	23 19 49.83	&	+00 34 41.90	&	October 16, 2011	&		7200		&	171 	\\		
FMOS	&	RCS2319P2			&	23 19 54.30	&	+00 35 11.60	&	October 16, 2011	&		7200		&	174 	\\
GMOS-N	&	RCS2319+0039 &	23 19:53.40	&	+00:38:13.80    &       September 30, 2003  &           10800       &       27     \\
FORS2	&	RCS231953+0038.0	&	23 19 53.00	&	+00 38 00.00	&	70.A-0378(B)	&	10800	&	29	\\
FORS2	&	R2319M01	&	23 19 53.00	&	+00 37 13.00	&	October 30, 2011	&	1800		&	21	\\
FORS2	&	R2319M01	&	23 19 53.00	&	+00 39 27.00	&	November 1, 2011	&	2700		&	20	\\
\tableline
\end{tabular}
\tablecomments{Units of right ascension are hours, minutes, and seconds and units of declination are degrees, arcminutes, and arcseconds.}
\end{center}
\end{table*}

The spectroscopic catalog emanates from six different observing runs with five different Multi-Object Spectrographs (MOS): Magellan-IMACS, VLT-VIMOS, Subaru-FMOS, Gemini-GMOS, and VLT-FORS2.  Table~\ref{tab:spec} summarizes the central coordinates of each of the spectroscopic masks, the dates or program ID of the observations, the total integration time, and the number of target slits per mask.   

The observation setup and data reduction techniques used for each instrument are detailed below.  Figure~\ref{fig:specLayout} shows the coverage of the IMACS, VIMOS and FMOS masks over the supercluster field.  The cluster positions are identified by their X-ray contours (see Figure 1 of Hicks et al.~2008 for a more detailed view of the X-ray flux distribution).  The GMOS and FORS2 fields of view are smaller and fall within the larger instrument masks, centered on cluster A.

\subsection{IMACS Data} \label{sec:imacs}

Wide field MOS was taken with the Inamori Magellan Areal Camera and Spectrograph (IMACS) on the Baade 6.5 meter Magellan telescope using the University of Toronto's share of Magellan time.  Three pointing positions of the 27' diameter IMACS field of view were used.   The data was taken as part of a large optical spectroscopic survey covering 42 RCS1 and 1 RCS2 clusters over eight observing runs from October 2004 to October 2007 (R.\,Yan et al.\,2012 in preparation).  Three masks were centered on the RCS\,2319+00 supercluster field (masks 231953+A, 231953+B, and 231953+C).  Two additional masks covering the nearby RCS\,231831+0034.2 (2319NE) and RCS\,232102+0050.5 (2319W) RCS 1 cluster candidates overlap the supercluster field region.  Each mask was observed in 4 $\times$ 1800s exposures with $\sim$ 650-750 slits per mask.  The g150 grism was used with the f/2 camera and 1 arcsecond slit widths, yielding a spectral resolution of R=555.  A custom band-limiting filter was used to restrict wavelength coverage to 6050-8580\,\AA~encompassing the [OII]$\lambda$3727 to G-band ($\sim$4304\,\AA) spectral features and slightly beyond at the supercluster redshift.  In total, there are 2138 unique IMACS spectroscopic targets in a 35 square arcminute region around the supercluster, with duplicates used to check the redshift accuracy of the IMACS sample (see R.\,Yan et al.\,2012, in preparation).   

The full details of the IMACS reduction and redshift finding are discussed further in the R.\,Yan et al.\,(in preparation) paper detailing the large IMACS RCS survey.  Briefly, the IMACS spectroscopy was reduced using a modified version of the DEEP2 DEIMOS reduction pipeline \citep{Davis2002}.  The redshift finding was done using the IDL-based $\mathsf{zspec}$ program, which uses chi-squared fitting to template spectra.  It was adapted from the DEEP2 software for use with the RCS IMACS spectroscopy sample by R. Yan.  All redshift assignments were checked by eye and assigned a quality flag: Q $\ge$ 3 for a secure redshift, Q = 2 for a potential redshift, Q = 1 for bad spectra,  and Q = -1 for stars.  Using high quality spectra of repeated targets, the redshift error for the IMACS sample was calculated to be $\Delta z\sim~0.000388$, or $\sim$60~km~s$^{-1}$ at $z \sim 0.9$. 

\subsection{VIMOS Data}\label{sec:vimos}

Optical spectroscopy was taken with the VIsible Multi-Object Spectrograph \citep[VIMOS;][]{LeFevre2003} on the 8.2 meter Unit Telescope 3 of the Very Large Telescope (VLT UT3).  VIMOS contains four detectors, each with a 7\,x\,8 arcminute fov, separated by 2 arcminute gaps between quadrants.  Four VIMOS pointings were designed to cover a $\sim$25 square arcminute region centered on RCS\,2319+00 cluster A (see Figure~\ref{fig:specLayout}).  At the time of the spectroscopic proposal, clusters B and C had not yet been identified.  No color or magnitude selection was applied to the mask design; however, the highest priority was given to sources with radio and submm counterparts \citep{Noble2012}.  The low resolution red grism (LR\_red), with a wavelength range of 5500-9500\,\AA, was used to cover the [OII]$\lambda$3727 to [OIII]$\lambda$5007 spectral features at $z \sim 0.9$.  This set-up resulted in a spectroscopic resolution of R=210 with a dispersion of 7.3\,\AA~pixel$^{-1}$.

The VIMOS spectroscopy was run through the standard VIMOS data reduction pipeline.  We decided to re-extract the 1D spectra from the 2D sky-subtracted, flat-fielded, wavelength-calibrated spectra in order to (a) verify that the correct primary object spectra was extracted when multiple sources appeared in a single slit, (b) improve the trace of faint continuum sources and (c) allow for better removal of contamination from cosmic rays, 0$^{th}$, -1$^{st}$ and 2$^{nd}$ order contamination and fringing in the spectra. The IRAF $\mathsf{noao}$ package was used to trace and extract the 1D spectra, with additional background subtraction performed and bad pixels replaced using model points.   Spectra of separate observations were then coadded with $\mathsf{imarith}$ and checked for any residual bad pixels, which were removed manually.  

To prepare the spectra for redshift finding the continuum was fitted and subtracted using $\mathsf{fit1d}$ and stacked with a weighted spectrum to remove the A-band absorption line.  Redshifts were found by first running the final 1D spectra through a cross-correlation program adapted for IDL \citep{Blindert2006} from the CNOC1 technique using five template galaxy spectra \citep{Yee1996} and then interactively checking and, if necessary, re-fitting the redshifts using an IDL-based redshift program written for use with the output of the cross-correlation program.  The redshifts were assigned the same quality flags as the IMACS data. The VIMOS redshifts carry errors of $\Delta z\sim0.002$ ($\sim$300~km~s$^{-1}$) based on the wavelength calibration and dispersion of the data, the widths of the correlation peaks and matched targets between the IMACS and VIMOS samples.  

\subsection{FMOS Data}\label{sec:fmos}

Near-Infrared spectroscopy was obtained with the Fiber Multi Object Spectrograph (FMOS) on the 8.2 meter Subaru telescope \citep{Kimura2010}.  FMOS contains 400 fibres with an accuracy of 0.2 arcsec RMS in a 30 arcminute diameter field of view.  Two pointings were centered on the supercluster field.  We used the Cross-Beam-Switching (CBS) mode, which uses nodding for better sky subtraction by allocating two fibre spines per target: one that is initially on target and one that is initially on the nearby sky.  This results in only $\lesssim$200 science targets per pointing but allows each target to be observed for 100\% of the exposure time and provides more accurate sky subtraction.  We used the Low Resolution mode (R=600, $\Delta\lambda\sim20$\,\AA) with a wavelength range of 0.9-1.8\mum, which allows coverage of H$_{\alpha}$ (6563\,\AA) and the SI ($\sim$6716/6730\,\AA) doublet features and beyond at the supercluster redshift.  Spectroscopic targets were chosen based on our CFHT WIRCam J and Ks catalogs, with priority going to mid-infrared 24\mum~counterparts from our Spitzer MIPS catalog (T.M.A.\,Webb et al.\,2012, submitted).  Previously identified (Q$\ge$3) and suspected (Q=2 redshifts) luminous infrared galaxy (LIRG) supercluster members from the IMACS and VIMOS redshift lists were targeted for further confirmation of their redshifts and for use in further studies of their spectral features.  Overall 345 targets, including 121 targets not previously observed by other instruments, were observed with FMOS. 

The FMOS spectroscopy was run through the FMOS preliminary data reduction and calibration package, FIBRE-pac: FMOS Image-Based REduction Package \citep{Iwamuro2011}, which uses IRAF scripts and C programs based on CFITSIO to produce 1D spectra.  The reduction includes flat-fielding, bad pixel removal, spatial distortion correction, spectral distortion correction, background subtraction, bias correction, wavelength calibration and flux calibration.  The resulting spectra have a dispersion of 5\,\AA~pixel$^{-1}$.  Final 1D spectra were prepared in the same way as the VIMOS spectra for use with the cross-correlation and redshift confirmation programs, with redshifts having the same quality flags and errors as the previous data.  The redshift errors were again taken to be $\Delta z \sim 0.002$ to account for the wavelength dispersion and correlation peak widths.  For any target with a previous, low confidence spectra from IMACS or VIMOS, the estimated FMOS redshift was checked with the lower wavelength range spectra in an effort to confirm the FMOS redshift and/or increase the confidence of the previous redshift.  As FMOS is still in shared risk mode and the FIBRE-pac reduction package is still in development, the data were noisier than anticipated, despite photometric observing conditions, with only $\sim$30\% redshift success.  

\subsection{GMOS North Data}\label{sec:gmos}

Spectra from the 5.5 arcminute field of view Gemini Multi-Object Spectrograph (GMOS) on the 8.1 meter Gemini North Telescope was obtained in an effort to identify the redshift for the strong lensing arc in cluster A \citep{Gilbank2008}. The data was observed in the band-shuffle nod-and-shuffle mode, resulting in 27 targets in a $\sim$1 square arcminute region.  The R150\_G5306 red grism was used with a band-limiting filter (GG455 G0305) and 2x2 detector binning for a wavelength range of $\sim$4500-10000\,\AA, allowing coverage of [OII]$\lambda$3727 to [OIII]$\lambda$5007 at $z \sim 0.9$ and giving a resolution of 11.48\,\AA~and a dispersion of 3.564\,\AA~pixel$^{-1}$.  

The GMOS spectroscopy was reduced in IRAF with the help of the $\mathsf{gemini}$ package tasks to bias-subtract, flat-field, and wavelength-calibrate the data.  The positive and negative spectra for each object from the nod-and-shuffle observations were coadded and 1D spectra were extracted using the IRAF $\mathsf{apall}$ task with residual bad pixels removed manually.  The spectra were prepared and redshifts found as in the VIMOS and FMOS data, with the same errors and quality flags assigned.   

\subsection{FORS2 Data}\label{sec:fors2}

Spectroscopy taken with the FOcal Reducer and low dispersion Spectrograph (FORS2) of the Very Large Telescope was obtained from two separate observation runs.  In the first observation run (mask RCS231953+0038.0) spectra of 29 target galaxies was taken as part of the larger spectroscopic follow-up of RCS-1 clusters.  FORS2 was used in MXU multi-object spectroscopy mode with the 300I grism in conjunction with the OG590 filter to give a dispersion of 108\,\AA~mm$^{-1}$ and wavelength coverage from 6000-11000\,\AA.  The target galaxies were selected primarily using a fixed magnitude limit.  The observations totalled three hours per mask, including overheads, split into individual observations to improve cosmic ray rejection.  Redshifts were found using cross-correlation with the IRAF $\mathsf{rvsao}$ package and interactively checked and, where necessary, refitted using the $\mathsf{emsao}$ task.  Quality flags have been adjusted to be consistent with our larger redshift catalog.  

The second FORS2 data set consisted of a total of 43 slits over two masks (R2319M01 and R2319M02) taken as a follow up of Herschel selected sources from the Herschel Lensing Survey \citep{Egami2010}.  The data were taken using the 300V grism combined with the GG435 filter to give a wavelength coverage of 445-865\,\AA~with a 112\,\AA~mm$^{-1}$ dispersion.  Redshifts were found manually using the peak of the [OII]$\lambda$3727 emission line.  The targets in this run comprised mainly background galaxies lensed by cluster A of the RCS2319+00 supercluster, however six fell within the supercluster redshift range, with three having no previous spectroscopy.  

\section{The RCS\,2319+00 Redshift catalog}\label{sec:Cat}
\subsection{The combined spectroscopic data set}\label{sec:redCat}

\begin{figure*}[hbt!]
\centering
\vspace{-0.25cm}
\subfigure{\includegraphics[height=17 cm, angle=90, trim=0.55cm 0.25cm 0.25cm 0.1cm, clip=true]{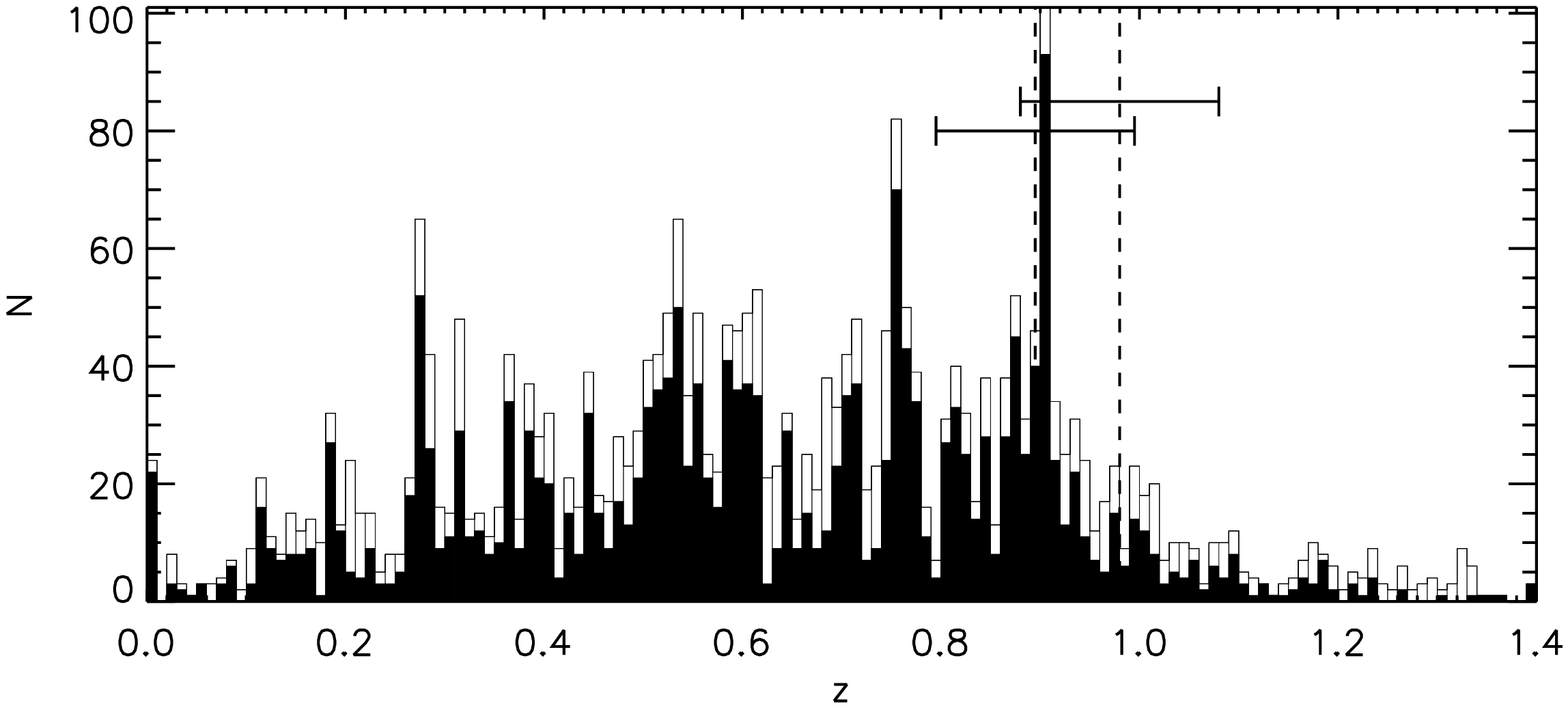}}
\hspace{-1cm}
\subfigure{\includegraphics[height=17 cm, angle=90,  trim=0.5cm 0.25cm 0.75cm 0.25cm, clip=true]{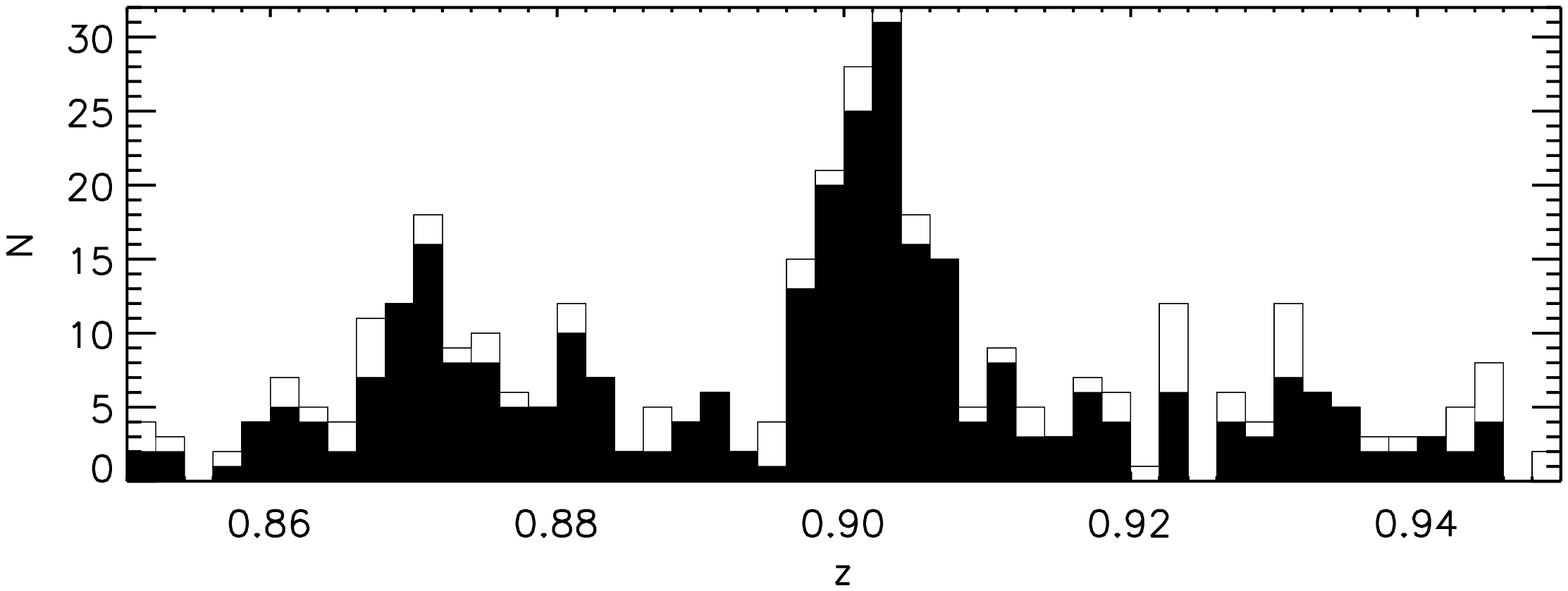}}
\caption{\small \textit{Top}: Spectroscopic redshift distribution in the 35\,x\,35 arcminute region around the RCS\,2319+00 supercluster field with a bin size of $z=0.01$ ($\sim1500$ km s$^{-1}$ at $z \sim 0.9$).  The filled histogram represents good confidence redshifts while the white histogram shows all redshift results.  Potential foreground structures are visible in the histogram, as well as a clear peak at $z \sim 0.9$.  The dashed lines indicate the photometric redshifts for the three component clusters of RCS\,2319+00 (A \& B: $z_{phot,RS}$ = 0.984, C: $z_{phot,RS}$ = 0.895; L.F.\,Barrientos et al.\,2012, in preparation).  The 1$\sigma$ error on the RCS-1 photometric redshifts at $z \sim 0.9$ is $\Delta z \sim 0.1$ and is shown by the horizontal error bars.  \textit{Bottom}: Zoomed in redshift histogram around the supercluster redshift of $z \sim 0.9$ with a bin size of $z=0.002$ ($\sim300$ km s$^{-1}$ at $z \sim 0.9$), consistent with the largest redshift error in our catalog.\label{fig:zhist}}
\end{figure*}

In total, our spectroscopic campaign in the 35\,x\,35 arcminute field around the RCS\,2319+00 supercluster contains 3305 unique spectroscopic targets.  Where there are duplicate targets present between masks or different instruments, the redshift with the higher quality factor is used.  If the quality factor is the same between spectrographs, then priority is given to IMACS redshifts since it is the instrument with the smallest error.  In total, there are 66 duplicate targets between the IMACS and VIMOS samples that have confident redshifts with both instruments.  Of these, 9\% had redshifts that did not fall within 3$\sigma$ of each other.  This implies that, in the worst case scenario, 8 of the VIMOS structure members listed in Table~\ref{tab:zCat} have been mis-identified.  The average redshift error on the IMACS/VIMOS matched galaxies is $\Delta z \sim 0.002$.  The redshifts of overlapping targets between the combined IMACS/VIMOS spectroscopic catalog and the FMOS data were found with the aid of the previous spectra and consequently they all match within errors.  The redshifts of the duplicate sources in the FORS2 spectroscopy matched to the combined redshift list to within $\Delta z \sim 0.001$ and the two duplicates from the GMOS sample matched to within $\Delta z \sim 0.003$.

The full redshift distribution of the RCS\,2319+00 supercluster region is shown in the top panel of Figure~\ref{fig:zhist}.  Our catalog contains 2000 good confidence redshifts, with 1961 galaxy spectra and 39 stars.  There is a clear peak at the supercluster redshift of $z \sim 0.9$ along with potential foreground peaks at $z\sim$ 0.28, 0.45, 0.53 and 0.76.  The photometric redshifts of the RCS\,2319+00 clusters, as identified by overdensities of red galaxies in this field (L.F.\,Barrientos et al.\,2012, in preparation), are shown as dashed lines in the top panel of Figure~\ref{fig:zhist} (A \& B: $z_{phot,RS}$ = 0.984, C: $z_{phot,RS}$ = 0.895).  The red-sequence redshifts were found to carry errors of $\Delta z \lesssim 0.05$ over the majority of the RCS-1 survey range, increasing to $\Delta z \sim 0.1$ at the highest redshifts due to larger uncertainties in the photometric data and poorer sampling of the 4000\,\AA~break in the R$_c$ and z$'$ bands.  

\begin{figure}[htb]
\centering
{\includegraphics[width=9cm, trim=0.cm 0cm 0cm 0.5cm, clip=true]{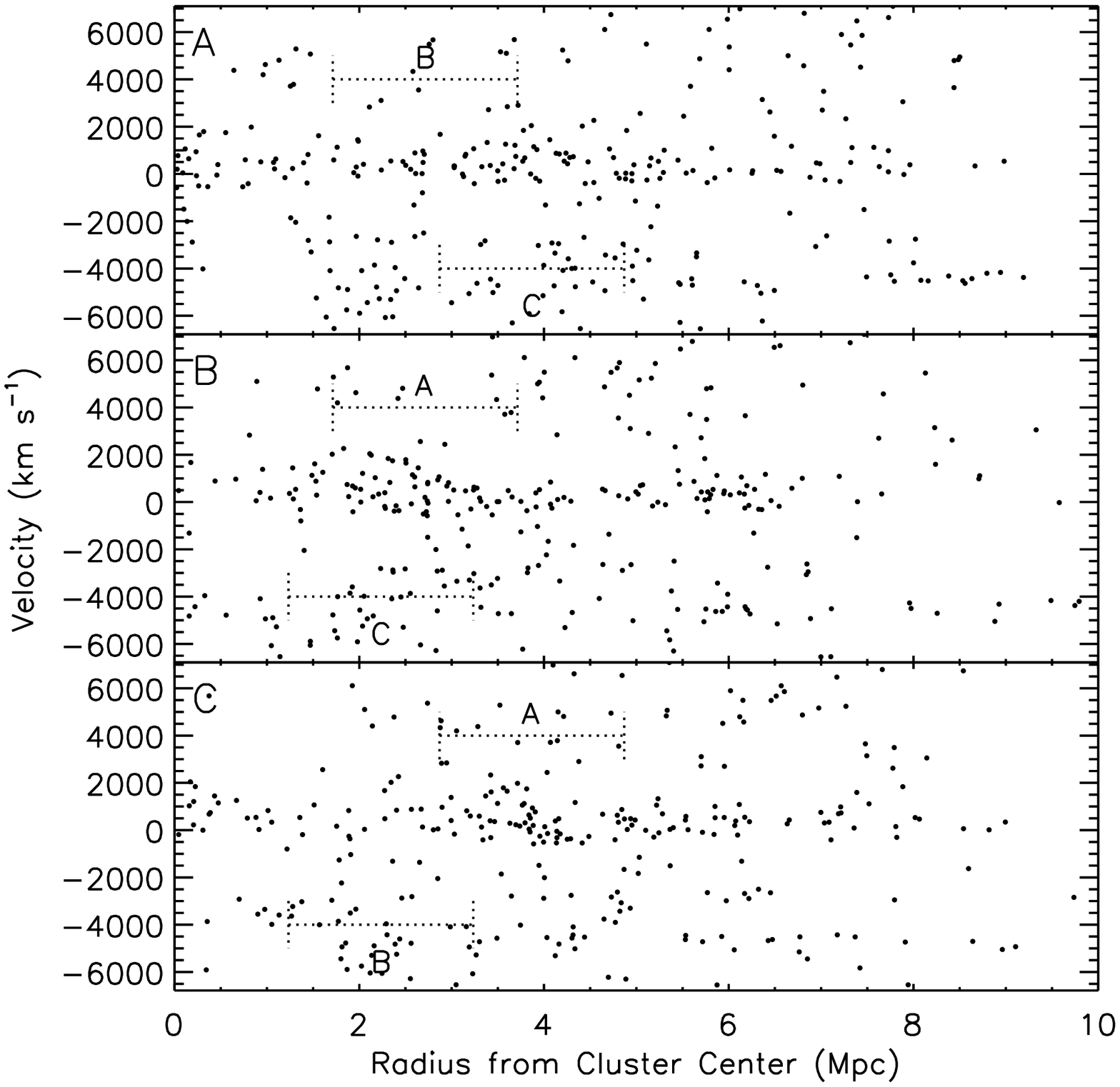}}
\vspace{-0.5cm}
\caption{\small Velocity versus radius of all spectroscopically confirmed galaxies in the RCS 2319+00 supercluster field within the redshift range $0.848 \le z \le 0.946$.  In each panel the galaxies' radial positions are plotted relative to the specified cluster center.  All velocities are plotted relative to an arbitrary $z = 0.9$.  The dotted lines denote the $\pm$ 1 Mpc radii for the neighboring cluster centers relative to their radial position from the center of the specified cluster  to demonstrate the difficulty in assigning cluster membership to specific cluster cores.  \label{fig:vel-rad_all}}
\vspace{0.1cm}
\end{figure}

The bottom panel of Figure~\ref{fig:zhist} shows the zoomed in redshift histogram covering a redshift range of $0.85~\le~z~\le~0.95$.  Three distinct redshift peaks are apparent, separated by $\sim 4700$ km\,s$^{-1}$ ($\Delta z \sim 0.03$) from one another, $\sim 65$ Mpc assuming Hubble flow.  To determine membership in the RCS\,2319+00 supercluster field we use a broad membership range of $0.858~\le~z~\le~0.946$ based on the bottom panel of Figure~\ref{fig:zhist}.  This is done to encompass any infalling groups, filaments or walls in redshift space.  This is the range over which we search for substructure in the supercluster and may contain isolated galaxies not connected to the supercluster structures (see \S\ref{sec:structure}).  

A different view of this redshift range is shown in Figure~\ref{fig:vel-rad_all}, where each of the three panels represents the velocity versus radius of all spectroscopically confirmed galaxies from the center of each cluster relative to $z = 0.9$.  The dotted lines indicate the positions of the $\pm$ 1 Mpc radii from the neighboring cluster centers relative to their radial position from the center of the specified cluster.  The radial proximity of the clusters highlights the difficulty in assigning cluster membership over the field to specific cluster cores. The three separate velocity walls can be seen in all three panels at $z \sim 0.87, 0.90, 0.93$.  These redshift sheets may or may not belong to a single,  gravitationally bound structure.  For the purpose of investigating substructure around the supercluster region, we retain all spectroscopically confirmed galaxies within the broader redshift range.  

\begin{table}[tb!]
\centering
\scriptsize
\caption{Spectroscopic target numbers in the RCS 2319+00 supercluster field}
\begin{tabular}{cccccc}
\tableline
 Spectroscopic & Targets & Secure   & Stars & Galaxies &  Structure \\
 Source &   &   Redshifts  &   &   &  Members \\
\hline
\hline
IMACS &         2133 &        1325 &          33 &        1292 &         185 \\
VIMOS &          962 &         562 &           6 &         556 &          85 \\
FMOS &          181 &          87 &      0 &          87 &          25 \\
GMOSN &           14 &          11 &      0 &          11 &           4 \\
FORS2 &           15 &          15 &      0 &          15 &           4 \\
\tableline
\textbf{TOTAL} &         \textbf{3305} &        \textbf{2000} &          \textbf{39} &        \textbf{1961} &         \textbf{302} \\
\tableline
\end{tabular}
\label{tab:zCat}
\tablecomments{All numbers listed are for unique spectroscopic targets only.  Any duplicate targets have been attributed to the instrument with the highest redshift quality and/or the lowest redshift error.}
\end{table}

Table~\ref{tab:zCat} gives a summary of the number of unique spectroscopic targets and the good confidence redshifts for each spectrograph along with the number of good confidence stars, galaxies and overall structure members.  For the rest of this paper we use only secure spectroscopic redshifts.  We find 302 members in the broad supercluster range of $0.858~\le~z~\le~0.946$.  

\subsection{Spectroscopic completeness and success rates}\label{sec:specComp}

\begin{figure*}[tb!]
\centering
\subfigure{\includegraphics[width=6.1 cm, trim=0.5cm 0cm 0cm 0cm, clip=true]{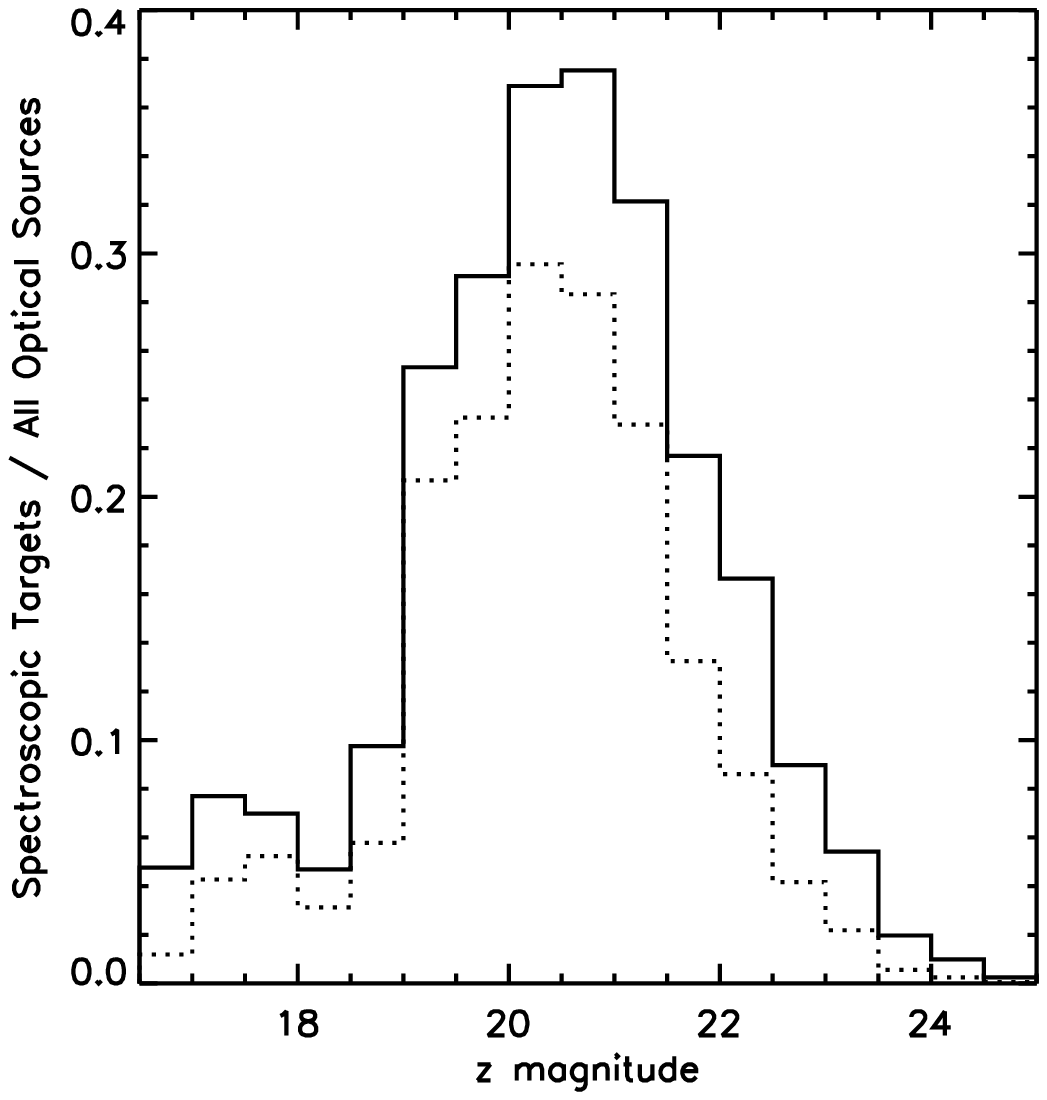}}
\hspace{-0.4 cm}
\subfigure{\includegraphics[width=6.1 cm, trim=0.5cm 0cm 0cm 0cm, clip=true]{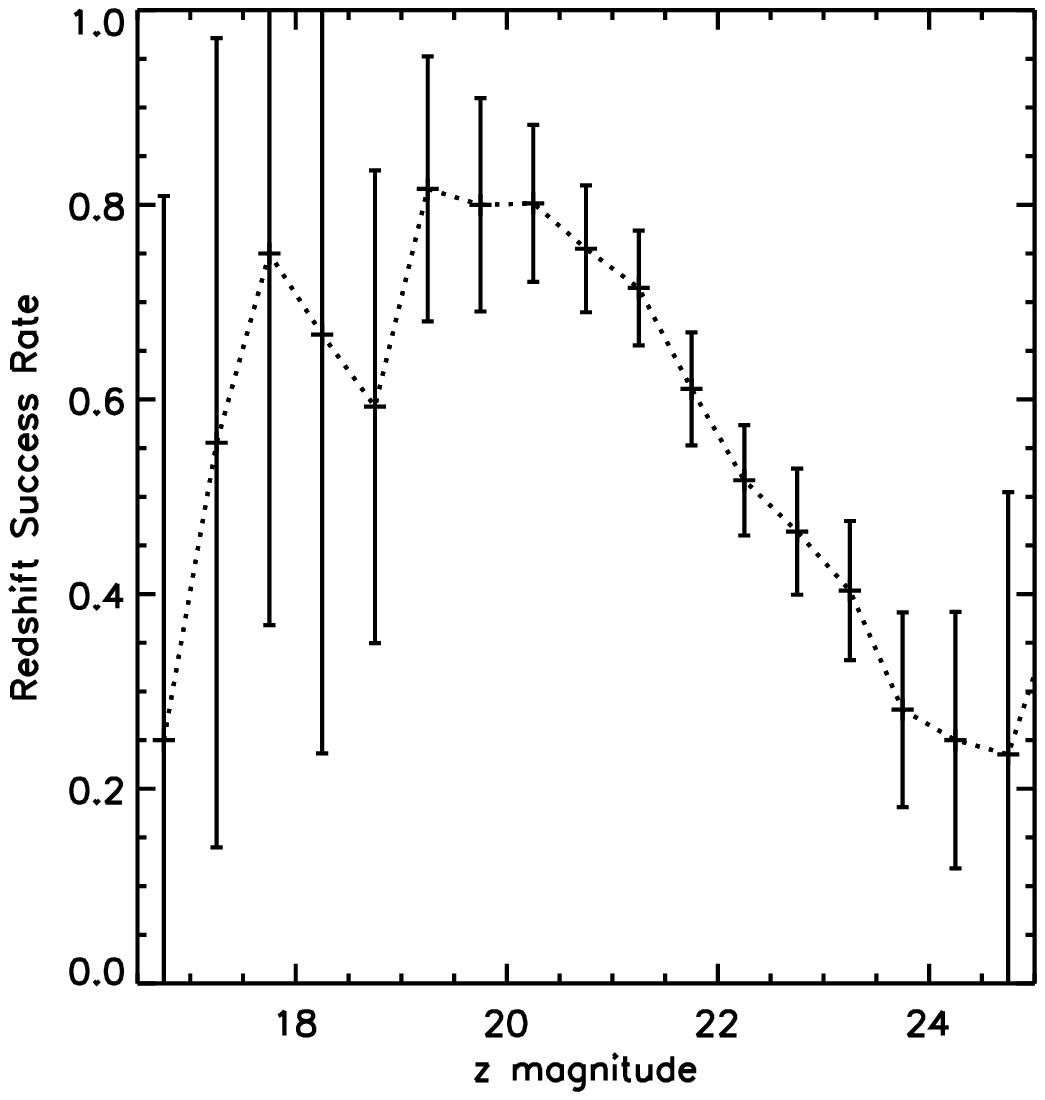}}
\hspace{-0.4 cm}
\subfigure{\includegraphics[width=6.1 cm, trim=0.5cm 0cm 0cm 0cm, clip=true]{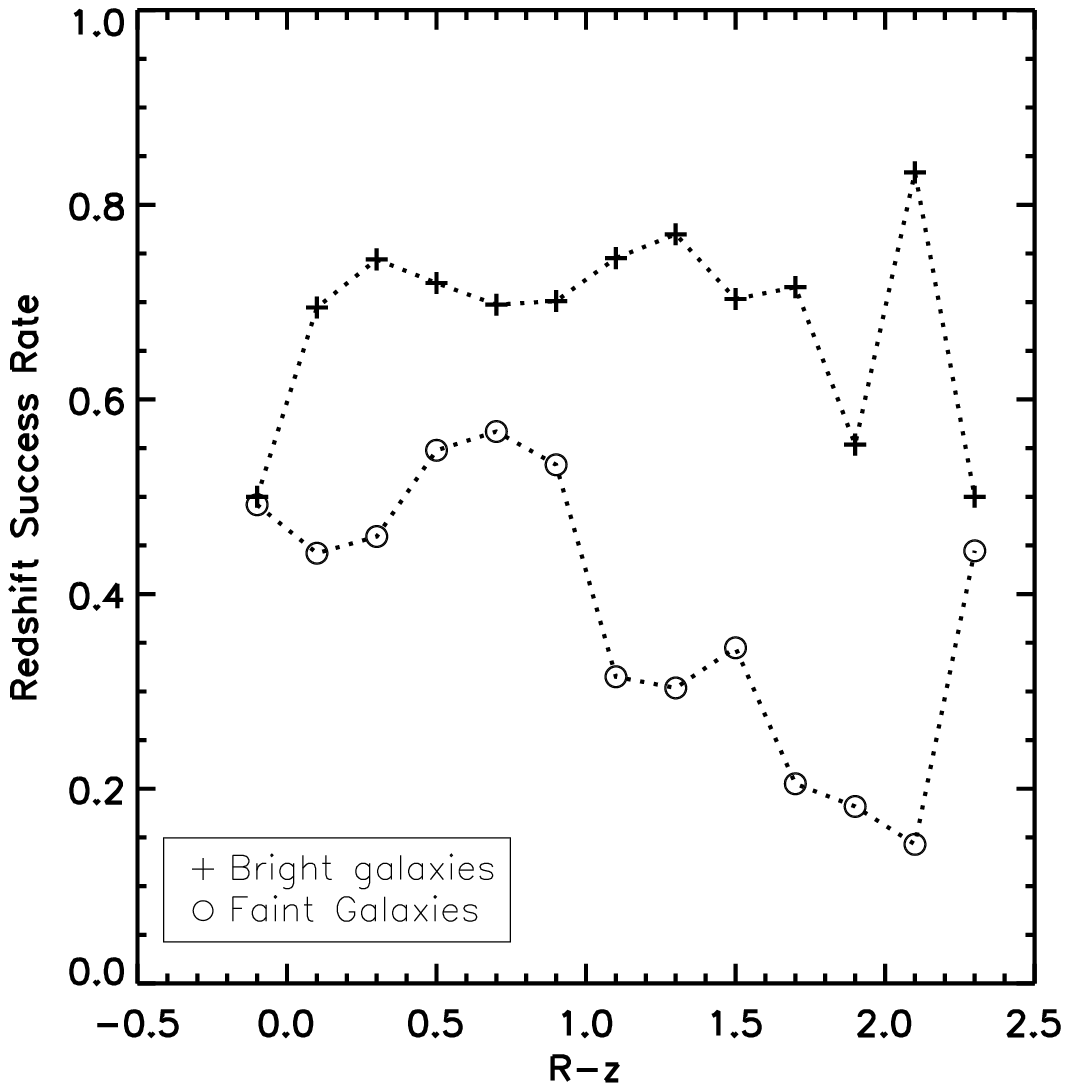}}
\caption{\small \textit{Left}: Spectroscopic target coverage completeness for the 35$\times$35 arcminute catalog region.  The solid line shows spectroscopic targets representing all slits, including those that did not yield reliable redshifts, compared to the full optical catalog with 5$\sigma$ depth of z$'$ = 24.05 mag.  The dotted line shows the coverage completeness when only the secure redshifts are taken into account.  \textit{Middle}: Spectroscopic redshift success for all targets based on z$'$ magnitude.  \textit{Right}: Redshift success based on R$_c$--z$'$ color for bright (z$'~\le~$22 mag) and faint (z$'~>~$22 mag) targets.\label{fig:specSuccess}}
\end{figure*}

Due to the large survey area of the supercluster field, targeting all sources with spectroscopy is unrealistic.  The 5$\sigma$ magnitude limits of the RCS survey are z$' = 24.05$ mag and R$_c = 24.7$mag \citep{Gladders2005}, resulting in costly time overheads to obtain high S/N spectroscopy for the faintest sources in our catalog.  From Figure~\ref{fig:specLayout} it can be seen that the spectroscopic coverage of the three cluster cores varies, with cluster A having the most mask coverage from all instruments including the additional GMOS and FORS2 data.  Cluster B has the least amount of spectroscopic coverage, with its core falling outside or along the edges of the VIMOS mask coverage.  

The completeness of the spectroscopic target coverage over the entire optical catalog for magnitude bins of 0.5 in z$'$ magnitude over the 35\,x\,35 arcminute area is shown in the left panel of Figure~\ref{fig:specSuccess}.  The solid line shows the target completeness for all spectroscopic targets, regardless of redshift success, while the dashed line shows only those targets that yielded a secure redshift. Despite our 3305 independent spectroscopic targets, our highest target completeness, for 20--21 magnitude sources, is only $\sim$37\%, decreasing to $\sim$30\% coverage for secure redshifts.   

Spectroscopic selection functions and weights were calculated for each source with a spectroscopic redshift following the method used in the two Canadian Network for Observational Cosmology (CNOC) surveys (see \citet{Yee1996,Yee2000} for a detailed explanation).  Briefly, the magnitude selection function ($S_m$) of each spectroscopically confirmed supercluster member is found by taking all galaxies within the photometric catalog in a magnitude bin centered around the desired galaxy and calculating the ratio of galaxies in that bin having a measured redshift to those without a redshift.  The geometric   distribution selection function ($S_{xy}$), used as a correction to the primary magnitude selection function, was also computed.  The geometric selection function corrects for different target densities over the larger field and uses only sources within 2 arcmins of the desired galaxy, within $\pm 0.5$mag bins to compute a local magnitude selection function.  This is then divided by the previously found $S_m$. The spectroscopic weight of each galaxy is the inverse of its total selection function ($S = S_mS_{xy}$). 

The spectroscopic redshift success rates for the targeted sources is shown in the middle panel of Figure~\ref{fig:specSuccess} and represents all instruments combined, with $\sim$80\% of the brighter targets yielding confident redshifts and success rates dropping to only $\sim$30\% at our 5$\sigma$ depth of z$'$ = 24.05mag. The S/N, wavelength dispersion and reduction methods all effect the quality of the final spectra and therefore the redshift success rate.  Most spectra were not taken under photometric conditions and the S/N varies between different spectra from the same instrument based predominantly on the weather conditions throughout the observations.  The success rate of bright (z$'~\le~$22mag) and faint (z$'~>~$22 mag) sources, based on R$_c$--z$'$ color is also shown (Figure~\ref{fig:specSuccess}, right).  For bright sources, the redshift success is fairly consistent at 70-80\% for all colors, with the lowest and highest color bins affected by small numbers.

\section{Cluster Properties}\label{sec:clusterProps}

\subsection{Determining the proper cluster centers}\label{sec:center}

\begin{figure*}[tb!]
\centering
{\includegraphics[width=19 cm, angle=0, trim=0.25cm 1cm 0.25cm 0.25cm, clip=true]{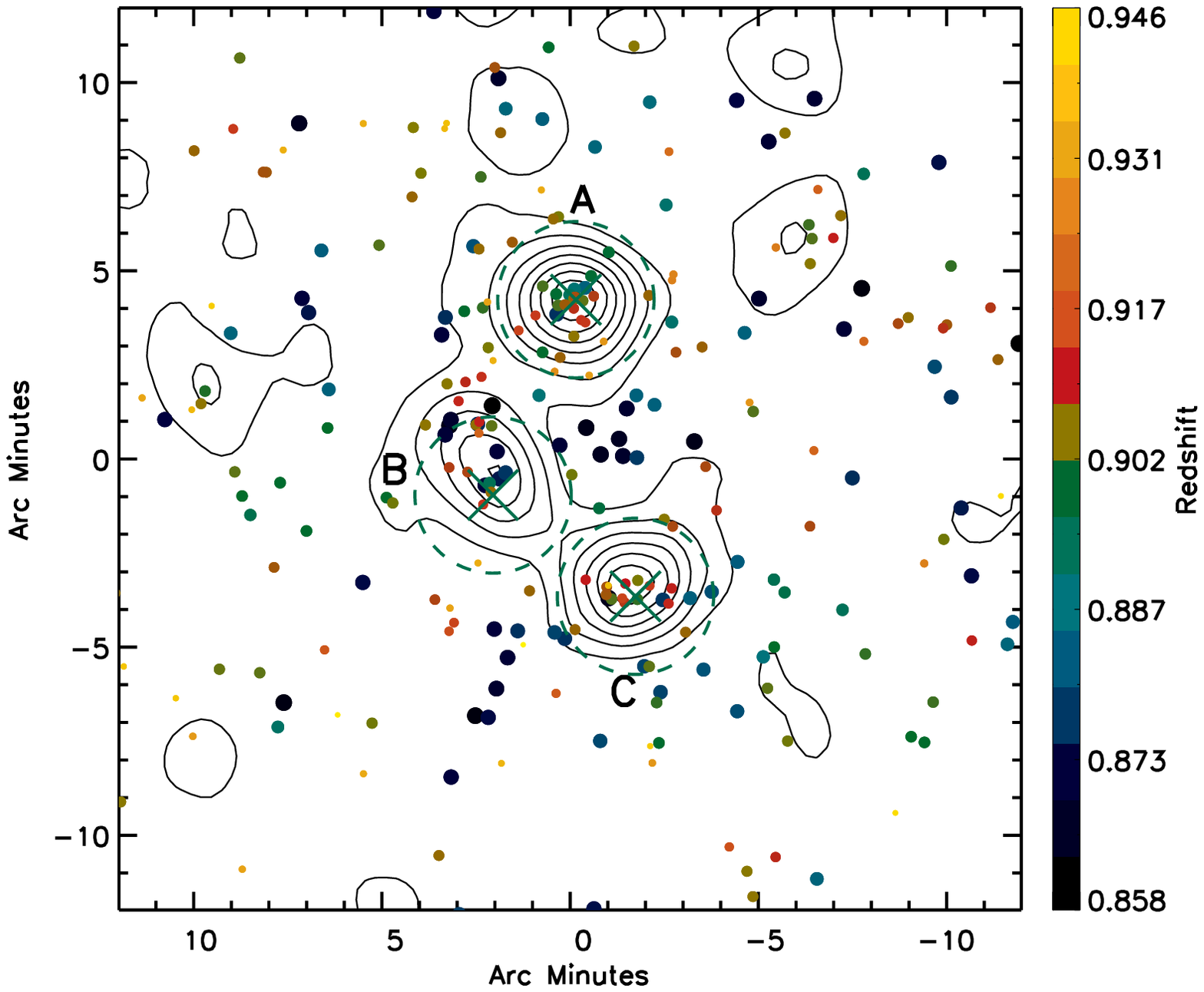}}
\caption{\small Redshift distribution of RCS\,2319+00 supercluster members ($0.858~\le~z~\le~0.946$) centered on $23^\textup{h}19^\textup{m}52^\textup{s}.8 +00^{\circ}34'12''$ and plotted over the significance contours of the red-sequence galaxy density profile at a photometric redshift of $z \sim 0.9$.  Spectroscopically confirmed members are plotted as circles, the size and color varying with redshift, with circle size decreasing with redshift and the colors varying from black through to yellow with increasing redshift.  The dashed green circles denote the 1.0 Mpc radii around each clusters X-ray center, indicated with large green X marks and seen to be offset from the red-sequence significance peaks.\label{fig:zDist_all}}
\end{figure*}

Figure~\ref{fig:zDist_all} shows the distribution of galaxies within the RCS\,2319+00 supercluster field over the $0.858 \le z \le 0.946$ range discussed above, plotted over the density profiles of the red-sequence galaxies map at a photometric redshift of $z \sim 0.9$ (L.F.\,Barrientos et al.\,2012, in preparation).  The symbol size and color varies with the redshift of the galaxies: larger, darker symbols are blue shifted relative to the mean redshift; smaller, lighter symbols are red shifted.  The X-ray contours seen in Figure~\ref{fig:specLayout} \citep[from][]{Hicks2008} are not plotted but their peaks are shown with large green X marks and show an offset to the red-sequence peaks (see Table~\ref{table:offset} for coordinates and offset distances).  These offsets are most likely due to the contamination from the lower redshift wall seen in Figures~\ref{fig:zhist} and \ref{fig:vel-rad_all} causing the cluster peak positions in the red-sequence density profile, which spans a broad redshift bin, to shift.  The RCS\,2319+00 supercluster is already visible in the red-sequence density profile maps with central redshifts as low as $z \sim 0.758$. 
As our spectroscopic coverage is not complete, the galaxies in this lower redshift peak, whose spatial distribution can be seen in Figure~\ref{fig:zDist_all}, represent a lower limit on the red-sequence density profile contamination.  

We look to the brightest cluster galaxy (BCG) in each cluster core to aid in isolating the correct cluster center.  The BCGs were found by first isolating all galaxies less than 0.5 Mpc from the red-sequence cluster centers in the photometric catalogue.  From the reduced source lists, the brightest $z'$-band galaxy within $\pm0.2$ of the $R_c-z'$ red-sequence line for each cluster was selected as the BCG. The identified BCGs for the three clusters are all found  $\le$6.5 arcseconds from the cluster X-ray peaks, closer than the 13.5-32.5 arcseconds between the X-ray and red-sequence cluster center positions.  We therefore use the X-ray centers in calculating the dynamical properties of the clusters.      

\subsection{Determining the cluster core members}\label{sec:component}

\begin{figure*}[tb!]
\centering
\subfigure{\includegraphics[width=5.25 cm, trim=1.1cm 1.25cm 1.25cm 0cm, clip=true]{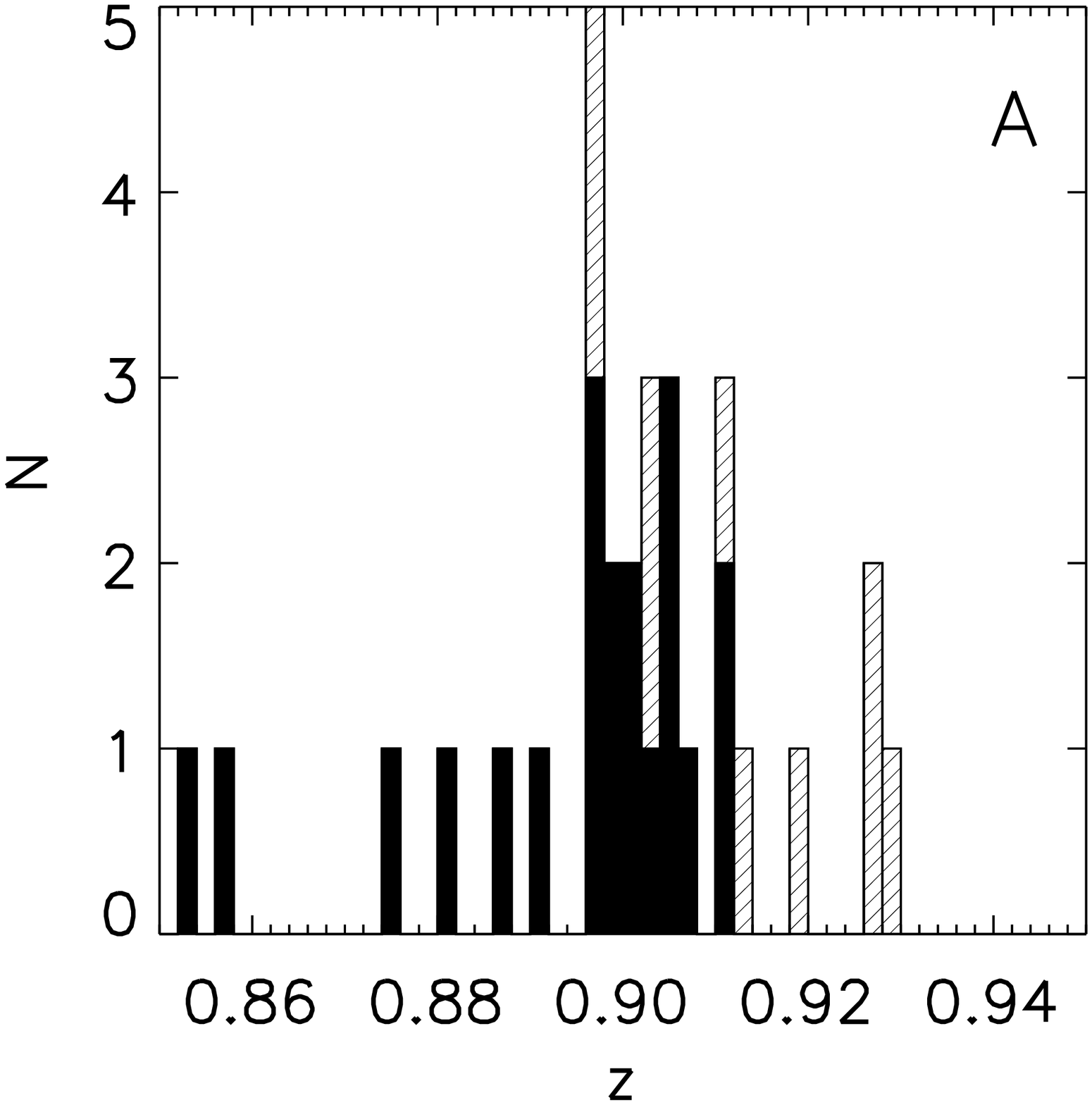}}
\hspace{0.25 cm}
\subfigure{\includegraphics[width=5.25 cm, trim=1.1cm 1.25cm 1.25cm 0cm, clip=true]{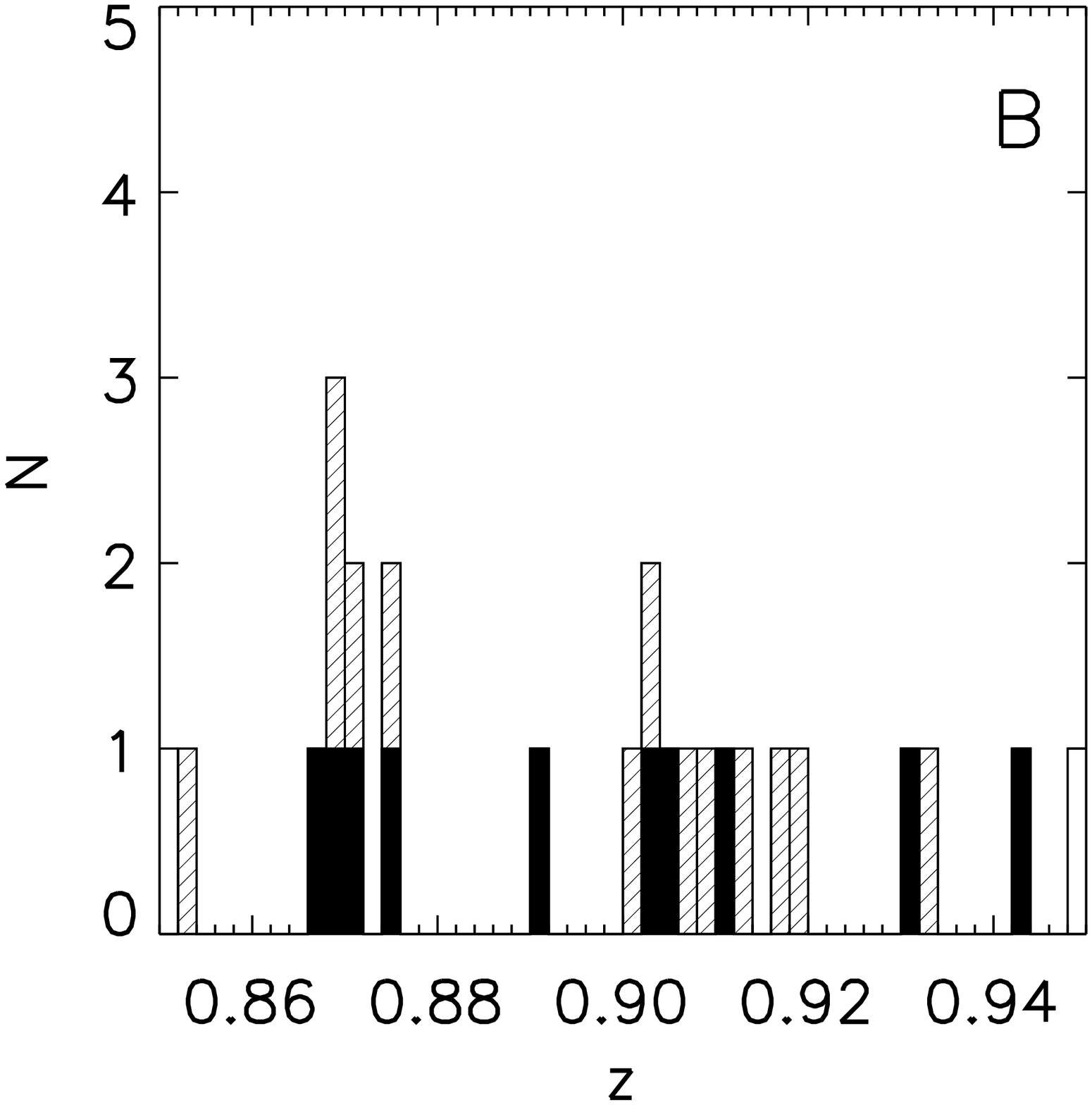}}
\hspace{0.25 cm}
\subfigure{\includegraphics[width=5.25 cm, trim=1.1cm 1.25cm 1.25cm 0cm, clip=true]{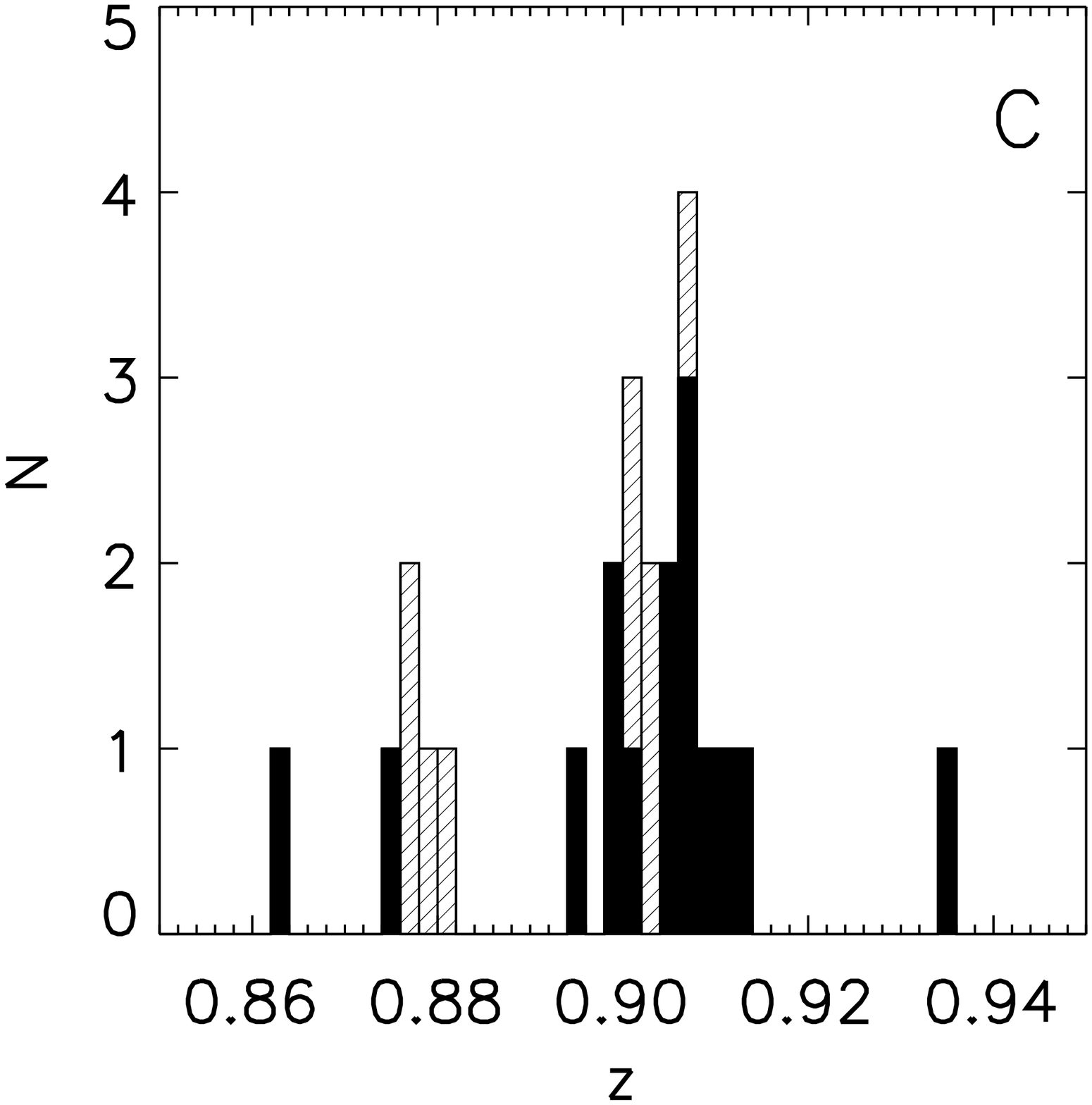}}
\caption{\small Histograms of the spectroscopic redshifts within a 1.0 Mpc radius of each cluster's X-ray peak.  Solid histograms show confident redshift within 0.5 Mpc of the cluster centers and hatched histograms extend the radius to 1.0 Mpc.\label{fig:coreHists}
}
\end{figure*}

To calculate the velocity dispersions and spectroscopic redshifts of the component clusters (\S \ref{sec:reds}) we use all confident redshifts within radii of 0.5 and 1.0 Mpc of the X-ray peak positions in each cluster.  Due to the close separation distance between the three component cluster centers, a 1.0 Mpc radius is the largest used to avoid double counting of galaxies between component clusters.  The green dotted circles in Figure~\ref{fig:zDist_all} shows the 1.0 Mpc radii from the X-ray centers. An initial cut of $0.85~\le~z~\le~0.95$ around the supercluster redshift peak was used and the redshift histograms for the three cluster cores are shown in Figure~\ref{fig:coreHists}.  All secure redshifts within 1.0 Mpc of the center positions are shown as hatched histograms, with the solid histograms including only those galaxies within 0.5 Mpc of the centers.  The histograms were examined for obvious outliers not consistent with a Gaussian curve that could falsely broaden the velocity distribution.  

\begin{table*}[tb!]
\caption{Cluster center positions}
\label{table:offset}
\centering
\scriptsize
\begin{tabular}{ccccccccc} 	
\tableline
  & \multicolumn{2}{c}{Red-sequence (RS) center} 	&  \multicolumn{2}{c}{X-Ray center} & \multicolumn{2}{c}{BCG position} & RS/X-ray Offset & X-ray/BCG Offset\\
 & & & & && & (arcsec) & (arcsec) \\
\tableline
A  & 23 19 52.8 &  +00 38 03.5 & 23 19 53.2 & +00 38 12.5 & 23 19 53.42 & +00 38 13.65 & 13.9 & 3.5\\
B & 23 20 02.5 & +00 33 29.6  & 23 20 02.1 & +00 32 57.6 & 23 20 02.33 & +00 33 03.10 & 32.5 & 6.5\\
C & 23 19 46.1  & +00 30 37.5 & 23 19 46.8 & +00 30 14.3 & 23 19 46.60 & +00 30 09.54 & 25.5 & 5.6\\
\tableline
\end{tabular}
\end{table*}

For cluster A, no sources were removed as the redshift distribution, while broad, does not show any obvious irregularity and an iterative interloper clipping method is used in \S\ref{sec:reds} when determining the cluster redshifts and velocity dispersions.  We again look to the BCG of cluster A for comparison to ensure that the redshift distribution in the region around the cluster core is tracing the proper cluster redshift.  The BCG is found to have a confident spectroscopic redshift of 0.901, consistent with the approximate peak of the redshift distribution in the left-hand panel of Figure~\ref{fig:coreHists}.

Clusters B and C seem to show double peaks in their redshift distributions, with lower redshift peaks at z $\sim$ 0.868 and z $\sim$ 0.877 respectively, and higher redshift peaks consistent with the overall supercluster redshift peak at z $\sim$ 0.9 visible in all three cluster cores.  This could possibly be due to under sampling of the cluster cores, especially for B, which falls at the edges or in the gaps of the VIMOS spectroscopy masks (Figure~\ref{fig:specLayout}).  However, if this were correct and the true redshift of the cores falls between the two peaks, the resulting velocity dispersions would lead to cluster virial masses much too large to be consistent with the measured X-ray masses \citep{Hicks2008}.   Another, more probable possibility is that the lower redshift peaks in clusters B and C belong to filaments and groups of the potential foreground structure seen in Figure~\ref{fig:vel-rad_all}.  We investigate this further by looking at the spatial distribution of all members having redshifts consistent with the separate peaks in the 1.0 Mpc regions around clusters B and C.  As well, we use the spectroscopic redshifts of the BCG in each cluster core to corroborate our choice of approximate cluster redshift.  

\begin{figure*}[tb!]
\centering
\subfigure{\includegraphics[width=6 cm, trim=0.7cm 0cm 0cm 0cm, clip=true]{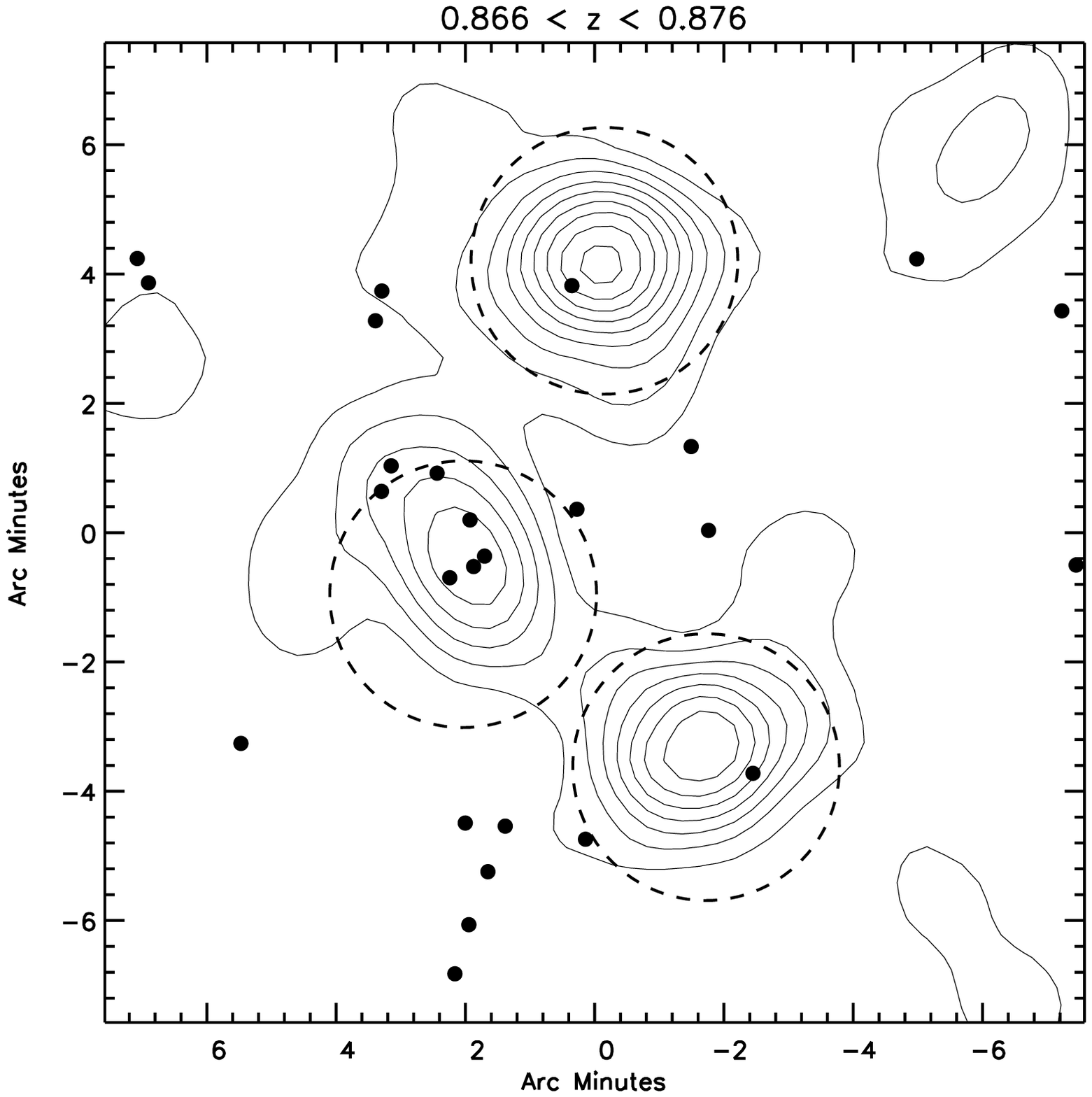}}
\hspace{-0.25 cm}
\subfigure{\includegraphics[width=6 cm, trim=0.7cm 0cm 0cm 0cm, clip=true]{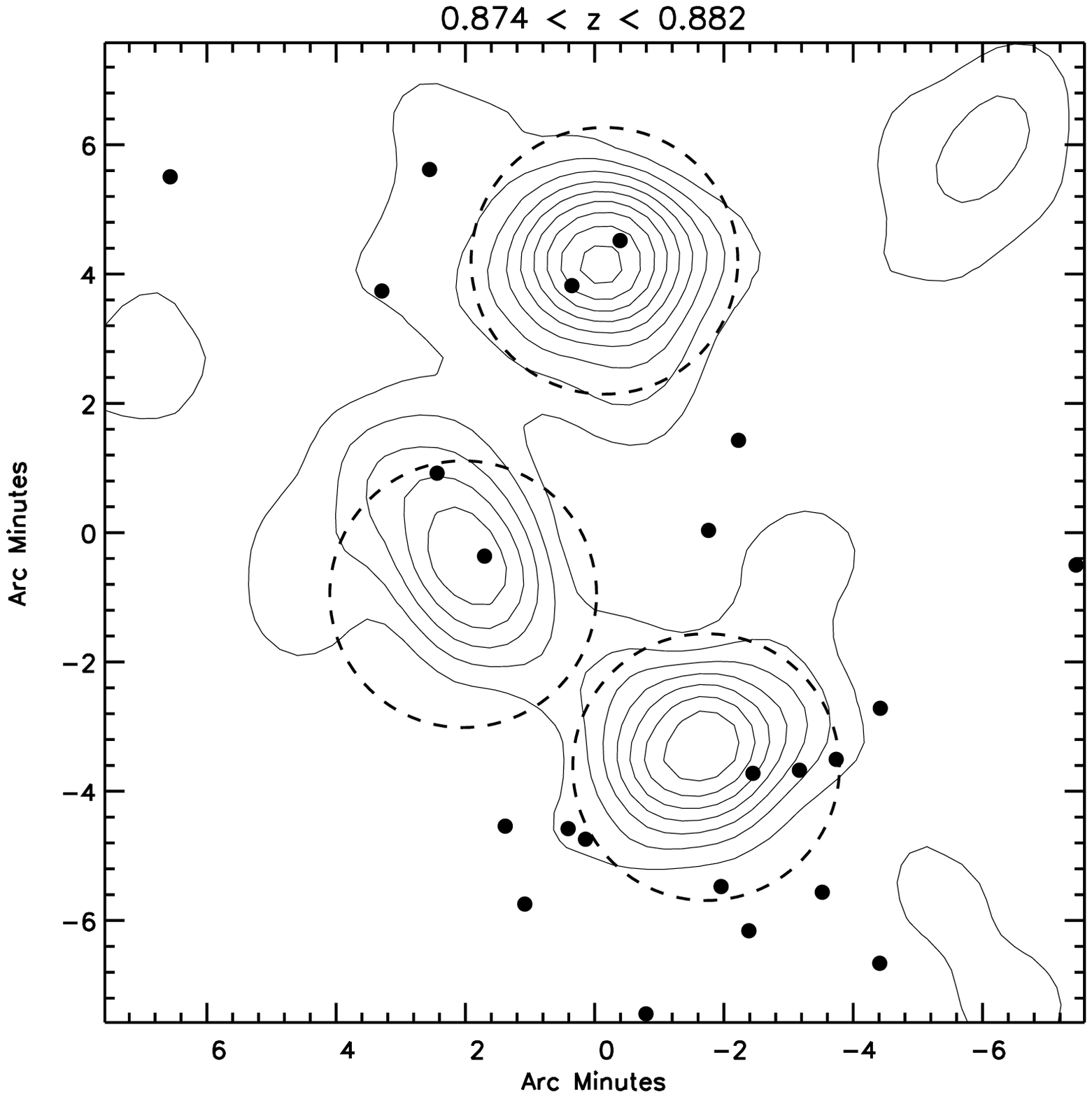}}
\hspace{-0.25 cm}
\subfigure{\includegraphics[width=6 cm, trim=0.7cm 0cm 0cm 0cm, clip=true]{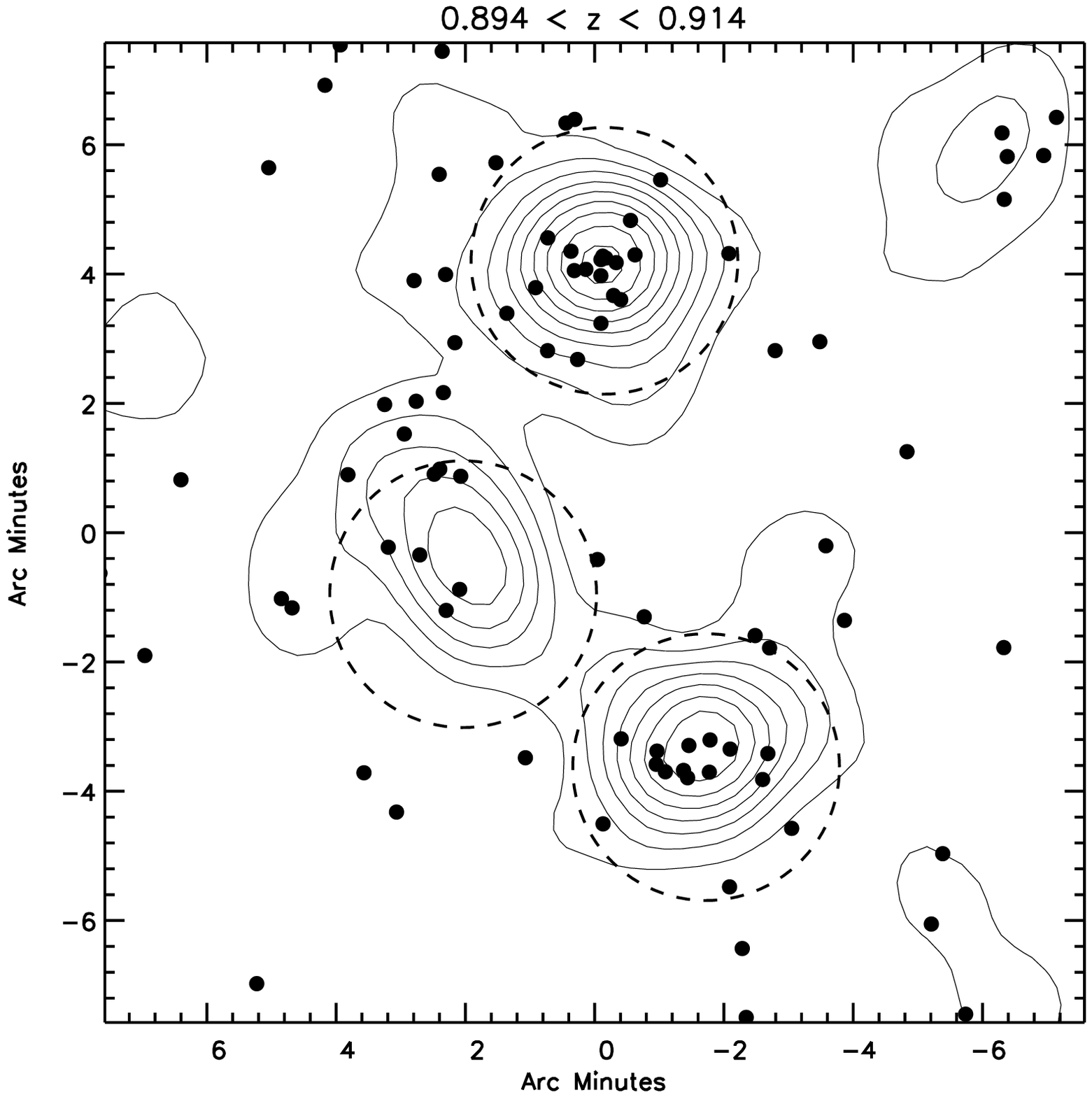}}
\vspace{-0.25 cm}
\caption{\small Distribution of galaxies in the three component cluster cores within three redshift slices corresponding to the lower redshift peaks in the histograms of Cluster B (\textit{left}) and C (\textit{middle}) and the higher redshift peak present in all three clusters (\textit{right}).  The dashed circles denote the 1 Mpc radii from the X-ray peaks.\label{fig:cutDists}}
\end{figure*}

For cluster C, the redshift peak of 0.874 $< z <$ 0.882 (Figure~\ref{fig:cutDists}, center) shows that the members of this peak are inconsistent with the red-sequence significance contours of the cluster core. However, the members of the higher redshift peak of 0.894 $< z <$ 0.914 (Figure~\ref{fig:cutDists}, right), present in all three cluster cores, are consistent with the red-sequence density contours.  The BCG in cluster C also validates the choice of the higher redshift peak as the correct one, with a spectroscopic redshift of 0.899.  

The case of cluster B is more difficult to disentangle.  While there are similar numbers of galaxies in the two redshift peaks, neither the higher, broader peak nor the lower, sharper peak at 0.866 $< z <$ 0.876 (Figure~\ref{fig:cutDists}, left) seem to have members significantly more clustered within the red-sequence over-density contours.  Again, this is most likely due to the spectroscopic undersampling of cluster B compared to the other cluster cores.  We rely on the cluster BCG, with a spectroscopic redshift of 0.903, to isolate the correct redshift peak for this cluster as the higher of the two at $z\sim0.9$, consistent with both clusters A and B.  

We manually remove the lower redshift peaks from the histograms of clusters B and C so as not bias the cluster dynamics determinations.  
	
\subsection{Cluster Redshifts and Velocity Dispersions}\label{sec:reds}	
	
The reduced galaxy redshift lists from the cluster core histograms, with the lower redshift peaks in clusters B and C removed as described, were used to determine the cluster redshifts and velocity dispersions from within the 0.5 and 1.0 Mpc radii.  We follow the method described by \citet{Postman1998}, using the biweight mean estimate \citep{Beers1990} to iteratively calculate the mean spectroscopic redshift and velocity dispersion of the cluster cores.  At each iteration, all galaxy rest frame relative velocities to the biweight mean cluster redshift are calculated as $\Delta\nu~=~c(z-\bar{z})/(1+\bar{z})$.  Any galaxy falling outside of the mean dispersion by more than $3\sigma$ or 3500~km~s$^{-1}$ is removed from the list and the biweight mean is recalculated.  The biweight mean is used as it is less susceptible to outlying interlopers that may otherwise disproportionally shift the value of the mean.  

\begin{table*}[bt!]
\caption{Spectroscopic redshifts and velocity dispersions within two different radii of the cluster centers.}
\label{table:props}
\centering
\begin{tabular}{cccccccc}
\tableline
 & \multicolumn{3}{c}{0.5 Mpc} & & \multicolumn{3}{c}{1.0 Mpc} \\
\cline{2-4}
\cline{6-8}
 & N & $z_{spec}$ & $\sigma$ &  &  N & $z_{spec}$ & $\sigma$  \\
  & & & (km~s$^{-1}$) &   &  & & (km~s$^{-1}$)  \\
\tableline
A & 17 & 0.900~$\pm$~0.008 & 1253~$\pm$~306 &  & 23 & 0.901~$\pm$~0.008 & 1202~$\pm$~233  \\
B & -- & -- & --  & & 9 & 0.905~$\pm$~0.007 & 1154~$\pm$~367  \\
C & 11 & 0.906~$\pm$~0.005 & 759~$\pm$~203 & & 15 & 0.905~$\pm$~0.005 & 714~$\pm$~180  \\
\tableline
\end{tabular}
\end{table*}

The results of the mean redshift and velocity dispersion measurements within the two different radii from the cluster centers are listed in Table~\ref{table:props}. The larger number of cluster galaxies used for the dynamical property measurements within 1 Mpc of the cluster centers result in smaller errors and more reliable values.  For cluster B, there are not enough galaxies within a 0.5 Mpc radius of the center coordinate to calculate the dynamical properties.  We therefore use the values derived from the 1 Mpc radii.  Cluster B again has the fewest galaxies within this radius with only 9 galaxies remaining after the iterative interloper clipping, making the velocity dispersion unreliable, as seen by the large error.  The 15 galaxies remaining within 1Mpc of cluster C offer a better estimate of the properties but would still benefit from additional spectrocopic members for a more dependable measurement.  Cluster A has 23 galaxies remaining after interloper clipping, though there remains a significant error of $\sim$20\% due to the broad distribution.  This may indicate that we are not sampling a virialized population in this cluster and are instead witnessing a cluster merger.  We discuss this possibility below when looking for substructure both in the multiwavelength and spectroscopic data (see sections \ref{sec:comparison} and \ref{sec:structure}).

\subsection{Virial Radius and Mass}\label{sec:virial}

Using the measured velocity dispersions we can estimate the dynamical mass and radius of the component cluster cores.  The cluster virial mass can be approximated by calculating the mass at which the overdensity of the cluster is 200 times the critical density ($\rho_c$).  We use the method of \citet{Carlberg1997} to calculate M$_{200}$ from the velocity dispersion using,
\begin{equation}
\label{eq:M200}
M_{200} = \frac{3\sqrt{3}\,\sigma^{3}}{H(z)\,G}.
\end{equation}

\noindent where $H(z) = H_o\sqrt{\Omega_{\Lambda}+\Omega_{M}(1+z)^3}$ and $\sigma$ is the one dimensional velocity dispersion.  The virial radius is then approximated from $M_{200} = (4/3) \pi r^3_{200} \times 200\rho_c$, which can be simplified as,

\begin{equation}
\label{eq:r200}
r_{200} = \frac{\sqrt{3}}{10}\frac{\sigma}{H(z)},
\end{equation}

The calculated $r_{200}$ and $M_{200}$ values are found in Table~\ref{table:r200-m200}.

Another method of determining the mass of clusters within $r_{200}$ from their velocity dispersions is to employ the method of \citet{Evrard2008}, who used hierarchical clustering simulations to determine a virial scaling relation for massive dark matter halos of 

\begin{equation}
\label{eq:M200_DM}
M_{200} = \frac{10^{15}\,M_{\odot}}{h(z)}\left(\frac{\sigma_{DM}}{\sigma_{15}}\right)^{1/\alpha},
\end{equation}
where the logarithmic slope is $\alpha = 0.3361~\pm~0.0026$ and the normalization at a cluster mass of $10^{15}\,M_{\odot}$ is $\sigma_{15} = 1082.9~\pm~4.0$ km s$^{-1}$.  The dark matter halo velocity dispersion is related to the velocity dispersion of galaxies by the velocity bias, $b_\upsilon = \sigma_{gal}/\sigma_{DM}$.  The current best estimate is $\left<b_\upsilon\right> = 1.00~\pm~0.05$.  We assume the non-bias case of $b_\upsilon = 1$, as done in previous cluster studies \citep[e.g.][]{Brodwin2010,Sifon2012}.  From the derived mass, we can compute the related $r_{200}$ value for each cluster.  The results of the dynamical mass and radius using this method are listed in Table~\ref{table:r200-m200}.

The results for the cluster masses using the two methods agree within errors; however, the masses derived using the virial theorem are $\sim50$\% larger than those found using the dark matter scaling relation.  \citet{Carlberg1996} found that masses measured using the virial theorem are dependent on the boundary within which the cluster properties are determined and should be corrected with a surface pressure term, which requires a knowledge of the mass profile of the cluster.  For cases where only the interior of the cluster's profile is being measured with no surface pressure correction, as is the case here, they found that the overestimation of the cluster mass should be less than 50\%.  

With this in mind, and the results for the mass and virial radius found using the method of \citet{Evrard2008} being $\sim50$\% lower than those found with the virial theorem, we estimate that the masses of the clusters are closer to those calculated using the dark matter scaling relation.  However, as the velocity dispersion measured for cluster B is based on only 9 spectroscopic galaxy redshifts it, along with the resulting cluster properties, must be corroborated with the values determined for this cluster from additional supporting data before it can be deemed as a reliable estimate.  In the following section we compare the cluster parameters found using the spectroscopic data to estimates from multiwavelength data.

\begin{table}[b!]
\caption{Dynamical properties of the component clusters found with two different methods from spectroscopic velocity dispersions}
\label{table:r200-m200}
\centering
\begin{tabular}{cccccc}
\tableline
 & \multicolumn{2}{c}{\citet{Carlberg1997} method} &  &\multicolumn{2}{c}{\citet{Evrard2008} method}  \\
\cline{2-3}
\cline{5-6}
 & $r_{200}$ & $M_{200}$&  & $r_{200}$ & $M_{200}$\\
 & $(\textup{Mpc})$  &$(10^{14}~ M_{\odot})$ & & $(\textup{Mpc})$  &$(10^{14}~ M_{\odot})$\\
\tableline
A & 1.82~$\pm$~0.35 & 18.4~$\pm$~10.7 & & 1.58~$\pm$~0.30 & 11.9~$\pm$~6.9   \\
B & 1.75~$\pm$~0.55 & 16.2~$\pm$~15.5 &  &  1.51~$\pm$~0.48 & 10.5~$\pm$~10.0\\
C & 1.08~$\pm$~0.27 & 3.8~$\pm$~2.9 & &  0.94~$\pm$~0.24 & 2.5~$\pm$~1.9  \\
\tableline
\end{tabular}
\end{table}

\subsection{Comparison With Other Mass Estimators}\label{sec:comparison}

\begin{table*}[htb]
\caption{Dynamical properties from multiwavelength data}
\label{table:otherProps}
\centering
\begin{tabular}{cccccccc}
\tableline
 & \multicolumn{3}{c}{X-Ray\footnote{X-ray mass estimates ($M_{200,X}$) were extrapolated to $r_{200}$ by \citet{Gilbank2008}.  We use the X-ray temperatures reported in \citet{Hicks2008} and the $\sigma-T_X$ relation from \citet{Xue2000} of $\sigma = 10^{2.49\pm0.02}\,T_X^{0.65\pm0.03}$ to predict the velocity dispersions of the three clusters.}} & & \multicolumn{3}{c}{Weak Lensing\footnote{Parameters are derived through weak lensing analysis of HST ACS/WFC data by Jee et al.\,(2011) and not through dynamical methods.}} \\
\cline{2-4}
\cline{6-8}
 & $T_X$ & $M_{200,X}$ &$\sigma$ & & $\sigma$ & $r_{200}$ & $M_{200}$   \\
  &  keV & $(10^{14}~ M_{\odot})$ & (km~s$^{-1}$) & & (km~s$^{-1}$) & $(\textup{Mpc})$  & $(10^{14}~ M_{\odot})$ \\ 
  \tableline
A & 6.2$^{+0.9}_{-0.8}$ &  6.4$^{+1.0}_{-0.9}$ &1012$^{+107}_{-97}$  &     & 898$^{+67}_{-71}$ & 1.22$^{+0.15}_{-0.13}$ & 5.8$^{+2.3}_{-1.6}$\\
B &  5.9$^{+2}_{-1}$ & 4.7$^{+0.9}_{-1.4}$ & 980$^{+102}_{-93}$ &  & --  & --  & --\\
C & 6.5$^{+1}_{-1}$ &  5.1$^{+0.8}_{-0.8}$  & 1043$^{+112}_{-101}$   &  & -- & -- & -- \\
\tableline
\end{tabular}
\end{table*}

We can compare our results for the velocity dispersions and mass estimates of the three clusters to those found by other methods where available. We use the X-ray temperatures reported in \citet{Hicks2008} and the $\sigma-T_X$ relation from \citet{Xue2000} of $\sigma = 10^{2.49\pm0.02}\,T_X^{0.65\pm0.03}$ to predict the velocity dispersions of the three clusters.  The cluster masses extrapolated directly from the X-ray data are also available.  \citet{Jee2011} calculated weak lensing values for cluster A from HST ACS/WFC data.  Using a SZ scaling relation, \citet{Gralla2011} found an $M_{500}$ value for cluster A of $\sim 0.5\times$ the X-ray derived $M_{500,X}$ value.  They noted that the SZ derived masses are lower than those previously published for a number of the clusters in their sample from various multiwavelength methods and therefore we do not include the SZ mass estimate for cluster A in our discussion.  Table~\ref{table:otherProps} lists the known multiwavelength measurements of the masses and velocity dispersions of the three cluster cores.

The velocity dispersions measured from the spectroscopic data and those predicted from the X-ray temperatures agree well within errors for clusters A and B.  The values for cluster C are less consistent, varying by about 1.5$\sigma$ between the two values.  This would suggest that the dynamical mass estimates based on the measured velocity dispersions should be reasonable, even for cluster B with it's low spectroscopic galaxy numbers.  In fact, the dynamical masses measured using the \citet{Carlberg1997} virial mass method, which is known to overestimate the cluster mass by up to 50\% \citep{Carlberg1996}, overestimates the X-ray extrapolated mass by only $1.12\sigma$ and $0.74\sigma$ for A and B and underestimates the mass by $0.45\sigma$ for cluster C from the dynamical mass errors only.  The masses estimated using the dark matter scaling relation of \citet{Evrard2008} are all consistent with the X-ray masses, which are also based on a scaling relation to extrapolate the mass based on the temperature of the cluster cores \citep{Hicks2008}.  The X-ray data therefore confirms that the dynamical properties found from the spectroscopic data using the \citet{Evrard2008} dark matter scaling relation, with their large errors, are reasonable estimates for the cluster cores' mass and radius.  For cluster A, the velocity dispersion found from the weak lensing analysis is lower than that found using the spectroscopic and X-ray data, though just consistent within errors of the two measurements.  The weak lensing mass estimate is consistent with both dynamical mass estimates as well as with the X-ray mass.  From the multiwavelength data currently available we estimate the masses of the three clusters to be $\sim 10^{14.5} - 10^{14.9}$. 

The X-ray data also provide insight into the possibility of merging activity in the cluster cores.  There is evidence in the X-ray data that cluster B may contain significant substructure or have undergone a recent merger, as the peak of the X-ray emission is not consistent with the center of the extended X-ray emission \citep[see Figure 1 in][]{Hicks2008}.  The X-ray data for cluster A also seems to indicate the presence of some possible substructure in the radial surface brightness profile, but not enough to show evidence of a significant merger or verify if the population we are sampling is virialized or not \citep[see Figure 2 in][]{Hicks2008}.  

\section{The Substructure of the RCS\,2319+00 Supercluster}\label{sec:structure}

Due to the close proximity of the three cluster cores, both in angular separation and redshift space, designating galaxies as unique members of one of the particular cluster cores is difficult, as seen in Figure~\ref{fig:vel-rad_all}.  We therefore chose to study the cluster member properties as a function of their distribution within the overall structure and their apparent local density environment within the supercluster. The spectroscopically confirmed supercluster members have allowed us to begin tracing the supercluster structure and identifying spatial and dynamical sub-environments within it.  

\subsection{The Dressler-Shectman Test for Substructure}\label{sec:DStest}

As an initial test to identify the distribution of the underlying structure in the RCS\,2319+00 supercluster field we performed the \citet{Dressler1988} (DS) test.  The DS test looks for inconsistency between the locally measured velocity dispersion and velocity mean with the parameters measured over the whole structure field.  Deviations from the global velocity parameters of the cluster indicate the likely presence of substructure.  The DS test was used with the RCS\,2319+00 data to begin tracing the locations of possible groups and filaments within the supercluster before attempting to assign specific galaxies to these potential groups.  We follow \citet{Dressler1988} to compute, for each galaxy in our structure, the deviation $\delta$ from the global mean as 
\begin{equation}
\label{eq:DStest}
\delta^2= \left(\frac{1+N_{nn}}{\sigma^2}\right)[(\bar{\upsilon_{local}}-\bar{\upsilon})^2+(\sigma_{local}-\sigma)^2],
\end{equation}
where $N_{nn}$ is the number of nearest neighbors used to compute the local variables.  We use 10 nearest neighbors, as done by \citet{Dressler1988} in their cluster study.  The $\Delta$ statistic is defined as the sum of all deviations $\delta_i$ and a cluster is said to have substructure if $\Delta/N_{members} > 1$.  To test for false positives, we shuffle the redshifts between galaxy positions and perform 25000 Monte Carlo (MC) tests.  If the substructure detected in the real data is indicative of true substructure, then the probability that the shuffled data has a higher $\Delta$ statistic than the real data will be low.  The probability of false structure detection is 
\begin{equation}
\label{eq:Pstat}
P = \frac{N (\Delta_{shuffle} > \Delta_{real})}{N_{shuffle}},  
\end{equation}
 where $N_{shuffle}$ is the number of MC shuffling tests performed.  
 
The DS test was done both on the full redshift range of our larger structures and on the three separate redshift walls with ranges of $\pm2000$ km s$^{-1}$ for each to avoid significant overlap.  For the full redshift range, we find $\Delta/N_{members} = 2.02$,  while the three redshift peaks yield values of $\Delta/N_{members} = 2.00, 2.76$ and 2.64 for the redshift walls centered on $z \sim 0.87, 0.90,$ and 0.93, indicating the presence of substructure in each case.  The Monte Carlo shuffled tests find P-values of $< 0.006$ for the full redshift range, and $<$0.05, 0.0002 and 0.003 for the three separate redshift walls, leading us to believe there is real substructure in each of the redshift peaks in our supercluster field and prompting a search for the specific sub-environments within our spectroscopic galaxy catalogue in the following sections.
 
\begin{figure}[tb!]
\centering
\subfigure{\includegraphics[width=4 cm, trim=0.75cm 0.5cm 0.75cm 0.3cm, clip=true]{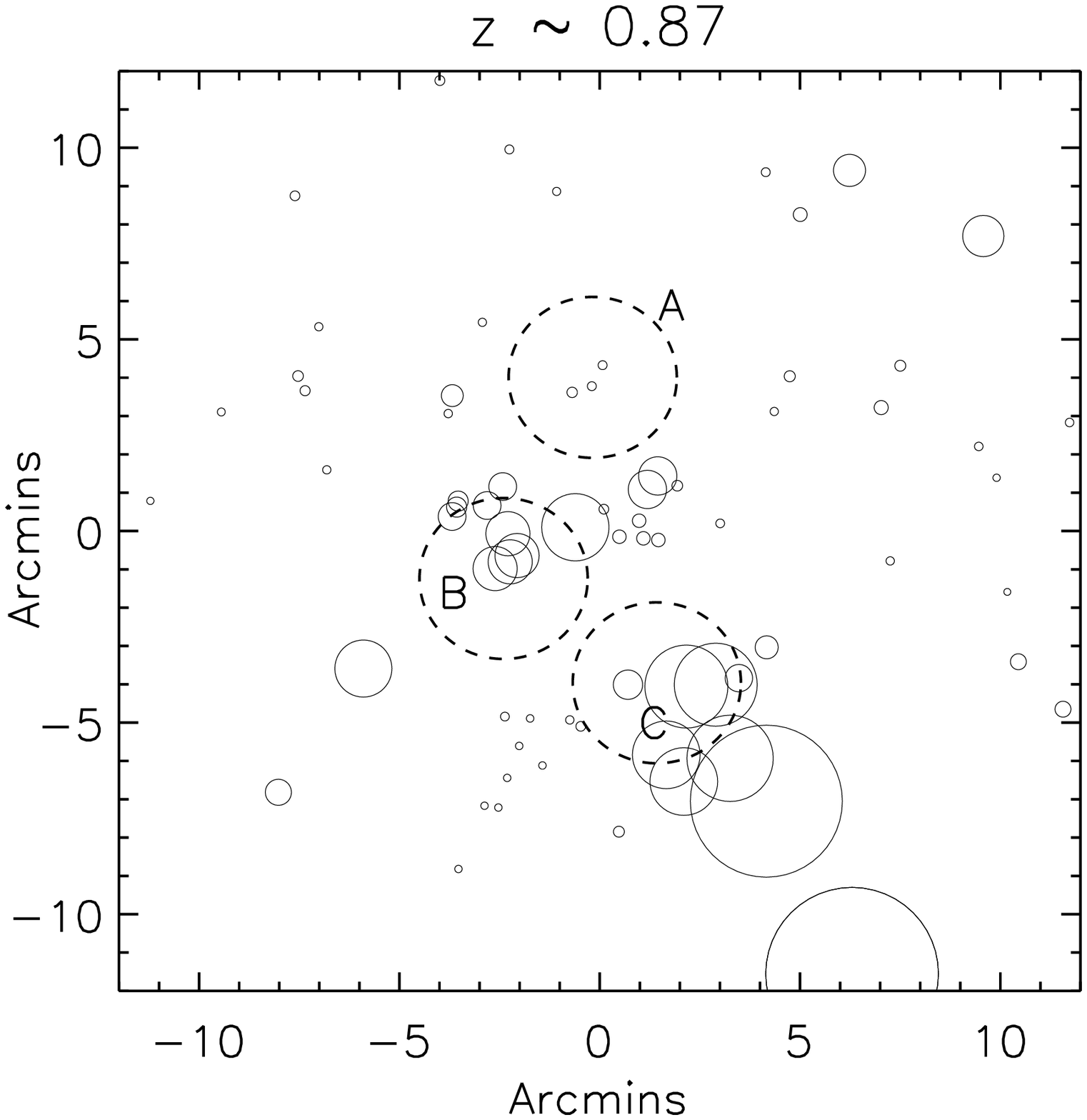}}
\hspace{0.25 cm}
\subfigure{\includegraphics[width=4 cm, trim=0.75cm 0.5cm 0.75cm 0.3cm, clip=true]{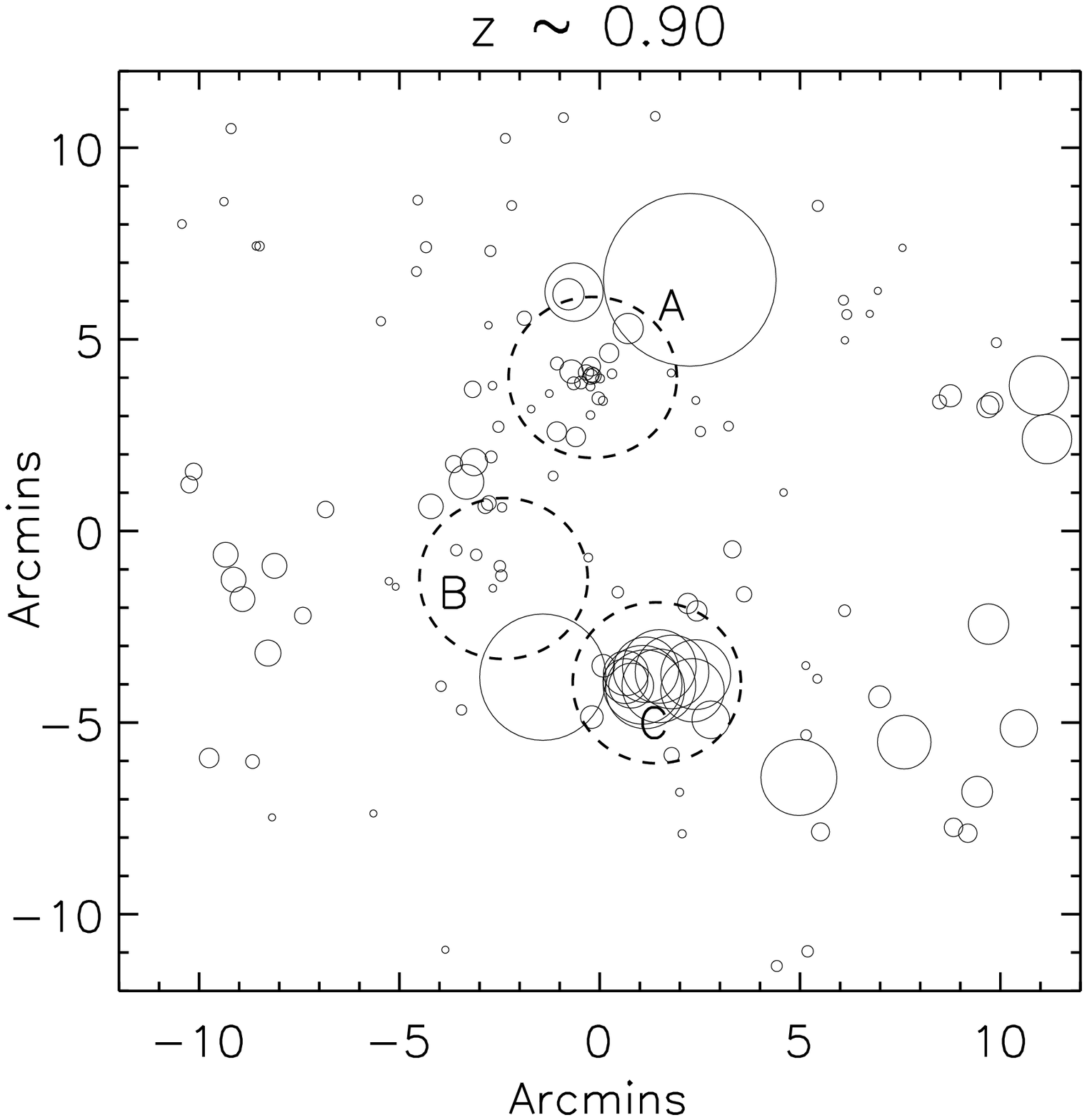}}
\caption{\small DS test bubble diagrams with each galaxy in our structure catalogue denoted by a circle whose radius scales with $e^\delta$.  Dashed circles show the 1Mpc radius around each cluster center. Results from the DS test on the $\pm$2000 km s$^{-1}$ region around the lower $z \sim 0.87$ redshift wall (\textit{left}) and the $z \sim 0.9$ redshift structure (\textit{right}). \label{fig:DS}}
\end{figure}

The results of the test for the lower and central-redshift walls are shown in Figure~\ref{fig:DS} in DS test bubble diagrams, with each galaxy represented by a circle whose size scales with $e^\delta$ for that galaxy.  The three cluster cores are indicated by the 1Mpc radius dashed circles on the each of the panels.  
Though the highest redshift slice has a high $\Delta/N_{members}$ of 2.64, it has the lowest number of galaxies, causing the 10 nearest neighbours to be spread over a large region in most cases and so we do not find any evidence of substructure in concentrated regions.  

In the $z \sim 0.87$ redshift slice, we find that both clusters B and C show deviations from the overall mean parameters, with substructure extending from the centers of both cluster core positions.   Deviations are found to extend from the center of cluster B slightly to the north-east as well as a more elongated deviation north-west towards the middle of the three clusters.  These areas of broad redshift distribution can be seen in the top panel of the velocity versus radius plot in Figure~\ref{fig:vel-rad_all} as a wide grouping of galaxies, both in radius and velocity, $\sim1.5-2.5$ Mpc from the center of cluster A.  Whether the deviations that extend in the two directions outward from cluster B belong to the same substructure or to separate overdensities in the supercluster will be explored in the following section.   In cluster C, evidence of substructure is shown to extend south west of the cluster core, with the largest deviations from the mean present in this region.   

The redshift slice around $z \sim 0.9$ also shows indications of substructure in the area of cluster C, though in a more concentrated region and with lower deviation than seen in this cluster in the lower redshift slice.  There are also regions of potential substructure east of cluster B and west of clusters A and C.  An interesting area to note is the region between clusters A and B, where there seems to be some evidence of substructure consistent with a previously confirmed infrared bright filament found in the Herschel 250-500\mum~Spectral and Photometric Imaging REceiver (SPIRE) map by \citet{Coppin2012a}.  The complementary detection of this structure in the IR provides support for our claim of a substructure detection in this area.

Both clusters B and C were seen to show two redshift peaks in Figure~\ref{fig:coreHists} within the central 1 Mpc around the clusters, one in each of the lower and middle redshift slices.  For cluster B, the signs of deviation from the mean parameters in the core region in the lower redshift wall at $z \sim 0.87$ in contrast to the lack of deviation in the DS test of the central, overall supercluster redshift slice around $z \sim 0.9$ give further evidence that the cluster core lies in the higher $z \sim 0.9$ redshift slice since we expect cluster core galaxies to show little deviation from the mean parameters.  In the case of cluster C, the DS test on both redshift slices show signs of substructure emanating from the center, with the larger deviations present in the lower redshift slice.  In \S\ref{sec:component} we motivated the choice of the $z \sim 0.9$ peak in the cluster C plot of Figure~\ref{fig:coreHists} as the one corresponding to the cluster core given its agreement with both the spectroscopically confirmed BCG and the distribution of the red-sequence galaxies significance contours.  The deviations seen in the core of cluster C in this redshift slice could therefore be an indication that there is a line-of-sight group falling towards the cluster core \citep{Einasto2012, Pinkney1996}.  The following search for individual groups will explore this possibility in more detail. 

\subsection{Friends-of-Friends Search for Substructure in \\RCS\,2319+00}\label{sec:fof}

The distribution of possible substructure in our supercluster field found with the DS test lead us to attempt to recover the substructure in a more tangible way.  We used a modified friends-of-friends (FOF) algorithm \citep{Huchra1982} with linking lengths of $D_L$ and $V_L$, representing distance and velocity separations, to identify associations of three or more linked galaxies.  Briefly, the algorithm uses each of our spectroscopically confirmed cluster members and searches for friends within the two linking lengths.  The test galaxy and its friends are removed from the main source list, and each friend is then tested against the remaining galaxies in the list.  This is done iteratively until there are no new friends found.  If the final candidate friends list has three or more members then it is considered a FOF association, or else the galaxies are returned to the isolated galaxy list.  We chose three member associations to compensate for the incompleteness of the spectroscopic data in an effort to identify regions that may be undersampled spectroscopically but still contain signs of structure.  Monte Carlo Simulations are performed in \S\ref{sec:MC} to help determine the membership number above which we believe the FOF associations are tracing real structures.

The linking lengths were determined through trial and error, with the aim of recovering structure while not linking all three cluster cores together.  Monte Carlo Simulations (\S\ref{sec:MC}) were also used to test the FOF linking lengths, again to ensure that linking lengths were large enough to recover structure while not linking broad redshifts walls across the test region.  A more thorough method of determining the linking lengths would have been to construct a mock catalog from N-body simulations to test our linking lengths based on completeness and recovery purity, as done by \citet{Farrens2011} in their search for groups and clusters in the 2SLAQ catalog (2dF-SDSS Large Red Galaxy and QSO survey).  However, a search of the existing Millenium Simulation light cones \citep{Lemson2006} for a similar structure to the RCS\,2319+00 supercluster did not yield any comparable structures and attempting to model this superstructure with our own simulations falls outside the scope of this paper.  

We found that the spatial linking length, $D_L$, was more influential in linking all three cluster cores as each cluster is separated by $<$3 Mpc from its nearest neighbor and $D_L$ was thus limited to 0.5 Mpc.  The velocity separation, $V_L$, had more subtle effects on the group recovery rate and membership.  We tested several different velocity separations, from 600-1500~km~s$^{-1}$ ($\sim2$--$5\times$ our spectroscopic redshift error of $\Delta z$ = 0.002).  For the lowest velocity separations, the core groups of the clusters are recovered as several separate, small groups, even for members of the prominent peak in cluster C.  We found that in our real data and Monte Carlo tests (\S\ref{sec:MC}) the difference in group recovery between $V_L$ lengths of 1000 and 1500~km~s$^{-1}$ were minimal, recovering the same groups on average, with only one or two members missing from two of the larger groups.  Since we suspect recent merger activity is present in some regions of the supercluster field and therefore expect our substructures may not be virialized, we choose to use the larger linking length of 1500~km~s$^{-1}$ to look for extended structures. 

\begin{figure*}[tb!]
\centering
{\includegraphics[width=17 cm, trim=0.5cm 2cm 0cm 0cm, clip=true]{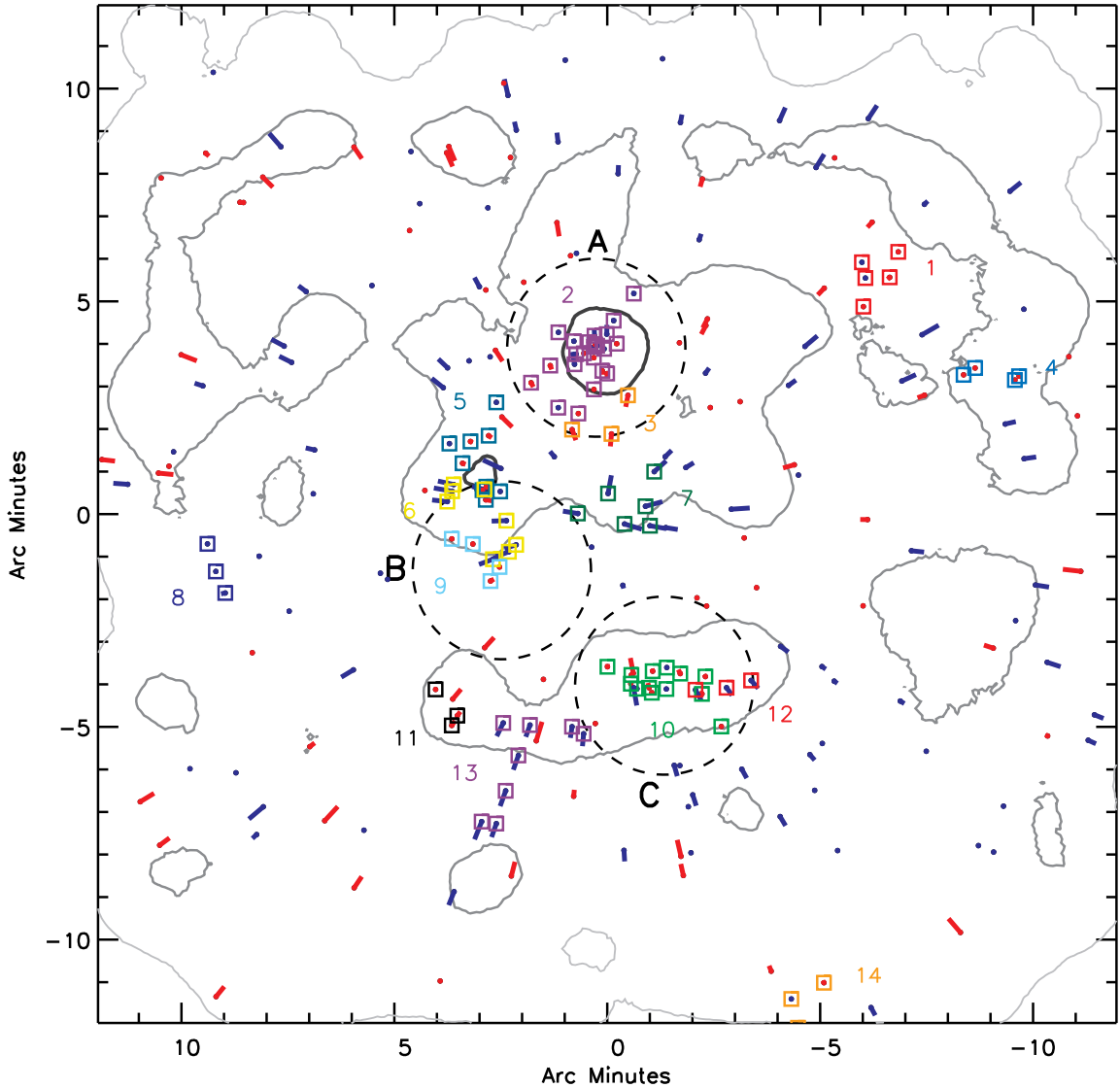}}
\caption{\small Friends-of-friends associations identified with the linking lengths of 0.5 Mpc and 1500~km~s$^{-1}$ are surrounded by like-colored boxes.  Red and blue arrows represent red/blue shifted velocities of spectroscopically identified cluster members relative to a midpoint redshift of 0.897 in the $0.858 \le z \le 0.946$ range and directed towards an arbitrary center position.  Dotted circles represent 1 Mpc radii around the X-ray centres of the three cluster cores.  Grey contours show the spectroscopic target density relative to the full optical source catalog, with higher target densities having thicker, darker grey contours (i.e.\, the $\sim$0.5 Mpc region around cluster A is a high target density region).\label{fig:structure}}
\end{figure*}

\begin{figure}[tb!]
\centering
\subfigure{\includegraphics[width=9 cm, trim=1.25cm 0.4cm 0cm 0.5cm, clip=true]{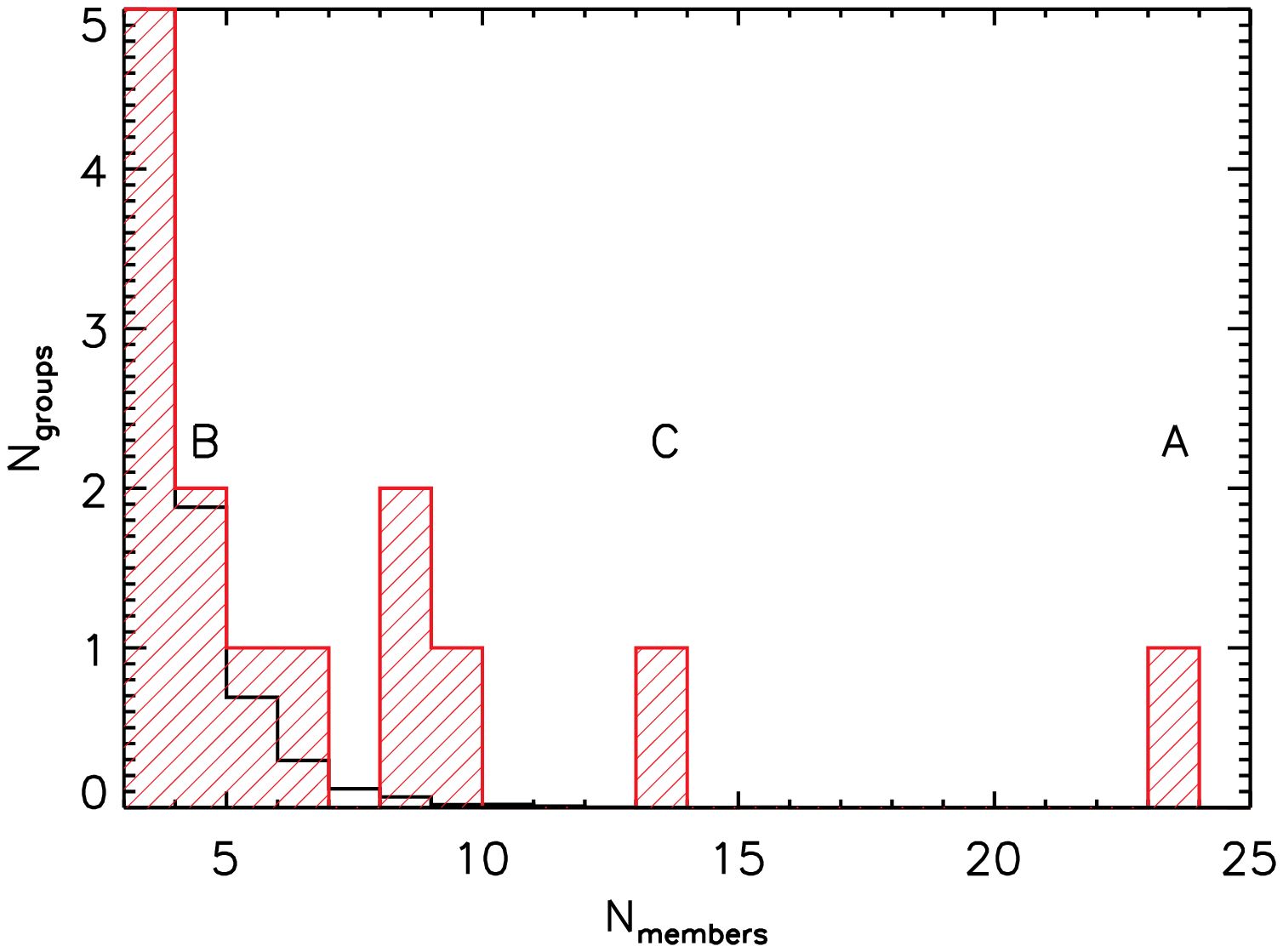}}
\caption{\small The number of FOF associations found for each membership bin for the real supercluster data (red hatched histogram) and the MC test data described in \S\ref{sec:MC} (averaged over all tests - black outlined histogram) recovered using the FOF method.  The letters indicate the bins in which the groups corresponding to the three cluster cores are found. \label{fig:MC}}
\end{figure}

Figure~\ref{fig:structure} shows the results of the friends-of-friends test using the linking lengths of 0.5 Mpc and 1500~km~s$^{-1}$, with linked FOF associations identified by same colored boxes.  The red and blue lines on the plot indicate all supercluster member positions with the color and length of the lines representative of the relative velocity of the galaxies to a median redshift of z=0.897.  The lines are oriented arbitrarily towards a center position on the plot, with the galaxy located at the end pointing towards the center for blue-shifted galaxies and at the end pointing away from the center for red-shifted galaxies.  This was done to retain some redshift information in order to better visualize the layout of the FOF associations within supercluster field.  The grey contours represent the smoothed spectroscopic target density relative to the full optical catalog, with the higher target densities having darker, thicker contours.  Due to an incomplete spectroscopic sample, some of our isolated superstructure `field' galaxies may in fact belong to infalling groups or filaments.  We note, however, that FOF associations are still found in regions of lower spectroscopic density and some high target density regions contain few cluster members and no FOF associations.  As such we do not suspect that our recovered structure is simply tracing the areas of high spectroscopic targeting.   We identify 14 FOF associations, with 7 having more than three members.  Figure~\ref{fig:MC} shows the number of groups for each membership number (hatched histogram). 

\subsection{Monte Carlo Simulations}\label{sec:MC}

To test the veracity of our group finding method and look for spurious groups in our identified FOF associations we performed Monte Carlo simulations.  All 1961 confident galaxy redshifts were first randomly shuffled between their positions and then the new supercluster field members in the redshift range of $0.858 < z < 0.946$ were selected, now with their new positions.  We then performed the same friends-of-friends test as done on our non-shuffled data.  This was repeated 1000 times and the average number of FOF associations and membership number per association was compared with our real data.  

The average number of FOF associations returned in the MC tests was 8.8, compared to the 14 recovered overdensities in the real data.  Figure~\ref{fig:MC} shows the number of FOF associations in each membership bin for the real data (red hatched histogram) versus the average numbers over all shuffled MC tests (black outlined histogram).  The bins containing the FOF associations identified as the main cluster cores (see \S\ref{sec:recoveredStructure} for details) are labeled with their corresponding cluster ID.  This shows that larger FOF associations are recovered only in the real data, including the cores of clusters C and A at 13 and 23 members, but not in the randomly shuffled data, leading us to believe that FOF associations with $\ge5$ members are tracing real group and core structures.  Also, since the positions of the spectroscopic sampling are consistent between the real data and the MC test data, we are not recovering large structures simply due to highly targeted regions.  

\subsection{Recovered Structures in the RCS\,2319+00 Supercluster Field}\label{sec:recoveredStructure}

Taking into account the results of the MC simulations, we designate all FOF associations with $\ge5$ members recovered from the FOF test on our real data as a lower estimate on the true substructure of the RCS\,2319+00 supercluster, noting that the simulations show the possibility that one of the recovered 5-8 member groups may still be a spurious detection (see Figure~\ref{fig:MC}).  Several structural features are immediately discernible in Figure~\ref{fig:structure}.  The cluster cores all comprise groups that seem to extend in an elongated fashion pointing to the neighbouring core, with 23 and 13 members in the core groups in clusters A and C.  The galaxies in the core group of cluster A (Group 2) span a redshift range of $\sim$6000~km~s$^{-1}$ and have a dispersion of 1404~$\pm$~249~km~s$^{-1}$, with no interloper clipping.  This is slightly higher, though within errors, than the dispersion we calculated in \S\ref{sec:reds} from the spatial distribution and interloper clipping.  In cluster C, the main core group (Group 10) contains galaxies spanning $\sim$960~km~s$^{-1}$ with an estimated dispersion of 722~$\pm$~182~km~s$^{-1}$, very close to the dispersion calculated in \S\ref{sec:reds} with interloper clipping.

\begin{figure}[tb!]
\centering
{\includegraphics[width=9cm, trim=0.25cm 0cm 0cm 0.5cm, clip=true]{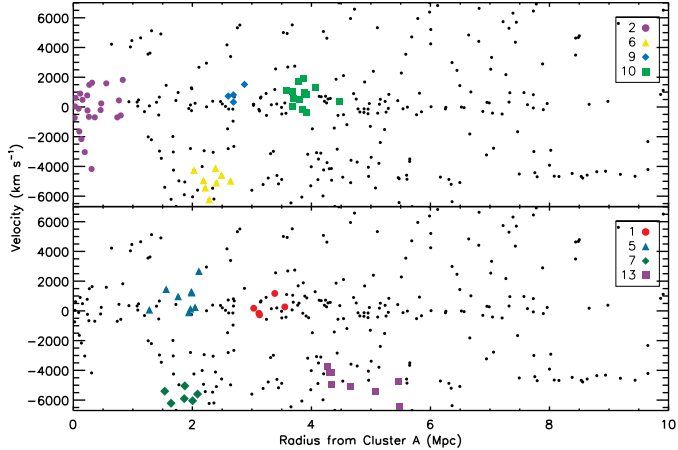}}
\caption{\small Velocity versus radius for all confirmed cluster members in the RCS\,2319+00 supercluster relative to cluster A.  \textit{Top panel}:  Groups that correspond to the spatial positions of the three cluster cores are highlighted: cluster A - group 2, cluster B - groups 6 and 9, cluster C - group 10.  The two groups near the spatial center of cluster B lie on two of the distinct redshift walls that extend through the supercluster region.  We assign group 4 as the cluster core as it falls along the same velocity space as the two other cluster cores and corresponds to the red-sequence peak redshift in this region.  \textit{Bottom panel:}  Groups having $\ge$5 members outside of the cluster cores are highlighted.  Two larger groups are seen to be potentially falling into the cluster from the lower redshift peak.  The filamentary structure between clusters B and C (group 5) is shown to lie within the same redshift wall as the core groups.  \label{fig:vel_rad}}
\end{figure}

Figure~\ref{fig:vel_rad} shows the relative velocity of the galaxies in groups from the average redshift of cluster A as a function of the spatial distance from cluster A, with the identified $\ge5$ member groups highlighted against all confirmed members.  The numbers and colors plotted for each group correspond to the same numbers and colors  as Figure~\ref{fig:structure}.  The top panel highlights the groups consistent with the spatial centers of the cluster cores and shows that the core group of cluster A is elongated in velocity space as well as in its spatial distribution, while the core group of cluster C is more tightly bound.  The lower redshift galaxies in the core group of cluster A may be interlopers from the lower redshift structure, from which they seem to extend.

Once again, cluster B presents some difficulties.  The two redshift peaks that were apparent in Figure~\ref{fig:coreHists} fall into two separate groups with 8 members at an average redshift of 0.870 and 4 members in a group $\sim$5700~km~s$^{-1}$ away at z=0.906.  Again, we suspect that more galaxies belong to the higher redshift group in this core, corresponding to the red-sequence density, X-ray contours and BCG, that have yet to be targeted with spectroscopy.  Figure~\ref{fig:vel_rad} again shows that the smaller group in cluster B (group 9) lies along the same velocity space as clusters A and B.  It can also be seen in Figure~\ref{fig:structure} that this higher redshift group falls in the region of lower spectroscopic target density.  In light of this, we assign the 4 member group 9 as the core group of cluster B and the 8 member group 6 is considered a potentially infalling group distinct from the core of the cluster.  We attempt to correct for spectroscopic completeness in Section~\ref{sec:LF}.  

An interesting structure recovered through the FOF algorithm is an apparent 9 member filamentary structure (Group 5) stretching from cluster B towards cluster A, first hinted at in the results of the DS test around the $z \sim 0.90$ redshift wall (\S~\ref{sec:DStest}, Figure~\ref{fig:DS}).  This structure has an average redshift of $z = 0.906$ with galaxies spanning $\sim$2800~km~s$^{-1}$ towards higher redshift (bottom panel, Figure~\ref{fig:vel_rad}).  It corresponds well to the previously confirmed Herschel infrared bright filament that \citet{Coppin2012a} found may be forming stars with a SFR~$\simeq$~900~M$_{\odot}$~yr$^{-1}$ and building up significant stellar mass.  

Other notable groups include a 6 member group centered between the three component cluster cores (group 7) as well as an 8 member filamentary group falling in towards cluster C (group 13).  Both of these groups appear to be falling into the clusters from the lower redshift wall (bottom panel, Figure~\ref{fig:vel_rad}) and may represent infalling groups. 

Comparing our recovered groups from the FOF method with the substructure detected with the DS test (see the DS test bubble diagram, Figure~\ref{fig:DS}), we see that the central group (7), the filamentary structure (group 5), and the 8 member group in the line of sight of cluster B (group 6) correspond to regions of deviation in the DS test.  Cluster C has one large core group (10), which is relatively compact in velocity space, and a small, three member FOF association (12) found in the same line-of-sight.  This small overdensity falls within the lower redshift wall and accounts for the deviation seen in the DS test of this redshift slice when tied with the other blue-shifted galaxies seen south-west of the cluster core in Figure~\ref{fig:structure}.  This area and other areas of possible substructure in the DS test found to have small, 3-4 member overdensities with the FOF method (groups 4 and 8), lack adequate members in the immediate area to confirm substructures.  The addition of further spectroscopic redshifts may serve to confirm the DS substructures. 

We find that of our 302 cluster members in the supercluster field $\sim$13\% are found in the core cluster groups and $\sim$18\% belong to the potential infalling groups.  We note that these are rough estimates, since our supercluster `field' galaxies may in fact be tied into groups given further spectroscopy.  Areas of deviation from the global properties in the DS test that are not recovered with the FOF method may belong to structures whose linking members have not yet been confirmed spectroscopically, leaving them out of the FOF group list and identified as part of our superstructure `field' population.  Conversely, we may find a higher proportion of isolated galaxies given further spectroscopy.  At low redshift \citet{Small1998} found that approximately one third of the galaxies associated with the Corona Borealis supercluster do not belong to the component clusters.  At higher redshifts we would expect the number of galaxies in superclusters without specific cluster associations to increase as the clusters are less relaxed.  This has already been seen by \citet{Gal2008} in their study of the structure of the CL1604 supercluster at $z \sim 0.9$, where they found that a large portion of galaxies were associated with the connecting large-scale filaments and not with the individual clusters and groups in their supercluster.  As such, while we do expect that additional spectroscopically confirmed supercluster members will result in more discernible structures, the percentage of unassociated supercluster members will most likely remain high.  

\section{The Color-magnitude relation}\label{sec:CMR}

A color-magnitude relation (CMR) for elliptical field galaxies was first identified by \citet{Baum1959}, with bright ellipticals having redder colors then their fainter counterparts.  In clusters, the CMR for early-type galaxies, the red-sequence, has been shown to have a tight, well-defined slope at low redshift \citep[eg][]{Visvanathan1977, Bower1992, Lopez-Cruz2004} and the cluster cores are dominated by early-type galaxies \citep{Dressler1980}.  At higher redshifts however, the dynamically younger clusters have shown an increasing scatter in their CMR and a deficit of red-sequence galaxies at the low-luminosity end of the slope \citep[e.g.][]{DeLucia2004, Tanaka2005, Mei2009, Lemaux2012}.  This is seen as a sign that the cluster red-sequence is still in the process of being built up at $z \sim 1$.  \citet{Rudnick2012} propose that the lack of low-luminosity red-sequence galaxies may be due to the accretion of faint blue galaxies whose star-formation is suppressed in groups or in the cluster and experience mergers with other red-sequence galaxies, increasing their mass and moving them along the red-sequence.  Recent studies of clusters a $z > 1$ differ over the basis for the scatter in the red-sequence at higher redshifts, with clusters at $z =$ 1.46 and 1.62 found to be consistent with models of a single burst of star-formation \citep[][respectively]{Hilton2009,Papovich2010} while more diverse star formation histories are needed to explain the width of the CMR in clusters at $1.0 < z <1.5$ \citep{Snyder2012}.

\subsection{Color-magnitude diagram}\label{sec:col-mag}

\begin{figure}[tb!]
\centering
{\includegraphics[width=9cm, trim=0.75cm 0.25cm 0.cm 0.5cm, clip=true]{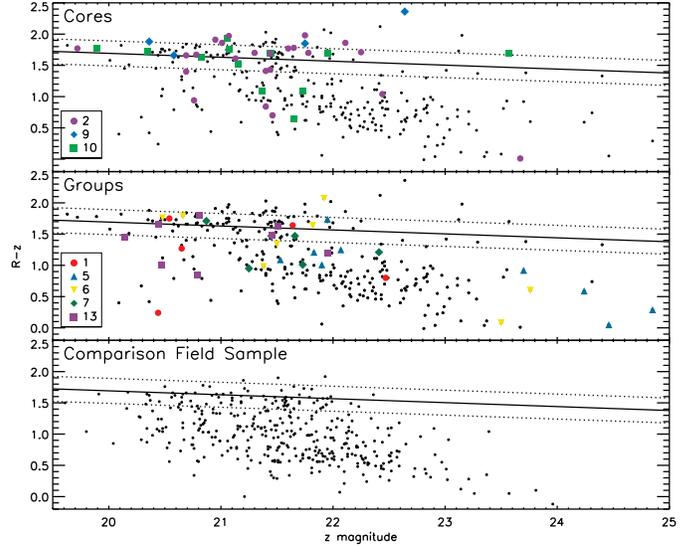}}
\caption{\small $R_c-z'$ versus $z'$ color-magnitude diagram.  Black points represent all spectroscopic galaxies in the redshift range $0.858 \le z \le 0.946$.  The top two panels plot all members of the RCS\,2319+00 structure with colored symbols highlighting galaxies belonging to the numbered groups identified with the FOF algorithm as shown in Figure~\ref{fig:structure}, divided into the core groups (top panel) and $\ge5$ member groups (middle panel).  The bottom panel shows the comparison field sample from the large RCS IMACS spectroscopic campaign (R. Yan et al.\,2012, in preparation).  Lines represent the red-sequence slope at z=0.9 \citep[solid line, from][]{Gilbank2008a} and $\pm$0.2 of the red-sequence slope (dotted line).  Note that no correction for redshift success has been applied to the color-magntude diagram. \label{fig:col-mag}}
\end{figure}

In this section we investigate the color-magnitude relation in the RCS\,2319+00 supercluster.  The R$_c$--z$'$ versus z$'$ color-magnitude diagram for all spectroscopically confirmed galaxies within the supercluster redshift range of $0.858 \le z \le 0.946$ is shown in Figure~\ref{fig:col-mag}, with the red-sequence slope at $z~=~0.9$ \citep[from][]{Gilbank2008a} plotted for reference as a solid line, and dotted lines denoting $\pm 0.2$ mag from the the red-sequence.  The errors on the photometric catalog range from 0.01~mag for the brightest galaxies, increasing to 0.62~mag for the faintest, $z' = 24.85$ magnitude cluster member and 0.73~mag for the faintest $R_c = 25.73$ magnitude member, with an average error of $\sim 0.08$~mag in both filters.  

The two top panels of Figure~\ref{fig:col-mag} both show all spectroscopically confirmed RCS\,2319+00 supercluster members plotted as black circles.  The colored symbols in each panel represent members of the identified groups from our friends-of-friends algorithm in \S\ref{sec:structure} with the numbers and colors plotted for each group corresponding to the same numbers and colors on the structure plot (Figure~\ref{fig:structure}) separated into the two panels to highlight the cluster core groups (top panel) and the $\ge5$ member groups (middle panel).   The bottom panel shows the color-magnitude diagram of a comparison field sample, detailed below.  It should be noted that the redshift success rate of the spectroscopic sample is a function of both the magnitude and color for galaxies fainter than $z'\sim22$ mag (see Figure~\ref{fig:specSuccess}), and is not accounted for in the color-magnitude diagram

Due to the incompleteness of our spectroscopic sample, we attempt a rough completeness correction when looking at the overall population distribution in the color-magnitude diagram.  Since the spectroscopic correction weights do not take into account the color of the galaxies, we instead use the redshift success rates for bright ($z' \le$ 22 mag) and faint ($z' >$ 22 mag) galaxies found in the right panel of Figure~\ref{fig:specSuccess}.  We set the success rate of the bright galaxies conservatively at 70\% from our 70-80\% redshift success rate for bright galaxies.  For the faint galaxies, we define a color cut at $R_c-z' = 1$ and find that bluer galaxies have a success of $\sim$50\% while the redder galaxies have a redshift success of only $\sim$25\%.  We can use these rates to correct the faint galaxy numbers in Figure~\ref{fig:col-mag} up to the redshift success rate of the bright galaxies in order to better compare the populations.  This is an oversimplification of the completeness correction, however it is sufficient given our small number statistics.  

Using all spectroscopic structure members corrected to the redshift success of bright galaxies and defining a rough limit for the red versus blue galaxies at  -0.2 mag of the red-sequence line, we find 36$\pm$3\% of the galaxies are red and 64$\pm$4\% of the structure members are blue galaxies, with Poisson errors.  We can compare these populations to the field in the same redshift range using a spectroscopic sample of galaxies derived from observations whose target cluster falls outside this redshift range in the RCS IMACS spectroscopic campaign (R. Yan et al.\,2012, in preparation).  We find 422 spectroscopically confirmed field galaxies that fit this criteria in the IMACS sample.  The field sample is plotted in the bottom panel of Figure~\ref{fig:col-mag} with the red-sequence line again at $z = 0.9$ again plotted for reference.  Those galaxies populating the red-sequence make up only 18$\pm$2\% of the sample while the blue cloud is populated by 82$\pm$4\% of the comparison field sample.  The limit on the depth of the spectroscopy is brighter for the field sample at $z'\sim24$ mag ($M_{z'} \sim -18.35$) compared to the limit of the RCS\,2319+00 sample at $z' \sim 25$ mag ($M_{z'} \sim -17.35$).  As such, we can set the depth of the RCS\,2319+00 structure to that of the field sample for better comparison.  The fraction of red and blue galaxies in the structure to this brighter depth remains unchanged.  One caveat to the field sample is that, though it is drawn from spectroscopic observations whose main target cluster lies outside of the supercluster range, there may exist groups and low-richness clusters within the supercluster range that have not yet been identified.  Despite this, our supercluster galaxies show an excess red-sequence over the field population, as expected in clusters.  To further investigate this excess we look at the identified structure from \S\ref{sec:structure}.

It can be seen that the cluster cores (groups 2, 9, and 10) show a more prominent red-sequence than the groups, with the core groups having 77$\pm$13\% red galaxies and only 23$\pm$7\% in the blue cloud, for all identified members down to the fainter $z' \sim 25$ mag limit.  The large errors are due to the small numbers, even after correcting for the redshift success.  The groups show a lower fraction of red-sequence galaxies at 37$\pm$10\%, with the percentage of blue cloud galaxies almost three times higher than those in the cores at 63$\pm$13\%.  Though we do not expect to see significant differences between the populations given the small numbers of galaxies in our sample and the large errors, the populations are still not consistent within errors, only crossing at $\sim2\sigma$.  From this we infer that the two group categories contain galaxy populations that may be in slightly different stages of evolution.  In their study of the Cl1604 supercluster at $z \sim 0.9$, \citet{Lemaux2012} found that the fraction of their cluster and group galaxies in the red-sequence was identical at 47\%, suggesting a similar level of processing in the two environments.  They also found that 70\% of the red-sequence galaxies in their supercluster belong to the clusters and groups.  Combined, the core and group members of the RCS\,2319+00 supercluster account for only 38$\pm$5\% of the corrected red-sequence supercluster members, which again hints that some fraction of the isolated supercluster `field' galaxies may belong to larger, as yet unidentified structures.  In the supercluster isolated `field' population, which accounts for 226 of our 302 supercluster members, 29$\pm$3\% of the corrected galaxies are red while the remaining 71$\pm$5\% are in the blue cloud region, consistent with the group fractions.  

The interesting filamentary structure between clusters A and B (group 5) has 8 of its 9 identified members in the blue cloud and includes some of the faintest identified members so far, two of whom are also among the bluest galaxies currently identified in the overall structure.  This is not surprising as \citet{Coppin2012a} have shown that this region is undergoing an excess of star formation activity.  If we remove this unique structure from our group category, the remaining $\ge$5 member groups contain 43$\pm$12\% red-sequence galaxies, still lower than the fraction of red-sequence galaxies in the cluster cores, though now consistent within $<1.5\sigma$.  This slightly higher group red-fraction is in closer agreement with the results of \citet{Lemaux2012} for both their groups and cluster galaxies.  On average then, we believe the groups in our structure have undergone some processing, though less than the cluster core galaxies, with the filamentary structure standing out as a unique environment of blue, star-forming galaxies currently undergoing intense activity.

Recalling that the Monte Carlo simulations run in \S\ref{sec:MC} showed the possibility of one spurious 5-8 member FOF group, we look to see what effect this might have on the red-sequence fractions of the populations.  The validity of the core groups is supported by the optical color data and the X-ray detections and we are therefore confident that the FOF overdensities represent real groups.  The filamentary structure is detected in the infrared and again we assume that it's corresponding FOF group can therefore be assumed to trace a real structure.  This leaves only one of the four remaining $\ge5$ member groups as the possible spurious structure detection.  If we remove the galaxies belonging to a single group and reassess the fraction of red vs blue galaxies remaining for the group population, repeating the process until each group has been removed once, we find that the fraction of red and blue galaxies does not change significantly, with the fraction of red galaxies varying from 41-45\% with errors of $\le14\%$, consistent with the previous results.    

With the aid of the color information, we can revisit the assignment of group 9 as the core group of cluster B over group 6, the lower redshift group also consistent with the spatial position of the red-sequence overdensity and the X-ray emission.  In Figure~\ref{fig:col-mag} it can be seen that all of the galaxies currently identified in group 9 are red, while 50\% of the identified galaxies in group 6 are blue, closer to the fraction seen for the groups.  This lends more credence to our claim that group 9 is the core group of cluster B while group 6 is a possible infalling group from the lower redshift wall.

\subsection{Color Distribution In Different Environments}\label{sec:colHist}

\begin{figure}[tb!]
\centering
\subfigure{\includegraphics[width=9cm, trim=0.25cm 0.25cm 0.5cm 0.5cm, clip=true]{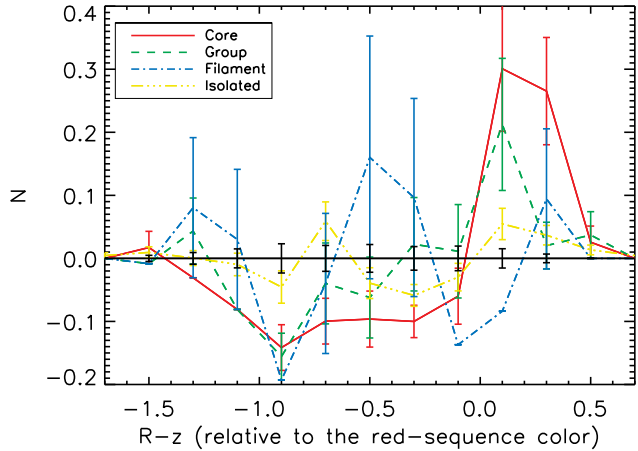}}
\caption{\small Number of galaxies in each environment normalized to one versus $R_c-z'$ color for the spectroscopically confirmed supercluster members plotted as a difference over the normalized comparison field sample.  Errors given are Poisson errors found for each population.  The solid line shows the color distribution in the cluster core galaxies over the field, the dashed line represents the $\ge5$ member, non-core groups with the filament removed, the dash-dot line shows the filament galaxies and the dash-dot-dot-dot line encompasses all remaining isolated supercluster `field' members.\label{fig:colHist}}
\end{figure}

To further investigate the populations in the different environments and look for possible signs of the build-up of the red-sequence we look to the distribution of galaxy colors in the different density regions.  Here we do not attempt to correct for the redshift success based on magnitude or for the galaxies mass or luminosity differences, which will be the subject of a follow up paper.  Instead, we normalize the number counts in the different populations to one in order to look at the relative color distributions. In Figure~\ref{fig:colHist}, we plot the normalized number counts of galaxies in the different environments, binned in color and plotted as a difference over the normalized comparison field sample discussed above.  We first k-correct the apparent magnitudes using the spectroscopic redshifts and version 4.2 of the $\mathsf{kcorrect}$ code by \citet{Blanton2007} to better compare the samples.  The solid line shows the color distribution in the cluster core galaxies over the field, the dashed line represents the $\ge5$ member, non-core groups with the filament removed, the dash-dot line shows the filament galaxies and the dash-dot-dot-dot line encompasses all remaining isolated supercluster `field' members.  The Poisson errors plotted are based on the parent population only, with no scaling applied when the difference between each population with the field population is found. 

The isolated supercluster galaxies possess the expected bimodal distribution of blue and red galaxies over the field, with a distinct, underpopulated `green valley', though $\sim38$\% of our spectroscopic cluster galaxies fall in this intermediate color area.  The core and group galaxies both show peaks in the red end of the histogram, as expected, with few blue galaxies and a deficit of `green' galaxies compared to the field.  The group galaxies do show a slight excess of galaxies at the bluest end of the distribution, though the large errors render the peak consistent with the comparison field at this color.  

The filament between clusters A and B shows three peaks above the field: a small peak at the red sequence end; one populating the green valley; and a final excess at the bluest end of the color distribution.   The small number of filament galaxies currently in our spectroscopic sample (9 members) results in large errors and therefore the three filament peaks are consistent with the field within the errors.   Though we do not have statistically significant data with such small sample numbers in the green galaxy population from the filament, we propose that, if this peak is real, these intermediate color galaxies may in fact belong to a transient population in the process of migrating to the red-sequence, similar to those seen by \citet{Balogh2011} in their study of seven spectroscopically confirmed galaxy groups at $0.85 < z < 1$.  They found that 30\% of their confirmed group members populated this intermediate `green' color region and propose that these galaxies represent a transient population who have recently ceased their star formation and are in the midst of migrating from the blue cloud to the red-sequence.  Currently, two thirds of the filament galaxies are found in this intermediate area of the color-magnitude diagram versus the $\sim45\%$ of field galaxies at this color.  Further studies with multiwavelength data on the RCS\,2319+00 supercluster is needed to determine the properties of these galaxies.

\section{Luminosity Functions}\label{sec:LF}

We wish to put a rough estimate on the average mass of our identified groups in order to better trace the buildup of our supercluster and to place our groups in context with theories of hierarchical formation from semi-analytic models \citep[eg.][]{McGee2009, Berrier2009, DeLucia2012, Cohn2012}.  The majority of our groups have too few spectroscopic members to compute reliable velocity dispersions and, as we have seen in \S\ref{sec:virial}, masses derived from velocity dispersions can carry large errors even with a larger sample.  Instead, we attempt to estimate the mean mass of our groups by fitting luminosity functions (LF) to two stacked group categories; the cluster cores and the potential infalling groups.   

Luminosity functions measure the number of galaxies per volume in a given luminosity bin and can be used to test theories on the formation of galaxies, their evolution, and what role the environment plays on these factors. The faint-end of the luminosity function is populated by a mix of faint red and blue galaxies and slight variations in its slope can be used to help distinguish between different structure formation theories.  A steeper faint-end slope may be indicative of the accretion of faint galaxies onto the group or cluster while a flatter slope would be expected in a relaxed cluster dominated by early-type galaxies \citep[eg.][] {Goto2002, Propris2003,Goto2005, Zenteno2011}.  LFs can also be used to estimate the number of galaxies in each group or cluster by integrating the function to given luminosity limits.  While cluster luminosity functions are generally constructed from background subtracted photometric catalogs, we attempt to measure the luminosity functions of the stacked clusters and groups in our supercluster through the recovered spectroscopic membership from the friends-of-friends search for substructure in the previous section.  

We use the functional form of the Schechter function \citep{Schechter1976} to fit the luminosity function,
\begin{equation}
\label{eq:LF}
\small{\phi(M) = 0.4\,\textup{ln}(10) \phi^{\ast} (10^{0.4(M^{\ast}-M)})^{1+\alpha}~\textup{exp}(-10^{0.4(M^{\ast}-M)})},
\end{equation}
where $\alpha$ is the slope of the faint-end power-law, $M^{\ast}$ is the characteristic magnitude and $\phi^{\ast}$ is the normalization of the Schechter function.  

\subsection{Luminosity Function of the RCS\,2319+00 Supercluster Field}\label{sec:2319LF}

We first construct a LF for the full catalog of 302 spectroscopically confirmed supercluster members in our 35\,x\,35 arcminute region to estimate the value of $M_{z'}^{\ast}$ and look at the shape of the faint-end slope.  First, we convert the photometry of all cluster members to absolute magnitude $M_{z'}$ using,
\begin{equation}
\label{eq:LF}
M_{z'} = m_{z'} - 5 \textup{log}\left[\frac{D_L(z)}{10~\textup{pc}}\right] - K(z),
\end{equation}
where $D_L(z)$ is the luminosity distance to our average supercluster redshift and $K(z)$ is the k-correction computed using version 4.2 of the $\mathsf{kcorrect}$ code by \citet{Blanton2007}.  Our apparent magnitudes ($m_{z'}$) have already been corrected for galactic dust extinction.  The spectroscopic magnitude weights calculated in \S \ref{sec:specComp} are applied to each confirmed member with no geometric correction as we are looking at the statistics over the whole field and not at individual galaxy properties.  The completeness weighted spectroscopic members were then binned by absolute magnitude in 0.25 mag bins and the bins for each core or $\ge5$ member group were stacked.  Fractional errors were added to the stacked, weighted number counts in each magnitude bin.  The Schechter function was fit using the Levenberg-Marquardt algorithm for non-linear least squares fitting using the $\mathsf{mpfit}$ routine for IDL \citep{Markwardt2009}.  

\begin{figure}[tb!]
\centering
{\includegraphics[width=8.5cm, trim=0.5cm 0.5cm 0cm 0cm, clip=true]{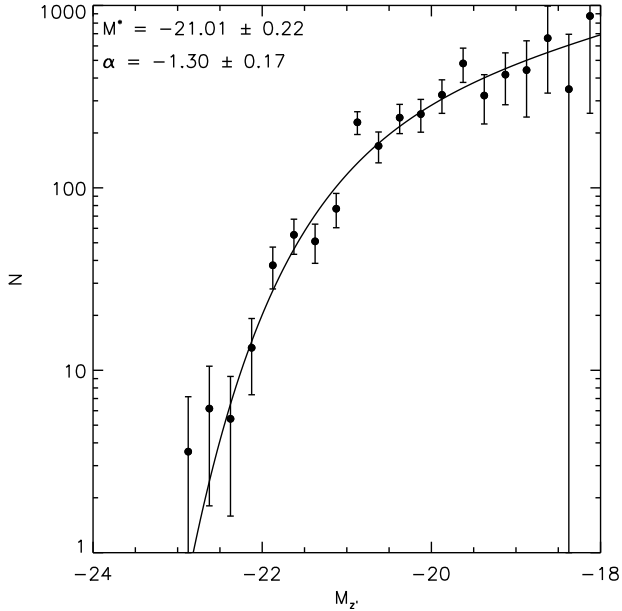}}
\caption{\small LF of the best-fit Schechter function to all completeness corrected spectroscopic supercluster members in the 35\,x\,35 arcminute region with the BCGs from the three cluster cores removed so as not to influence the bright-end fit of the LF. \label{fig:cmLF}}
\end{figure}

Figure~\ref{fig:cmLF} shows the LF for all weighted, spectroscopically confirmed members in the RCS\,2319+00 supercluster.  The BCGs from each of the three cluster cores were removed from the fitting so as not to influence the bright-end fit of the LF.  The solid line shows the best fit Schechter function.  The characteristic magnitude $M^{\ast}_{z'}$ is estimated to be $M^{\ast}_{z'}=-21.01~\pm~0.22$.  This value is slightly lower than the value derived by \citet{Gilbank2008a} of $M^{\ast}_{z'} = -21.39$ at $z = 0.9$ using the passively evolving model on a fit to the LF of composite RCS clusters at $z = 0.4$.  We note that our fitted value of the faint-end slope of $\alpha = -1.30 \pm 0.17$ is at the high-end of the values found with other galaxy clusters in the literature.  The fit to the composite RCS clusters of \citet{Gilbank2008a} resulted in a faint-end slope of $\alpha = -0.94 \pm 0.04$.  \citet{Lin2004} used a composite K-band LF of 93 X-ray selected groups and clusters from the Two Micron All Sky Survey and found a faint-end slope is fit by a slope ranging as $-0.84 \le \alpha \le -1.1$.  The closest agreement to our estimate of $\alpha$ was found by \citet{Zenteno2011}, who measured an average faint-end slope of $\alpha \approx -1.2$ for their four Sunyaev-Zel'Dovich selected clusters in the $r, g, i,$ and $z$-bands.  

To construct our supercluster LF, we use the weighted number counts of all spectroscopically confirmed cluster members over the broad supercluster region rather than using background subtracted number counts from photometry in the much smaller radii of $r_{500}$ \citep{Lin2004}, $0.5r_{200}$ \citep{Gilbank2008a} or $r_{200}$ \citep{Zenteno2011} around groups and cluster cores as done in the previous studies.  The steeper than average faint-end slope we find for RCS\,2319+00 may therefore be reflective of the presence of infalling structures and individual galaxies in the lower density regions of the supercluster field.  The addition of these lower density regions to our LF may also be the cause of the slightly lower characteristic magnitude found here compared to that found by \citet{Gilbank2008a}.  However, we must again note that our fit is based on an incomplete, completeness corrected spectroscopic sample, which may also affect the shape of the LF.  For these reasons, we choose to use the values for the $M^{\ast}_{z'}$ and $\alpha$ parameters found for the composite RCS galaxy clusters by \citet{Gilbank2008a} when looking at the LFs of the core and $\ge5$ member groups in the following section as they appear to be more representative of the denser cluster cores and groups than our fitted values from the broad supercluster field.  

\subsection{Luminosity Functions of the Cluster Cores and Groups}\label{sec:groupLFs}

To construct the LFs of our stacked $\ge5$ member groups and the cluster cores recovered from the friends-of-friends output we follow the same methods described above for the whole supercluster member catalog with the exception of the spectroscopic completeness weights, which were calculated using the photometric catalog in a smaller radius around the mean group position (see \S\ref{sec:specComp} for details on the spectroscopic weight calculations) rather than the completeness weight for the full 35\,x\,35 arcminute field of view.  This was done to simulate a smaller group radius of $\sim r_{200}$, which we initially set at 1.5 Mpc for the cluster cores, based on the values found for the clusters using the \citet{Evrard2008} relation (see Table~\ref{table:r200-m200}) and an arbitrary radius of 1 Mpc for the $\ge5$ member groups.  Since the LFs of our recovered groups and cluster cores are based on a small number of completeness weighted spectroscopic sources per group, we find that fitting all three Schechter parameters simultaneously is not reliable. Instead, we fix the characteristic magnitude and faint-end slope to those measured for the composite RCS clusters by \citet{Gilbank2008a} of  $M^{\ast}_{z'} = -21.39$ and $\alpha=-0.94$, as motivated above.  For the stacked LF of the three core groups (2, 9 and 10) the BCGs were again removed from the fitting so as not to influence the bright-end fit.  

\begin{figure}[tb!]
\centering
\subfigure{\includegraphics[width=8.5 cm, trim=0.5cm 0.5cm 0cm 0cm, clip=true]{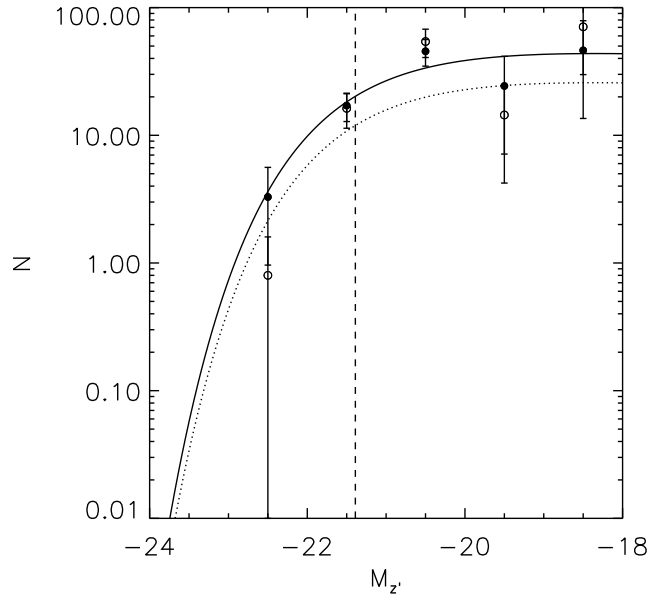}}
\caption{\small Stacked LFs for the three cluster cores  in the supercluster from 1.5 Mpc completeness corrected spectroscopic members found using the FOF algorithm (closed circles, solid line) and for the 1 Mpc completeness weighted members in the five $\ge 5$ member recovered groups (open circles, dotted line) in magnitude bins of $\pm$0.5 mag.  The curves show the best-fit Schechter functions using  $M^{\ast}_{z'} = -21.39$ (vertical dashed line) and $\alpha = -0.94$ from \citet{Gilbank2008a}. \label{fig:stackedLFs}}
\end{figure}

The stacked luminosity functions are shown in Figure~\ref{fig:stackedLFs}.  The LF of the stacked cluster cores is shown as closed circles with the best-fit function plotted as a solid line.  The open circles and dotted line show the results for the stacked $\ge5$ member groups.  The dashed vertical line indicates the position of $M^{\ast}_{z'}$ for reference.  The luminosity functions with fixed $M^{\ast}_{z'}$ and $\alpha$ were fit to $\phi^{\ast}$, integrated to estimate the number of galaxies in the stacked groups, $N_{gals}$, and then divided by the number of groups in the stacked LF as a proxy for the mean richness of the groups.  We used two methods of estimating the group mass from the number of galaxies. 

The first method employs the halo occupation number (HON), based on the halo occupation distribution (HOD); the probability distribution of finding $N$ galaxies in a halo of mass $M$ \citep[eg.][]{Seljak2000, Benson2001, Berlind2002, Lin2004, Lin2006}.  In the HOD formalism, the HON is defined as $\langle N \rangle (M)$, the mean number of galaxies per halo, and can be estimated by integrating the LF.  \citet{Lin2004} used a sample of 93 X-ray selected groups and clusters in the nearby universe to form a composite K-band LF and found a $N_{500}-M_{500}$ scaling relation.  Extending their work to $z = 0.9$, \citet{Lin2006} found no signs of evolution in the scaling relation.  \citet{Zenteno2011} converted this relation to $N_{200}-M_{200}$ for their work with Sunyaev-Zel'Dovich selected clusters and found that the \citet{Lin2004} relation becomes:
\begin{equation}
\label{eq:N-M}
N_{200} = (36 \pm 3) \left(\frac{M_{200}}{10^{14}\,h^{-1}_{70}\,\textup{M}_\odot}\right)^{0.87\pm0.04}.
\end{equation}
The $N-M$ relation used is based on the number of galaxies within the viral radius of the group or cluster.  Since we do not have that information available and our LFs are constructed from weighted spectroscopic samples with small numbers of known galaxies, we retain our 1.5 and 1 Mpc completeness corrections initially for these estimates for the cores and $\ge 5$ member groups respectively.  The total number of galaxies $N_{200}$ is estimated from the integration of the luminosity function.  Here the integration is done to a limit of $M^{\ast} + 2$.  

Using the estimated $N_{200}$ and its resulting mass, a new estimate of the virial radius $r_{200}$ can be found and a new spectroscopic weighting for the galaxies can be performed to reflect this new radius.  This is done iteratively until the resulting radius matches the input radius.  The mass estimates with the initial and final radii vary by $< 0.1 \times 10^{14} M_\odot$, with the core mass decreasing slightly and the group mass increasing for the final, smaller radii.  For the cluster cores, we find an average $N_{gals}=23.98$ and a mass estimate of $\sim 0.9 \times 10^{14} M_\odot$ using the \citet{Lin2004} $N_{200}-M_{200}$ scaling relation with the galaxies completeness weighted to an average virial radius of $\sim 0.7$ Mpc.  The mass estimated for the $\ge$5 member groups is over a factor of two lower at $\sim 0.4 \times 10^{14} M_\odot$ for $N_{gals}=11.55$ per group in a $\sim0.5$ Mpc radius.  

The second method we use to estimate the masses of our groups from the stacked LFs involves the cluster and group richness estimate used with the Sloan Digital Sky Survey maxBCG catalog \citep[e.g.][]{Hansen2005,Hansen2009,Becker2007,Johnston2007}.  In this case the richness, $N_{200}$, represents the number of red galaxies that are fainter than the BCG in an area that approximates $r_{200}$.  $N_{200}$ is first calculated for an arbitrary radius, which we take to be the 1.5 and 1 Mpc radii used above, and then calculated using the new $r_{200}$ from the maxBCG $N_{200}-r_{200}$ relations.  Richness-size relations have been found for the maxBCG catalog from both the radial density profiles of clusters \citep{Hansen2005} and weak lensing \citep{Johnston2007}, with the weak lensing relation scaling to the cluster halo mass.  

Due to our small numbers and completeness weighting, which is not dependant on color information, we can not readily extract the red and blue galaxies into separate LFs to find different $N_{gals}$ for the two populations.  Instead, we use the rough percentage of red-sequence galaxies in each of the group categories found in \S~\ref{sec:col-mag} of $\sim$77\% for the core groups and $\sim$33\% for the $\ge$5 member groups.  We used these percentages to scale the $N_{gals}$ recovered from the LFs to an approximate $N_{200}$ to be used with the maxBCG relations.  For the cores, we find $N_{200} \sim 41.2$ from $r_{200} \sim 1.1$ Mpc and subsequent mass estimates of $\sim 6 \times 10^{14} M_\odot$ per cluster.  For the $\ge$5 member groups we find $r_{200} \sim 0.6$ Mpc and $N_{200} \sim 6.7$, resulting in mass estimates of $\sim 5 \times 10^{13} M_\odot$ per group.

The average group mass estimates resulting from the different $N-M$ scaling relations are relatively consistent, with the cluster cores having masses $\sim 10^{14} M_\odot$, an order of magnitude more massive than the $\gtrsim 10^{13} M_\odot$ estimates for the potential infalling groups.  The mass estimate for the cluster cores of $\sim 0.9-6 \times 10^{14} M_\odot$ is consistent with those calculated from the spectroscopic velocity dispersions and the X-ray data of $10^{14.5}-10^{14.9} M_\odot$.  From this we reason that the average group mass from our LFs of $\sim 4 - 5 \times 10^{13} M_\odot$ is a reasonable initial group mass estimate.

As a test, the LFs were also fit with only the faint-end slope set and allowing $M^{\ast}_{z'}$ to vary.  This was done in order to check that any mass segregation between the denser cluster cores and the surrounding groups in the supercluster field did not cause the characteristic magnitude to shift drastically in the LFs of the stacked core/group galaxies (Figure~\ref{fig:cmLF}.  We found only slight changes in the value of $M^{\ast}_{z'}$ that did not result in any changes in the resulting average cluster core or group mass estimates.    

\subsection{Comparison to Theories of Hierarchical Structure Formation}

In order to put our $\sim 10^{13} M_\odot$ recovered groups in context, we look to studies which employ semi-analytic techniques to retrieve galaxy merger tree information from N-body simulations in order to trace the merger history of clusters.  In two recent studies by \citet{Berrier2009} and \citet{McGee2009}, slightly differing results were obtained as to the fraction of galaxies in clusters today that were accreted onto the cluster as a member of a $\sim 10^{13}$\,$M_\odot$ group.  \citet{Berrier2009} found that, of galaxies residing today in $10^{14.2}~h^{-1}$\,$M_\odot$ clusters, $<17$\% were accreted in groups of $\ge 5$ members, and $\sim$ 25\% from halos with masses of $\sim 10^{13}~h^{-1}$\,$M_\odot$.  For the same mass clusters, \citet{McGee2009} found a higher fraction of galaxies fell into the cluster from $\sim 10^{13}~h^{-1}$\,$M_\odot$ groups, at $\sim$32\%.  The discrepancy in the results of the two studies is likely due to the different methods used to populate the dark matter halos with galaxies.  

\citet{Cohn2012} also found that $\sim$ 30\% of galaxies in clusters with mass $\geq 10^{14}~h^{-1}$\,$M_\odot$ at $z=0.1$ had been accreted onto the final cluster as members of halos with $\geq 10^{13} h^{-1}$\,$M_\odot$, with the most massive cluster galaxies having a larger fraction coming from these groups.  This work also found that the largest subgroups accreted onto clusters were aligned along the clusters major axis, which we see in Figure~\ref{fig:structure} in groups 5 (the filamentary group between clusters A and B), 6 (large group in the line-of-sight of cluster B) and 13 (the large filamentary group seemingly falling into cluster C).  

Another study, by \citet{DeLucia2012}, also used galaxy merger trees to study the accretion of galaxies onto clusters.  While their study focuses primarily on the stellar mass of the galaxies accreted through different mass groups, they also find that a large fraction of the most massive cluster galaxies and about half of the low to intermediate stellar mass cluster members fell into the final cluster from $\sim 10^{13}$\,$M_\odot$ groups.   These studies all suggest that some pre-processing of cluster galaxies in groups is occurring prior to accretion onto the final cluster.  

The three clusters in RSC2319+00 are thought to be on a collision course to form a massive, $>10^{15}\,{M}_{\odot}$ cluster by $z=0.5$ \citep{Gilbank2008}.  If we consider cluster A to be the most massive halo in our supercluster (based on the X-ray masses) and assign it as the final host halo in the eventual merging of the RCS\,2319+00 supercluster, the spectroscopic members (as opposed to the completeness corrected members) of the remaining $\sim 10^{13}$\,$M_\odot$ groups and clusters currently account for 22\% of the spectroscopic cluster members.  We recall that our other identified groups ($<$ 5 members) as well as some of our isolated supercluster field galaxies may in fact be identified as larger groups given further spectroscopy and, on the other hand, we may find a larger isolated galaxy population in the supercluster field so this percentage may change.  

While comparing to these studies gives a general idea that our recovered groups likely represent real substructure in our supercluster that is destined to be accreted into the final cluster, a comparison to clusters at similar redshift and of similar final mass is more prudent.  \citet{McGee2009} extended their study to trace the history of accretion onto galaxy groups and clusters at different redshifts. They determined that at $z=1.0$ clusters with masses of $\sim10^{14.5}~ h^{-1}\,M_{\odot}$ had $\sim$36\% of their galaxies accumulated through $>10^{13}~ h^{-1}\,M_{\odot}$ halos.  For Coma-like clusters of $\sim10^{15}\,M_{\odot}$ at $z = 0.5$, which the RCS\,2319+00 supercluster is predicted to form, they found that $\sim 40$\% of the cluster galaxies had been accreted through large halos over the merger history.  Of more relevance is their result that by tracing the merger trees of the most massive M$_{clus}~\sim~10^{15.1}~h^{-1}$\,$M_\odot$ galaxy clusters today, they were found to never have more than 17\% of their galaxies in halos of mass $10^{13} - 10^{14}~h^{-1}$\,$M_\odot$, which occurred at a lookback time of $\sim$7 Gyrs, about 0.3 Gyrs more recently than the epoch at which we are viewing the RCS\,2319+00 supercluster.  In this context, our 22\% of galaxies in the supercluster currently residing in groups with masses of $10^{13} - 10^{14}$\,$M_\odot$, though slightly high, is comparable to the \citet{McGee2009} findings.  Again we reiterate our numbers are estimates based on incomplete spectroscopic coverage.  

\section{Summary}\label{sec:summary}

In this paper we have outlined the results of an extensive spectroscopic campaign to map out the substructure of the RCS\,2319+00 supercluster at $z \sim 0.9$.  
\begin{list}{$\bullet$}{\leftmargin=0.2in} 
\item From 1961 good confidence spectroscopic galaxy redshifts in a 35\,x\,35 arcminute region around the three main component clusters of the supercluster, we have found 302 structure members in a redshift range of $0.858  \le z \le 0.946$.  
\item The three clusters were found to lie at redshifts of 0.901, 0.905, and 0.905, a separation of only 630 km\,s$^{-1}$.  Combined with the $\le 3$ Mpc transverse separation of each component cluster to its nearest neighbor, the three clusters of RCS\,2319+00 represent the most compact supercluster found to date at this redshift.  
\item Three distinct walls in redshift space are apparent in the spectroscopic data, with the main supercluster peak surrounded by lower and higher redshift walls separated by $\sim$65 Mpc ($\Delta z \sim 0.03$).
\end{list}

\noindent We attempt to measure accurate velocity dispersions from the spectroscopically confirmed galaxies for the three main cluster components with the aim of determining the mass of each cluster.  
\begin{list}{$\bullet$}{\leftmargin=0.2in} 
\item \textit{Cluster A}: from 23 spectroscopically confirmed galaxies within 1 Mpc of the cluster center, we find a velocity dispersion of $1202 \pm 233$ km s$^{-1}$.  This results in a dynamical cluster mass of $\sim 10^{15} M_\odot$, consistent within errors with the lower mass estimates from the X-ray and weak lensing data of $\sim 6 \times 10^{14} M_\odot$.
\item \textit{Cluster B}: there are not enough cluster members left within 1 Mpc of the cluster center after interloper clipping to calculate a reliable velocity dispersion and we looked to the X-ray data to corroborate our dynamically derived properties.  The X-ray predicted velocity dispersion is consistent with the measured spectroscopic value of $1154 \pm 367$ km s$^{-1}$ and the resulting dynamical mass is consistent with the X-ray mass of $5.9 \times 10^{14} M_\odot$.  
\item \textit{Cluster C}: we believe that this cluster is in a relaxed state and that our measured velocity dispersion of $714 \pm 180$ km s$^{-1}$ and mass estimate of $2.5-3.8 \times 10^{14} M_\odot$ is accurate.  
\end{list}

\noindent The multiwavelength indications of substructure in the supercluster field along with the close proximity of the three clusters in redshift and angular space, which makes member assignment difficult, prompted the search for substructure in the supercluster.  
\begin{list}{$\bullet$}{\leftmargin=0.2in} 
\item We performed the Dressler-Shectman test to begin tracing the substructure of the supercluster by looking for deviations from the global mean parameters in the supercluster field.  We find evidence of substructure over the broad redshift range as well as within each separate redshift wall, with Monte Carlo tests showing a very low probability that the substructure is false.
\item To isolate specific structure, we used a modified friends-of-friends algorithm in an effort to associate supercluster galaxies with their parent clusters or groups.  
\item Comparing Monte Carlo simulations to the FOF recovered overdensities in the real data suggests that groups found to have $\ge$ 5 members in our real data are tracing legitimate structure.
\item We find evidence of possible infalling groups and extended structure in the main clusters, with five $\ge5$ member groups plus the three cluster core groups, which we believe are tracing true structures. 
\item We spectroscopically confirm the presence of a filamentary structure spanning the area between clusters A and B (group 5) previously detected in the infrared by \citet{Coppin2012a} to be forming stars at a SFR $\simeq 900 M_\odot$ yr$^{-1}$ and building up significant stellar mass.  The filament is detected in the DS test and using the FOF method.
\end{list}

\noindent An investigation of the color-magnitude relation and color distribution of the galaxies in the different density environments of the supercluster was performed.
\begin{list}{$\bullet$}{\leftmargin=0.2in} 
\item Of the galaxies identified as belonging to the three core cluster groups, 77$\pm$13\% are red-sequence galaxies, with blue galaxies making up only 23$\pm$7\% of the population. 
\item In the $\ge$5 member groups with the unique filamentary structure excluded, 43$\pm$12\% of the galaxies fall on the red-sequence with over half of the population remaining in the blue cloud.  This suggests that the group galaxies have undergone some processing to build up their red-sequence, though the remaining high fraction of blue galaxies would suggest less processing than the core galaxies.  
\item The filamentary structure between clusters A and B has 8 of its 9 members below the red-sequence, with three distinct peaks in color: one at the bluest colors; one in the green valley; and a small peak in the red-sequence of the color-magnitude diagram.  Despite the small numbers we propose the `green' galaxies may belong to the transient population identified by \citet{Balogh2011} to be migrating to the red-sequence.
\end{list}

\noindent We construct stacked LFs for the core and $\ge$5 member FOF groups from the spectroscopically weighted members to estimate average group richness and masses using two different methods: the Halo Occupation Distribution (HOD) method and the maxBCG richness estimators.
\begin{list}{$\bullet$}{\leftmargin=0.2in} 
\item The LF derived masses from the stacked core groups are $\sim 0.9-6 \times 10^{14} M_\odot$, consistent with the $\sim~10^{14.5}-10^{14.9} M_\odot$ cluster masses estimates from the multiwavelength data. 
\item The average $\ge$5 member group masses from the two methods are an order of magnitude less massive than the core groups, at $\sim 4-5 \times 10^{13} M_\odot$.
\item Assigning cluster A as the host halo of the eventual merged cluster, the remaining groups account for 22\% of the galaxies currently confirmed as supercluster members.  Compared to studies using semi-analytical derived merger trees on simulated cluster data, this fraction is slightly higher than the fraction seen at this epoch in massive, present-day clusters but not unrealistic given that all of the studies found a large fraction of $\sim$ 30\% of surviving galaxies in present day clusters spent some time as members of $\sim 10^{13}~h^{-1}$\,$M_\odot$ groups.  
\end{list}

The mapped structure of the RCS\,2319+00 supercluster defined in this paper provides an abundance of different density environments in which to study the effect of the environment on galaxy evolution, from the isolated supercluster field galaxies, along infalling groups and filaments to the dense cluster cores.  Our vast catalog of multiwavelength data on this field will be used in conjunction with the spectroscopic catalog defined in this paper to study the environmental effects on our member galaxies (e.g.~a study of the star-formation and stellar mass in RCS\,2319+00;\,Faloon et al.\,2012, in preparation).  Though again we caution that the structure defined here is a lower limit on the possible groups present in the RCS\,2319+00 supercluster field and `isolated' supercluster galaxies may not actually be isolated given further spectroscopy.  The addition of high-quality, multiwavelength photometric redshifts and supplementary spectroscopy will aid in limiting the mis-identification of the different density regions within the supercluster and add to our understanding of this unique, compact, high-redshift supercluster.  

\acknowledgments

We thank the anonymous referee for their constructive comments and suggestions which improved the paper.  We would like to thank the University of Colorado, Boulder redshift finding team for their aid in finding the IMACS redshifts: Tahlia De Maio, Rebecca Mickol, Garrett Lance Ebelke, and Sean Michael Peneyra.  We would also like to thank Kentaro Aoki from the Subaru Telescope for his invaluable assistance with the FMOS observations.  A.J.F.\,acknowledges the support of an NSERC PGS-D scholarship.  T.M.A.W.\,is supported by of the NSERC Discovery Grant and the FQRNT Nouveaux Chercheurs programs.  R.Y. acknowledges support by Ontario Postdoctoral Fellowship. D.G.G.\,acknowledges support from the National Research Foundation of South Africa. J.G.\,is supported by the NSERC Banting Postdoctoral Fellowship program. L.F.B.\,is partially funded by Centro de Astrof\'isica FONDAP and by proyecto FONDECYT 1120676. H.K.C.Y.\,acknowledges support from an NSERC Discovery Grant and a Tire 1 Canada Research Chair.

{\it Facilities:} \facility{Magellan:Baade (IMACS)}, \facility{VLT (VIMOS, FORS2)}, \facility{Subaru (FMOS)}, \facility{Gemini:Gillett (GMOS)}.

\bibliographystyle{apj}
\bibliography{RCS2319+00_structurePaper}

\begin{thebibliography}{88}
\expandafter\ifx\csname natexlab\endcsname\relax\def\natexlab#1{#1}\fi

\bibitem[{Bahcall(1988)}]{Bahcall1988}
Bahcall, N. 1988, ARA\&A, 26, 631

\bibitem[{Bahcall \& Soneira(1983)}]{Bahcall1983}
Bahcall, N., \& Soneira, R. 1983, ApJ, 270, 20

\bibitem[{Balogh {et~al.}(2011)Balogh, Mcgee, Wilman, Finoguenov, Parker,
  Connelly, Mulchaey, Bower, Tanaka, \& Giodini}]{Balogh2011}
Balogh, M.~L., Mcgee, S.~L., Wilman, D.~J., {et~al.} 2011, MNRAS, 412, 2303

\bibitem[{Baum(1959)}]{Baum1959}
Baum, W. 1959, PASP, 71, 106

\bibitem[{Becker {et~al.}(2007)Becker, Mckay, Koester, Wechsler, Rozo, Evrard,
  Johnston, Sheldon, Annis, Lau, Nichol, \& Miller}]{Becker2007}
Becker, M.~R., Mckay, T.~A., Koester, B., {et~al.} 2007, ApJ, 669, 905

\bibitem[{Beers {et~al.}(1990)Beers, Flynn, \& Gebhardt}]{Beers1990}
Beers, T., Flynn, K., \& Gebhardt, K. 1990, AJ, 100, 32

\bibitem[{Benson(2001)}]{Benson2001}
Benson, A.~J. 2001, MNRAS, 325, 1039

\bibitem[{Berlind \& Weinberg(2002)}]{Berlind2002}
Berlind, A.~A., \& Weinberg, D.~H. 2002, ApJ, 575, 587

\bibitem[{Berrier {et~al.}(2009)Berrier, Stewart, Bullock, Purcell, Barton, \&
  Wechsler}]{Berrier2009}
Berrier, J.~C., Stewart, K.~R., Bullock, J.~S., {et~al.} 2009, ApJ, 690, 1292

\bibitem[{Binggeli(1982)}]{Binggeli1982}
Binggeli, B. 1982, A\&A, 107, 338

\bibitem[{Blanton \& Roweis(2007)}]{Blanton2007}
Blanton, M., \& Roweis, S. 2007, AJ, 133, 734

\bibitem[{Blindert(2006)}]{Blindert2006}
Blindert, K. 2006, PhD thesis

\bibitem[{Bond {et~al.}(1996)Bond, Kofman, \& Pogosyan}]{Bond1996}
Bond, J., Kofman, L., \& Pogosyan, D. 1996, Nature, 380, 604

\bibitem[{Bower {et~al.}(1992)Bower, Lucey, \& Ellis}]{Bower1992}
Bower, R.~G., Lucey, J., \& Ellis, R. 1992, MNRAS, 254, 601

\bibitem[{Brodwin {et~al.}(2010)Brodwin, Ruel, Ade, Aird, Andersson, Ashby,
  Bautz, Bazin, Benson, Bleem, Carlstrom, Chang, Crawford, Crites, de~Haan,
  Desai, Dobbs, Dudley, Fazio, Foley, Forman, Garmire, George, Gladders,
  Gonzalez, Halverson, High, Holder, Holzapfel, Hrubes, Jones, Joy, Keisler,
  Knox, Lee, Leitch, Lueker, Marrone, McMahon, Mehl, Meyer, Mohr, Montroy,
  Murray, Padin, Plagge, Pryke, Reichardt, Rest, Ruhl, Schaffer, Shaw,
  Shirokoff, Song, Spieler, Stalder, Stanford, Staniszewski, Stark, Stubbs,
  Vanderlinde, Vieira, Vikhlinin, Williamson, Yang, Zahn, \&
  Zenteno}]{Brodwin2010}
Brodwin, M., Ruel, J., Ade, P. a.~R., {et~al.} 2010, ApJ, 721, 90

\bibitem[{Carlberg {et~al.}(1996)Carlberg, Yee, Ellingson, Abraham, Gravel,
  Morris, \& Pritchet}]{Carlberg1996}
Carlberg, R., Yee, H., Ellingson, E., {et~al.} 1996, ApJ, 462, 32

\bibitem[{Carlberg {et~al.}(1997)Carlberg, Yee, Ellingson, Morris, Abraham,
  Gravel, Pritchet, Smecker-Hane, Hartwick, Hesser, \& Others}]{Carlberg1997}
---. 1997, ApJL, 485, L13

\bibitem[{Cohn(2012)}]{Cohn2012}
Cohn, J.~D. 2012, MNRAS, 419, 1017

\bibitem[{Coppin {et~al.}(2012)Coppin, Geach, Webb, Faloon, Yan, O’Donnell,
  Ouellette, Egami, Ellingson, Gilbank, Hicks, Barrientos, Yee, \&
  Gladders}]{Coppin2012a}
Coppin, K. E.~K., Geach, J.~E., Webb, T. M.~A., {et~al.} 2012, ApJ, 749, L43

\bibitem[{Davis {et~al.}(2002)Davis, Faber, Newman, Phillips, Ellis, Steidel,
  Conselice, Coil, Finkbeiner, Koo, \& Others}]{Davis2002}
Davis, M., Faber, S., Newman, J., {et~al.} 2002, Arxiv preprint
  astro-ph/0209419

\bibitem[{{De Lucia} {et~al.}(2012){De Lucia}, Weinmann, Poggianti,
  Arag\'{o}n-Salamanca, \& Zaritsky}]{DeLucia2012}
{De Lucia}, G., Weinmann, S., Poggianti, B.~M., Arag\'{o}n-Salamanca, A., \&
  Zaritsky, D. 2012, MNRAS, 1292, 1277

\bibitem[{{De Lucia} {et~al.}(2004){De Lucia}, Poggianti, Aragon-Salamanca,
  Clowe, Hallidar, Jablonka, Milvang-Jensen, Pello, Poirier, Rudnick, Saglia,
  Simard, \& White}]{DeLucia2004}
{De Lucia}, G., Poggianti, B., Aragon-Salamanca, A., {et~al.} 2004, ApJ, 610,
  L77

\bibitem[{Dressler(1980)}]{Dressler1980}
Dressler, A. 1980, ApJ, 236, 351

\bibitem[{Dressler \& Shectman(1988)}]{Dressler1988}
Dressler, A., \& Shectman, S.~A. 1988, AJ, 95, 985

\bibitem[{Egami {et~al.}(2010)Egami, Rex, Rawle, P\'{e}rez-Gonz\'{a}lez,
  Richard, Kneib, Schaerer, Altieri, Valtchanov, Blain, Fadda, Zemcov, Bock,
  Boone, Bridge, Clement, Combes, Dessauges-Zavadsky, Dowell, Ilbert, Ivison,
  Jauzac, Lutz, Metcalfe, Omont, Pell\'{o}, Pereira, Rieke, Rodighiero, Smail,
  Smith, Tramoy, Walth, van~der Werf, \& Werner}]{Egami2010}
Egami, E., Rex, M., Rawle, T.~D., {et~al.} 2010, A\&A, 518, L12

\bibitem[{Einasto {et~al.}(2002)Einasto, Einasto, Saar, \&
  Tucker}]{Einasto2002}
Einasto, J., Einasto, M., Saar, E., \& Tucker, D. 2002, A\&A, 443, 425

\bibitem[{Einasto {et~al.}(2011)Einasto, Liivam\"{a}gi, Saar, Einasto, Tempel,
  Tago, \& Mart\'{\i}nez}]{Einasto2011}
Einasto, M., Liivam\"{a}gi, L.~J., Saar, E., {et~al.} 2011, A\&A, 535, A36

\bibitem[{Einasto {et~al.}(2007)Einasto, Einasto, Tago, Saar, Liivam,
  Liivam\"{a}gi, J\~{o}eveer, H\"{u}tsi, Hein\"{a}m\"{a}ki, M\"{u}ller, \&
  Tucker}]{Einasto2008}
Einasto, M., Einasto, J., Tago, E., {et~al.} 2007, ApJ, 464, 851

\bibitem[{Einasto {et~al.}(2012)Einasto, Vennik, Nurmi, Tempel, Ahvensalmi,
  Tago, Liivam\"{a}gi, Saar, Hein\"{a}m\"{a}ki, Einasto, \&
  Mart\'{\i}nez}]{Einasto2012}
Einasto, M., Vennik, J., Nurmi, P., {et~al.} 2012, A\&A, 540, A123

\bibitem[{Evrard {et~al.}(2008)Evrard, Bialek, Busha, White, Habib, Heitmann,
  Warren, Rasia, Tormen, Moscardini, Power, Jenkins, \& Gao}]{Evrard2008}
Evrard, A.~E., Bialek, J., Busha, M., {et~al.} 2008, ApJ, 672, 122

\bibitem[{Farrens {et~al.}(2011)Farrens, Abdalla, Cypriano, Sabiu, \&
  Blake}]{Farrens2011}
Farrens, S., Abdalla, F.~B., Cypriano, E.~S., Sabiu, C., \& Blake, C. 2011,
  MNRAS, 417, 1402

\bibitem[{Gal {et~al.}(2008)Gal, Lemaux, Lubin, Kocevski, Squires, \&
  Al}]{Gal2008}
Gal, R.~R., Lemaux, B.~C., Lubin, L.~M., {et~al.} 2008, ApJ, 684, 933

\bibitem[{Gal \& Lubin(2004)}]{Gal2004}
Gal, R.~R., \& Lubin, L.~M. 2004, ApJ, 607, L1

\bibitem[{Gilbank {et~al.}(2008{\natexlab{a}})Gilbank, Yee, Ellingson, Hicks,
  Gladders, Barrientos, \& Keeney}]{Gilbank2008}
Gilbank, D., Yee, H., Ellingson, E., {et~al.} 2008{\natexlab{a}}, ApJL, 677,
  L89

\bibitem[{Gilbank {et~al.}(2008{\natexlab{b}})Gilbank, Yee, Ellingson,
  Gladders, Loh, Barrientos, \& Barkhouse}]{Gilbank2008a}
Gilbank, D.~G., Yee, H. K.~C., Ellingson, E., {et~al.} 2008{\natexlab{b}}, ApJ,
  673, 742

\bibitem[{Gladders {et~al.}(2003)Gladders, Hoekstra, Yee, Hall, \&
  Barrientos}]{Gladders2003}
Gladders, M., Hoekstra, H., Yee, H., Hall, P., \& Barrientos, L. 2003, ApJ,
  593, 48

\bibitem[{Gladders \& Yee(2005)}]{Gladders2005}
Gladders, M., \& Yee, H. 2005, ApJS, 157, 1

\bibitem[{Goto {et~al.}(2002)Goto, Okamura, Mckay, Bahcall, Annis, Bernardi,
  Brinkmann, Gomez, Hansen, Kim, Sekigughi, \& Sheth}]{Goto2002}
Goto, T., Okamura, S., Mckay, T.~A., {et~al.} 2002, PASJ, 515

\bibitem[{Goto {et~al.}(2005)Goto, Postman, Cross, Illingworth, Tran, Magee,
  Franx, Benı, Bouwens, Demarco, Ford, Homeier, Martel, Menanteau, Clampin,
  Hartig, Ardila, Bartko, Blakeslee, Bradley, Broadhurst, Brown, Burrows,
  Cheng, Feldman, Golimowski, Gronwall, Holden, \& Infante}]{Goto2005}
Goto, T., Postman, M., Cross, N. J.~G., {et~al.} 2005, ApJ, 621, 188

\bibitem[{Gralla {et~al.}(2011)Gralla, Sharon, Gladders, Marrone, Barrientos,
  Bayliss, Bonamente, Bulbul, Carlstrom, Culverhouse, Gilbank, Greer, Hasler,
  Hawkins, Hennessy, Joy, Koester, Lamb, Leitch, Miller, Mroczkowski, Muchovej,
  Oguri, Plagge, Pryke, \& Woody}]{Gralla2011}
Gralla, M.~B., Sharon, K., Gladders, M.~D., {et~al.} 2011, ApJ, 737, 74

\bibitem[{Gregory \& Thompson(1978)}]{Gregory1978}
Gregory, S., \& Thompson, L. 1978, ApJ, 222, 784

\bibitem[{Gregory {et~al.}(1981)Gregory, Thompson, \& Tifft}]{Gregory1981}
Gregory, S., Thompson, L., \& Tifft, W. 1981, ApJ, 243, 411

\bibitem[{Hansen {et~al.}(2005)Hansen, McKay, Wechsler, Annis, Sheldon, \&
  Kimball}]{Hansen2005}
Hansen, S.~M., McKay, T.~a., Wechsler, R.~H., {et~al.} 2005, ApJ, 633, 122

\bibitem[{Hansen {et~al.}(2009)Hansen, Sheldon, Wechsler, \&
  Koester}]{Hansen2009}
Hansen, S.~M., Sheldon, E.~S., Wechsler, R.~H., \& Koester, B.~P. 2009, ApJ,
  699, 1333

\bibitem[{Hicks {et~al.}(2008)Hicks, Ellingson, Bautz, Cain, Gilbank, Gladders,
  Hoekstra, Yee, \& Garmire}]{Hicks2008}
Hicks, A., Ellingson, E., Bautz, M., {et~al.} 2008, ApJ, 680, 1022

\bibitem[{Hilton {et~al.}(2009)Hilton, Stanford, Stott, Collins, Hoyle,
  Davidson, Hosmer, Kay, Liddle, Lloyd-Davies, Mann, Mehrtens, Miller, Nichol,
  Romer, Sabirli, Sahl\'{e}n, Viana, West, Barbary, Dawson, Meyers, Perlmutter,
  Rubin, \& Suzuki}]{Hilton2009}
Hilton, M., Stanford, S.~A., Stott, J.~P., {et~al.} 2009, ApJ, 697, 436

\bibitem[{Huchra \& Geller(1982)}]{Huchra1982}
Huchra, J., \& Geller, M. 1982, ApJ, 257, 423

\bibitem[{Iwamuro {et~al.}(2012)Iwamuro, Moritani, Yabe, Sumiyoshi, Kawate,
  Tamura, Akiyama, Kimura, Takato, Tait, \& Others}]{Iwamuro2011}
Iwamuro, F., Moritani, Y., Yabe, K., {et~al.} 2012, PASJ, 64, 59

\bibitem[{Jee {et~al.}(2011)Jee, Dawson, Hoekstra, Perlmutter, Rosati, Brodwin,
  Suzuki, Koester, Postman, Lubin, Meyers, Stanford, Barbary, Barrientos,
  Eisenhardt, Ford, Gilbank, Gladders, Gonzalez, Harris, Huang, Lidman, Rykoff,
  Rubin, \& Spadafora}]{Jee2011}
Jee, M.~J., Dawson, K.~S., Hoekstra, H., {et~al.} 2011, ApJ, 737, 59

\bibitem[{Johnston {et~al.}(2007)Johnston, Sheldon, Wechsler, Rozo, Koester,
  Frieman, Mckay, Evrard, Becker, \& Annis}]{Johnston2007}
Johnston, D., Sheldon, E.~S., Wechsler, R.~H., {et~al.} 2007, Arxiv preprint
  arXiv: 0709.1159v1

\bibitem[{Kimura {et~al.}(2010)Kimura, Maihara, Iwamuro, Akiyama, Tamura,
  Dalton, Takato, Tait, Ohta, Eto, \& Others}]{Kimura2010}
Kimura, M., Maihara, T., Iwamuro, F., {et~al.} 2010, PASJ, 62, 1135

\bibitem[{Kolokotronis {et~al.}(2002)Kolokotronis, Basilakos, \&
  Plionis}]{Kolokotronis2002}
Kolokotronis, V., Basilakos, S., \& Plionis, M. 2002, MNRAS, 331, 1020

\bibitem[{Larson {et~al.}(2011)Larson, Dunkley, Hinshaw, Komatsu, Nolta,
  Bennett, Gold, Halpern, Hill, Jarosik, Kogut, Limon, Meyer, Odegard, Page,
  Smith, Spergel, Tucker, Weiland, Wollack, \& Wright}]{Larson2011}
Larson, D., Dunkley, J., Hinshaw, G., {et~al.} 2011, ApJS, 192, 16

\bibitem[{{Le Fevre} {et~al.}(2003){Le Fevre}, Saisse, Mancini, Brau-Nogue,
  Caputi, Castinel, D'Odorico, Garilli, Kissler-Patig, Lucuix, \&
  Others}]{LeFevre2003}
{Le Fevre}, O., Saisse, M., Mancini, D., {et~al.} 2003, Instrument Design and
  Performance for Optical/Infrared Ground-Based Telescopes (Eds. Masanori Iye
  AFM Moorwood) Proc. SPIE, 4841, 1670

\bibitem[{Lemaux {et~al.}(2012)Lemaux, Gal, Lubin, Kocevski, Fassnacht,
  McGrath, Squires, Surace, \& Lacy}]{Lemaux2012}
Lemaux, B.~C., Gal, R.~R., Lubin, L.~M., {et~al.} 2012, ApJ, 745, 106

\bibitem[{Lemson \& the Virgo~Consortium(2006)}]{Lemson2006}
Lemson, G., \& the Virgo~Consortium. 2006, Arxiv preprint arXiv:
  astro-ph/0608019v2

\bibitem[{Liivam\"{a}gi {et~al.}(2012)Liivam\"{a}gi, Tempel, \&
  Saar}]{Liivamagi2012}
Liivam\"{a}gi, L.~J., Tempel, E., \& Saar, E. 2012, A\&A, 539, A80

\bibitem[{Lin {et~al.}(2004)Lin, Mohr, \& Stanford}]{Lin2004}
Lin, Y.-T., Mohr, J., \& Stanford, S. 2004, ApJ, 610, 745

\bibitem[{Lin {et~al.}(2006)Lin, Mohr, Gonzalez, \& Stanford}]{Lin2006}
Lin, Y.-t., Mohr, J.~J., Gonzalez, A.~H., \& Stanford, S.~A. 2006, ApJ, 650,
  L99

\bibitem[{Lopez-Cruz {et~al.}(2004)Lopez-Cruz, Barkhouse, \&
  Yee}]{Lopez-Cruz2004}
Lopez-Cruz, O., Barkhouse, W.~A., \& Yee, H. K.~C. 2004, ApJ, 614, 679

\bibitem[{Lubin {et~al.}(2000)Lubin, Brunner, Metzger, Postman, \&
  Oke}]{Lubin2000}
Lubin, L. M.~L., Brunner, R., Metzger, M. M.~R., Postman, M., \& Oke, J. 2000,
  ApJL, 531, L5

\bibitem[{Markwardt(2009)}]{Markwardt2009}
Markwardt, C.~B. 2009, in ASPCS, Vol. 411, 251--254

\bibitem[{McGee {et~al.}(2009)McGee, Balogh, Bower, Font, \&
  McCarthy}]{McGee2009}
McGee, S.~L., Balogh, M.~L., Bower, R.~G., Font, A.~S., \& McCarthy, I.~G.
  2009, MNRAS, 400, 937

\bibitem[{Mei {et~al.}(2009)Mei, Holden, Blakeslee, Ford, Franx, Homeier,
  Illingworth, Jee, Overzier, Postman, Rosati, {Van der Wel}, \&
  Bartlett}]{Mei2009}
Mei, S., Holden, B.~P., Blakeslee, J.~P., {et~al.} 2009, ApJ, 690, 42

\bibitem[{Nakata {et~al.}(2005)Nakata, Kodama, Shimasaku, Mamoru, Furusawa,
  Hamabe, Kimura, Komiyama, Miyazaki, Okamura, Ouchi, Sekiguchi, Ueda, Yagi, \&
  Yasuda}]{Nakata2005}
Nakata, F., Kodama, T., Shimasaku, K., {et~al.} 2005, MNRAS, 357, 1357

\bibitem[{Noble {et~al.}(2012)Noble, Webb, Ellingson, Faloon, Gal, Gladders,
  Hicks, Hoekstra, Hsieh, Ivison, Lemaux, Lubin, O’Donnell, \&
  Yee}]{Noble2012}
Noble, A.~G., Webb, T. M.~a., Ellingson, E., {et~al.} 2012, MNRAS, 419, 1983

\bibitem[{Oort(1983)}]{Oort1983}
Oort, J. 1983, ARA\&A, 21, 373

\bibitem[{Papovich {et~al.}(2010)Papovich, Momcheva, Willmer, Finkelstein,
  Finkelstein, Tran, Brodwin, Dunlop, Farrah, Khan, Lotz, McCarthy, McLure,
  Rieke, Rudnick, Sivanandam, Pacaud, \& Pierre}]{Papovich2010}
Papovich, C., Momcheva, I., Willmer, C. N.~a., {et~al.} 2010, ApJ, 716, 1503

\bibitem[{Pinkney {et~al.}(1996)Pinkney, Roettiger, Burns, \&
  Bird}]{Pinkney1996}
Pinkney, J., Roettiger, K., Burns, J.~O., \& Bird, C.~M. 1996, ApJS, 104, 1

\bibitem[{Postman {et~al.}(1998)Postman, Lubin, \& Oke}]{Postman1998}
Postman, M., Lubin, L., \& Oke, J. 1998, AJ, 116, 560

\bibitem[{Propris {et~al.}(2003)Propris, Colless, Driver, Al, \& {De
  Propris}}]{Propris2003}
Propris, R.~D., Colless, M., Driver, S., Al, E., \& {De Propris}, R. 2003,
  MNRAS, 342, 725

\bibitem[{Rosati {et~al.}(1999)Rosati, Stanford, Eisenhardt, Elston, Spinrad,
  Stern, \& Dey}]{Rosati1999}
Rosati, P., Stanford, S., Eisenhardt, P.~R., {et~al.} 1999, AJ, 118, 76

\bibitem[{Rudnick {et~al.}(2012)Rudnick, Tran, Papovich, Momcheva, \&
  Willmer}]{Rudnick2012}
Rudnick, G.~H., Tran, K.-V., Papovich, C., Momcheva, I., \& Willmer, C. 2012,
  The Astrophysical Journal, 755, 14

\bibitem[{Sarazin(2002)}]{Sarazin2002}
Sarazin, C.~L. 2002, in Merging Processes in Galaxy Clusters, ed. G.~{Feretti,
  L. and Gioia, I. M. and Giovannini} (Springer Netherlands), 1--38

\bibitem[{Schechter(1976)}]{Schechter1976}
Schechter, P. 1976, ApJ, 203, 297

\bibitem[{Seljak(2000)}]{Seljak2000}
Seljak, U. 2000, MNRAS, 318, 203

\bibitem[{Sifon {et~al.}(2012)Sifon, Menanteau, Hasselfield, Marriage, Hughes,
  \& Al.}]{Sifon2012}
Sifon, C., Menanteau, F., Hasselfield, M., {et~al.} 2012, Arxiv preprint
  arXiv:1201.0991v1

\bibitem[{Small {et~al.}(1998)Small, Ma, Sargent, \& Hamilton}]{Small1998}
Small, T., Ma, C., Sargent, W., \& Hamilton, D. 1998, ApJ, 492, 45

\bibitem[{Snyder {et~al.}(2012)Snyder, Brodwin, Mancone, Zeimann, Stanford,
  Gonzalez, Stern, Eisenhardt, Brown, Dey, Jannuzi, \& Perlmutter}]{Snyder2012}
Snyder, G.~F., Brodwin, M., Mancone, C.~M., {et~al.} 2012, ApJ, 756, 114

\bibitem[{Swinbank {et~al.}(2007)Swinbank, Edge, Smail, Stott, Bremer, Sato,
  van Breukelen, Jarvis, Waddington, Clewley, Bergeron, Cotter, Dye, Geach,
  Gonzalez-Solares, Hirst, Ivison, Rawlings, Simpson, Smith, Verma, \&
  Yamada}]{Swinbank2007}
Swinbank, a.~M., Edge, a.~C., Smail, I., {et~al.} 2007, MNRAS, 379, 1343

\bibitem[{Tanaka {et~al.}(2009)Tanaka, Finoguenov, Kodama, Koyama, Maughan, \&
  Nakata}]{Tanaka2009}
Tanaka, M., Finoguenov, A., Kodama, T., {et~al.} 2009, A\&A, 505, L9

\bibitem[{Tanaka {et~al.}(2005)Tanaka, Kodama, Arimoto, Okamura, Umetsu,
  Shimasaku, Tanaka, \& Yamada}]{Tanaka2005}
Tanaka, M., Kodama, T., Arimoto, N., {et~al.} 2005, MNRAS, 362, 268

\bibitem[{Visvanathan \& Sandage(1977)}]{Visvanathan1977}
Visvanathan, N., \& Sandage, A. 1977, ApJ, 216, 214

\bibitem[{Wray {et~al.}(2006)Wray, Bahcall, Bode, Boettiger, \&
  Hopkins}]{Wray2006}
Wray, J.~J., Bahcall, N.~A., Bode, P., Boettiger, C., \& Hopkins, P.~F. 2006,
  ApJ, 652, 907

\bibitem[{Xue \& Wu(2000)}]{Xue2000}
Xue, Y.-J., \& Wu, X.-P. 2000, ApJ, 538, 65

\bibitem[{Yee {et~al.}(1996)Yee, Ellingson, Abraham, Gravel, Carlberg,
  Smecker-Hane, Schade, \& Rigler}]{Yee1996}
Yee, H., Ellingson, E., Abraham, R., {et~al.} 1996, ApJS, 102, 289

\bibitem[{Yee {et~al.}(2000)Yee, Morris, Lin, Carlberg, Hall, Sawicki, Patton,
  Wirth, Ellingson, \& Shepherd}]{Yee2000}
Yee, H., Morris, S., Lin, H., {et~al.} 2000, ApJS, 129, 475

\bibitem[{Zenteno {et~al.}(2011)Zenteno, Song, Desai, Armstrong, Mohr, Ngeow,
  Barkhouse, Allam, Andersson, Bazin, Benson, Bertin, Brodwin, Buckley-Geer,
  Hansen, High, Lin, Lin, Liu, Rest, Smith, Stalder, Stark, Tucker, \&
  Yang}]{Zenteno2011}
Zenteno, a., Song, J., Desai, S., {et~al.} 2011, ApJ, 734, 3

\end{thebibliography}

\end{document}